# Recommendation with Generative Models

**Alet Heezemans**

now publishers, Inc.

alet.heezemans@nowpublishers.com

**Mike Casey**

now publishers, Inc.

mike.casey@nowpublishers.com

now

the essence of knowledge

Boston — Delft

# Contents

# Recommendation with Generative Models

Yashar Deldjoo[1], Zhankui He[2], Julian McAuley[3], Anton Korikov[4],
Scott Sanner[5], Arnau Ramisa[6], Rene Vidal[7],
Maheswaran Sathiamoorthy[8], Atoosa Kasrizadeh[9], Silvia Milano[10]
and Francesco Ricci[11]

[1]*Polytechnic University of Bari, Italy; deldjooy@acm.org*
[2]*Google DeepMind, USA; zhankui@google.com*
[3]*University of San Diego, California, USA; jmcauley@eng.ucsd.edu*
[4]*University of Toronto, Canada; korikov@mie.utoronto.ca*
[5]*University of Toronto, Canada; ssanner@mie.utoronto.ca*
[6]*Amazon\*, USA; aramisay@amazon.com*
[7]*Amazon\*, USA; rvidal@amazon.com*
[8]*BespokeLabs.AI, USA; mahesh@bespokelabs.ai*
[9]*University of Edinburgh, UK; atoosa.kasirzadeh@ed.ac.uk*
[10]*University of Exeter and LMU Munich; s.milano@exeter.ac.uk*
[11]*Free University of Bozen-Bolzano, Italy; fricci@unibz.it*

---

[*]This work does not relate to the author's position at Amazon.

---

# 1

## Introduction

### 1.1 Context.

Generative models (GM) enable the construction of Artificial Intelligence applications that not only make decisions based on data but also "generate" new instances of a given data set by learning its statistical distribution and sampling from it. The use of generative technologies to generate new instances of data is particularly powerful and has led to applications in many fields of artificial intelligence (AI), such as image generation (Harshvardhan *et al.*, 2020), text synthesis (Li *et al.*, 2018c), and music composition (Mittal *et al.*, 2021; Yang and Lerch, 2020).

Recently, generative models have gained prominence in machine and deep learning, thanks to the introduction of new modeling approaches such as generative adversarial networks (GANs) (Goodfellow *et al.*, 2014), variational autoencoders (VAEs) (Kingma and Welling, 2013), diffusion models (Sohl-Dickstein *et al.*, 2015; Ho *et al.*, 2020), and transformer-based architectures such as GPT and other large language models (LLMs) (Wei *et al.*, 2022a; Bubeck *et al.*, 2023). These modeling approaches have enabled noteworthy breakthroughs or applications such as photo-realistic image generation (Diffusion) as in Saharia *et al.* (2022) or conversational bots based on natural language processing (Transformers) as in ChatGPT (Liu *et al.*, 2023b).

These advancements fall under the umbrella of deep generative models (DGMs), which combine conventional generative probabilistic models that capture underlying data distributions with deep neural networks (DNNs). DGMs have demonstrated remarkable capabilities in generating high-quality synthetic data, enhancing natural language understanding, and exhibiting emergent capabilities in in-context learning and few-shot generalization (Wei *et al.*, 2022a; Bubeck *et al.*, 2023). The core strength of DGMs lies in their ability to model and sample from the data distribution they are trained on and to use that for various inferential purposes.

Another important area of AI is Recommender Systems (RS). These may be embedded into eCommerce web sites, where they can interact with online users and suggest to them selected and personalized subsets of items, taken from a typically large catalog. They can help users to tame information overload and support more effective decision-making. They can also aid markets to promote sales and to discover new and less popular items in their catalog.

Generative models have been used in the development of RSs, to uncover relationships and patterns in items' consumption that generalize well to new data. This has enhanced the system's ability to provide accurate, diverse, and personalized recommendations. Recommender systems that integrate ideas from both generative AI and more traditional recommender systems are referred to as *Gen-RecSys* in this book (Deldjoo *et al.*, 2024a). Gen-RecSys can be distinguished on the basis of the nature of the outputs they produce:

- **Structured Outputs:** These systems generate recommendations that may be far more structured than a single item. Hence, the recommendation output may be structured as a vector, a bundle, or a sequence of items, providing users with a complementary set of recommendations (cf. Chapter 3).

- **Text Generation:** These systems may produce item recommendations in the form of textual content. Text can be conversational requests/responses or explanations. They overall produce more natural and engaging interactions with their users (cf. Chapter 4).

- **Image and Multimedia Generation:** These systems may generate visual content such as virtual try-ons, fashion designs, personalized visual advertisements, and contextual product visualizations. Furthermore, they can produce audio or video content that enhances the overall interaction experience of their users (cf. Chapter 5).

In practice our definition of Gen-RecSys is quite broad, and any system that combines generative AI and recommender systems falls under this category. In fact, the application goals of Gen-RecSys are very similar to those of traditional (non-generative) RSs, that is, to be more personalized, diverse, controllable, and engaging, by using new models that have the capability to generate output and not only to precisely filter the already available information contained in large item catalogues (as in traditional RSs).

## 1.2   Generative Models in Recommender Systems.

Throughout this book, we explore various ways in which generative models can enhance or improve RSs, which are summarized as follows. Readers are referred to Section 1.7 for a more fine-grained overview of GMs benefits and opportunities.

**Enhancing Core RS Tasks:**  Generative models have the potential to improve the quality of top-$k$ recommendations by approaching the recommendation problem through a *probabilistic* lens or by integrating knowledge-rich external data sources, such as LLMs. Hence GMs can address challenges in data-scarce scenarios, such as the cold start problem (Chae *et al.*, 2019; Yin *et al.*, 2023a; Zhou *et al.*, 2024; Wang *et al.*, 2022e).

For example, generative models like VAES (Kingma and Welling, 2013) have been shown to enhance top-$k$ recommendations by treating user-item interactions as probabilistic distributions within structured latent spaces (cf. Chapter 4). These models have been shown to outperform comparable non-generative collaborative filtering models, such as Matrix Factorization (MF) and Neural Matrix Factorization (NeuMF), in terms of the quality of recommendations. These latter models use fixed vectors to represent user-item interactions, which often fail to fully capture the complexity of user preferences.

Yet another example, Large Language Models (LLMs) can generate highly personalized recommendations by interpreting nuanced natural language descriptions of user preferences (Geng *et al.*, 2022; Sanner *et al.*, 2023; Zhou *et al.*, 2024). Moreover, emerging research by Rajput *et al.* (2024) introduces the concept of *generative retrieval,* where generative models enhance retrieval performance for items with little to no interaction history. This approach leverages semantic IDs and sequence-to-sequence models to autoregressively predict the identifiers of target items, improving recommendations in cold-start scenarios.

**Addressing Claimed but Not Fully Met RS Capabilities:** Generative models address tasks that conventional (non-generative) systems often claim to manage but do not handle as effectively, by enhancing user-system interactivity, flexibility and introducing innovative design concepts. For example, generative models enable more effective and personalized interactions by adapting in real-time to the variations of user preferences. This is particularly advantageous in *conversational recommendation scenarios* where non-generative models are (arguably) still unable to support an effective user/system dialogue. A case point is the GeneRec system by Wang *et al.* (2023c), which is capable of creating personalized micro-videos and re-purposing existing content into various styles and themes based on user instructions, delivering a highly customized user experience. Besides conversational scenarios, another promising application of generative models is in *cross-domain recommendation.* In fact, as discussed by Petruzzelli *et al.* (2024), LLMs can be leveraged to overcome data sparsity challenges, which have been addressed so far by cross-domain recommender systems but without producing tangible industrial applications. LLMs can more effectively help to tame the data sparsity problem by using their pre-existing knowledge to bridge different domains and generate personalized recommendations across them.

Moreover, traditional RSs have prioritized prediction accuracy over transparency. In contrast, generative models directly enable the genera-tion of explanations, such as motivations of the recommendations and counterfactual scenarios—illustrating what actions a user could take to receive different recommendations. For instance, by leveraging LLMs, these models can more easily support real-time critiquing (Amoukou and Brunel, 2022; Antognini and Faltings, 2021), allowing users to provide feedback that immediately influences the recommendations. As an example, Antognini and Faltings (2021) introduce the critique-explainable VAE, which uses user-generated key phrases and item preferences to deliver dynamic, personalized explanations that adjust in real time based on user critiques.

**Introducing Entirely New Capabilities:**  Generative models introduce groundbreaking capabilities to recommender systems, such as *on-demand content creation* and *whole-page generation*. With these models, a RS can generate new items based on user preferences or assemble coherent multi-item pages for enhanced interaction (cf. Chapter 3). For instance, a system could engage in a dialogue with a user to understand her preferences, and either generate a new item like a personalized micro-video or recommend a closely related existing item. Additionally, *multimodal capabilities* allow systems to understand and generate content across various formats like text, images, and videos. For example, a user can provide a visual input, such as a picture of a product, along with a textual modification (e.g., a dress like this but in red), and the system can either create the modified item or recommend the closest match (cf. Chapter 5). LLMs further enhance this by generating persuasive and context-aware interactions that simulate human opinions, making recommendations more engaging (cf. Chapter 4).

## 1.3   Types of Deep Generative Models (DGMS).

DGMs can be broadly used in RSs in the following operational modes:

**Directly trained models.**  This approach involves training generative models specifically for the recommendation task using commonly available standardized *user-item interaction* data (explicit or implicit), without relying on pre-trained large datasets. These models learn the probability distribution of items a user might prefer based on their previous interactions. These probability distributions can be conditioned on various features such as user preferences, demographic attributes, item features, temporal information, social connections, and geographical data to enhance the recommendations. Examples of directly trained models in recommender systems include VAE-CF (used for top-k recommendations), SVAE[1] (used for sequential recommendations), and IRGAN (used for negative sampling), see Chapter 3 for more examples.

**Pretrained Generative Models.** This approach employs models pretrained on diverse data (text, images, videos) to recognize complex patterns, relationships, and contexts, enabling the specialization to new tasks (Nazary *et al.*, 2023; Nazary *et al.*, 2024; Zhang *et al.*, 2024d). Examples of pre-trained models adapted for various data modalities include GPT-4 (Bubeck *et al.*, 2023), LLaMA (Gao *et al.*, 2023a), and CLIP (Radford *et al.*, 2021), which were initially designed for semi-supervised tasks. These models excel in:

1. **Zero- and Few-shot Learning**: Uses *in-context learning* for broad understanding without extra training, making LLMs effective in scenarios with minimal historical data, such as cold-start situations.

2. **Fine-Tuning**: Adjusts model parameters on specific data to provide personalized recommendations, such as fine-tuning a language model on user reading preferences to improve book recommendations.

3. **Retrieval-Augmented Generation (RAG)**: Combines information retrieval with generative modeling for contextually relevant outputs, enhancing recommendation accuracy.

4. **Feature Extraction for Recommendations**: Generates embeddings or token sequences to represent complex content for downstream recommendation tasks.

---

[1]SVAE: Sequential Variational Autoencoder

5. **Multimodal Approaches**: Integrates text, images, and videos to improve recommendations, such as a virtual shopping assistant using multiple data types to recommend clothing tailored to a user's preferences.

A few noteworthy case studies visited in this book are summarized in Table 1.2. A major observation about recommender systems with generative models (Gen-RecSys) is the **high diversity of techniques** introduced in recent years that operate under the generative paradigm. In fact, throughout this book, we mention more than 50 generative models. For instance, within the category of VAEs, there are numerous models (e.g., VAE-CF, C-VAE, RecVAE, ACVAE), each differing on the specific learned data distribution, e.g., whether or not the models are conditioned on factors such as user historical interactions, how priors are addressed, and the algorithm used for training (cf. Chapter 3). These diverse techniques have been developed to enhance the performance of Collaborative Filtering recommendations and address data sparsity issues among others. Moreover, they enable new applications, such as generating complex outputs like page-wise recommendations (Chen *et al.*, 2019c). Overall, the diversity of techniques within generative models holds significant potential for developing applications across various tasks.

## 1.4 Key definitions.

To help the reader to recall the key definitions used in this book, we have prepared the following table.

| Term | Definition |
|---|---|
| **Generative Model (GM)** | Probabilistic model that learns the data distribution of a dataset to generate (via sampling) new data samples that resemble the original data. |
| **Deep Generative Model (DGM)** | A subclass of generative models that use deep learning techniques to enhance generative capabilities, hence, capable of learning complex patterns in large datasets. Examples include GANs and VAEs. |

| Term | Definition |
|------|------------|
| **Large Language Model (LLM)** | Transformer-based models designed for natural language processing tasks, capable of generating coherent and contextually relevant text. Examples include OpenAI's GPT-4 and Google's PaLM, which are used for tasks such as text generation, summarization, and question answering. |
| **Vision- and Multimodal-Language Models (VLMs, MLMS)** | Models that integrate multiple data modalities, such as images and text, into a unified framework to perform tasks like image captioning, visual question answering, or text-to-image generation. Examples include CLIP, which aligns images and text for tasks like visual search, and DALL-E, which generates images from text prompts. |
| **Foundation Models (FMs)** | Large-scale models pre-trained on vast and diverse datasets that can be fine-tuned or adapted to perform specific tasks across different domains, such as natural language processing, vision, and beyond. These models exhibit generalization capabilities, enabling them to transfer knowledge across various applications. |
| **Recommender System (RS)** | Systems that analyze both explicit (e.g., ratings) and implicit (e.g., views, clicks) user feedback to suggest user relevant items, utilizing methods like collaborative filtering and content-based filtering. |
| **Recommender System with Generative Model (Gen-RecSys)** | Recommender Systems that incorporate generative models to enhance both predictive capabilities (e.g., top-k recommendations) and generative abilities (e.g., generating on-demand content). |

## 1.5 Guide to the Reader

This monograph is an **intermediate-level** guide on recommender systems using generative models, aimed at researchers, practitioners, students, and industry professionals. A basic understanding of recommender systems and machine learning concepts (e.g., collaborative filtering, supervised learning), as well as core concepts in NLP and Information Retrieval (IR) and multi-modal learning, might be helpful (e.g., dense retrieval, contrastive learning, variational autoencoders) for reading the technical Chapters 4 to 5). To support readers with varying expertise levels, Chapter 2 provides some basic definitions and results—from classical to generative models and their applications—and should be accessible by a large audience, making it also suitable for didactic purposes.

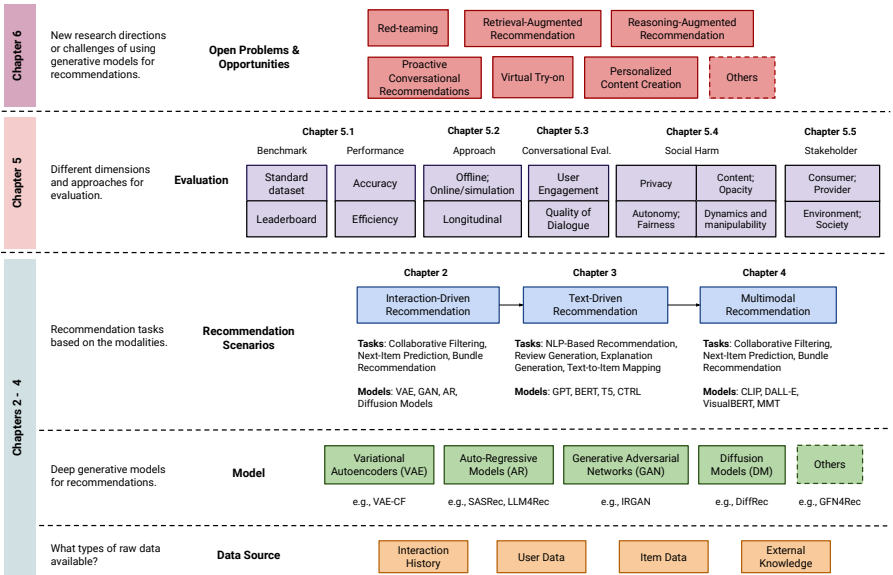

**Figure 1.1:** The scope of this book - Overview of the areas of interest in generative models in recommendation.

Nonetheless, this book is generally intended to be read by researchers interested to learn the state-of-the-art of generative recommender systems. While some techniques are still evolving, this book aims to capture the latest developments in the field. Readers are encouraged to read the chapters *sequentially* for a smooth understanding, except for Chapter 2, which can be skipped if the reader is already familiar with the most basic concepts. However, more advanced readers can also read each chapter independently based on their specific interests.

## 1.6 Scope

Several recent surveys have explored some aspects of generative models in recommender systems, such as GAN-based recommender systems by Deldjoo *et al.* (2021), examining training strategies for LLMs (Li *et al.*, 2023g), and discussing LLMs as recommendation engines (Wu *et al.*, 2023a). Additionally, Lin *et al.* (2023a) provide details on the adaptation of LLMs in recommendation tasks, while Fan *et al.* (2023b) provide an overview of LLMs emphasizing pre-training and fine-tuning strategies. In Huang *et al.*, 2024a the authors explore the use of foundation models, and in (Wang *et al.*, 2023c) the authors introduce GeneRec, a next-gen recommendation system leveraging AI generators. While the mentioned surveys offer important information, their scope is limited to specific methodological focuses, such as categories of models or training paradigms, like LLMs, and FMs, or specific models such as GANs.

With this book, we offer a wider perspective, treating various facets of Gen-RecSys and giving recommendations for implementing generative recommender systems, as illustrated in Figure 1.1 and Figure 1.2. In particular, we discuss these systems also from the perspectives of system design (Chapters 3, 4, and 5), evaluation for impact and harm (Chapters 6), and 7)), and highlight numerous tasks and applications.

## 1.7 Advantages and Opportunities in Gen-RecSys: Further Discussion

In this section, we provide more detailed description of the specific advantages of generative models for recommender systems, by referring to the rest of the book, as illustrated in Figure 1.2.

The content is structured around three topics: *Objectives and Application Scenarios*, *Diverse Outputs*, and *Model and Data Enhancement.*

### 1.7.1 Objective and Application Scenarios

#### Predictive Capabilities

Generative models can significantly enhance recommender systems by improving how they capture and utilize data. This book explores various approaches, from probabilistic methods such as VAEs, which model underlying data distributions for greater accuracy, to pre-trained models such as LLMs and multimodal foundation models (cf. Chapter 3 to 5). These models excel in optimizing complex outputs, including generating the entire recommendation lists or pages by directly learning item-item relationships. LLMs enable personalization by utilizing multi-turn dialogues where they can refine recommendations based on user feedback and preferences expressed in natural language. Additionally, multimodal generative learning can align and integrate diverse data types, enabling richer, personalized recommendations in visually-driven scenarios (e.g., food recommendation, fashion, e-commerce etc).

Another powerful capability enabled by generative models, particularly through LLMs, is their ability to adapt and personalize recommendations across different contexts rapidly. For instance, in an e-commerce scenario, an LLM can quickly be employed to tailor product recommendations based on a user's recent browsing history or real-time input during a chat interaction, without the need for extensive training. If a user initially searches for "summer dresses" and later shifts the focus to "formal evening wear," the LLM can seamlessly adjust its recommendations, offering relevant product options in both categories.

**Generative Capabilities**

Generative models offer capabilities that extend beyond recommendation prediction, enabling richer, more interactive user experiences.

These models can personalize individual items, create bundle or slate recommendations, and offer realistic previews like virtual try-on or in-context visualizations, as highlighted in Chapter 5. They also significantly enrich interactive and conversational recommendations, allowing for dynamic critiquing, preference negotiation, and multimodal interactive conversations, seamlessly integrating multiple types of user input and feedback (Chapter 4). Moreover, these models enable explanation and reasoning capabilities by generating personalized, factual, and visually enriched explanations, helping users understand and trust the recommender system.

**Supporting Functions**

Generative models also support persuasion and engagement by personalizing messages and interactions, thereby increasing conversion rates. These models offer user control, enabling users to modify preferences easily.

### 1.7.2 Output Space

In addition to the application scenarios discussed above, another perspective to consider for generative models is the nature of their output. Generative models extend beyond traditional (non-generative) recommendation systems by producing a wide range of outputs that vary in complexity, support for multi-modality, size, and interactivity. For example, in the context of personalized advertising (see Chapter 5 in the figure 1.2), generative models can create highly tailored ad content across different media formats, which traditional systems struggle to achieve due to their sensitivity to data limitations and modality constraints. Additionally, in multi-turn conversational recommendations (Chapter 4 and 5), generative models excel by offering dynamic, real-time interaction capabilities that are less prone to issues like data sparsity, making them useful for engaging and personalized user experiences.

### 1.7.3 Model and Data Enhancement

One of the key reasons for integrating generative models into recommender systems is their technical strength in model and data enhancement.

In terms of model enhancement, generative provide better latent representations by refining how the system captures and models user-item interactions. For instance, VAEs use probabilistic approaches to generate more expressive representations, improving both recommendation accuracy and the system's ability to handle sparse data scenarios. These models also play a role in regularizing and denoising data, which is critical in handling real-world noisy datasets, further ensuring that the recommendations generated are robust and contextually relevant.

On the data enhancement front, generative models augment existing

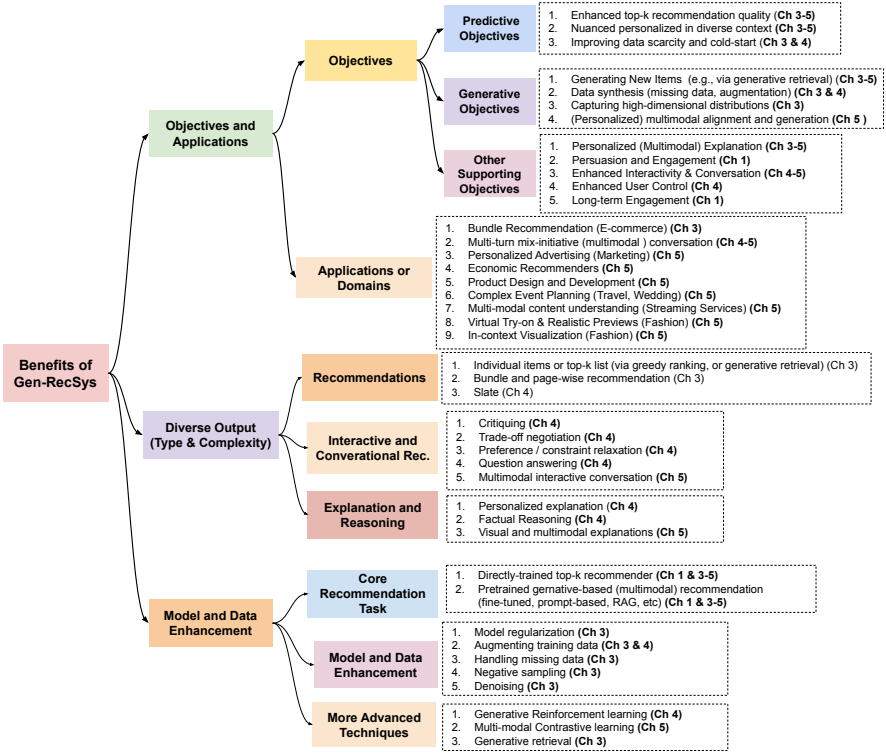

**Figure 1.2:** Advantages of recommender systems with generative models (Gen-RecSys) by considering: (i) Objectives and Applications, (ii) Diverse Outputs of Generative Models, (iii) Learning Mechanisms and Models

datasets, filling in missing data or generating synthetic interactions that can enhance training processes. As illustrated in Figure 1.2, these models can directly augment training data, making it more diverse and richer, leading to more accurate item predictions and better scoring functions for user-item interaction prediction, especially useful in cold-start scenarios, while enabling complex outputs like bundle and page-wise recommendations. Furthermore, the generative retrieval techniques employed by these models enable more complex output structures such as bundle and page-wise recommendations, which elevate the overall recommendation experience.

## 1.8 Organization of This Book

We can broadly look at the organization of this book as follows: introductory background topics (in Chapter 2), system design chapters (covering various models and techniques in Chapters 3, 4 to 5), and everything related to evaluation, risks, and harms (in Chapters 6, and 7).

- **Chapter 2: Foundations of Recommender Models**
  Provides an overview of traditional and generative recommender models, introducing key concepts and differences between discriminative and generative approaches. The chapter emphasizes the expanding role of generative models in creating complex outputs like personalized content, natural language explanations, and novel item designs. It also highlights foundational tasks within generative models, setting the stage for more advanced topics.

- **Chapter 3: ID-Based Models**
  Discusses generative models that utilize user-item interaction data to enhance recommendations. Topics include model architectures such as VAEs, GANs, and Diffusion models.

- **Chapter 4: LLM-Driven Models**
  Explores how large language models (LLMs) are leveraged for natural language recommendations, preference elicitation, and explanation generation.

- **Chapter 5: Multimodal Models**
  Focuses on models that integrate multiple data modalities (text, images, audio) for richer and more personalized recommendations.

- **Chapter 6: Evaluation Methods**
  Introduces new metrics and benchmarks specific to generative recommender systems, addressing challenges like output complexity and system-wide performance.

- **Chapter 7: Societal Harms and Risks**
  Analyzes the ethical considerations surrounding generative models, such as bias amplification, misinformation, and privacy concerns.

To help the reader further understand the practical applications of these generative models, we present a summary of select case studies in Table 1.2.

**Table 1.2:** Select Case Studies in Gen-RecSys: Recommender Systems using Generative Models

| **Collaborative Filtering** | |
| --- | --- |
| Standard Context-Based Collaborative Filtering | **Model:** VAE-CF<br>**Application:** Improves performance in sparse data scenarios using latent variables and normal priors to regularize autoencoders. |
| **Generative Retrieval** | |
| Generative Retrieval | **Model:** TIGER<br>**Application:** Represents items by semantic IDs and predicts sequences based on user history. |
| **Retrieval-Augmented Generation (RAG)** | |
| Retrieval-Augmented Generation | **Model:** RAG<br>**Application:** Combines retrieval systems with large language models for accurate, flexible recommendations. |
| **Query-Driven Recommendation** | |
| Natural Language Query-Driven Recommendation | **Model:** GPT-3<br>**Application:** Uses large-scale language models to interpret and respond to user queries in natural language. |
| **Conversational Assistants** | |
| Interactive Conversational Recommendation | **Model:** DialoGPT<br>**Application:** Enables dynamic and context-aware interactions with users through conversational agents. |
| Conversational Recommender Systems | **Model:** MLLM<br>**Application:** Integrates multimodal data handling capabilities in LLMs for rich conversational interactions. |
| **Virtual Try-On** | |
| Virtual Try-On for Fashion | **Model:** FashionGAN<br>**Application:** Uses GANs to visualize how different fashion items would look on users, enhancing the shopping experience. |

# 2

## Foundations

ABSTRACT

This chapter lays the foundational overview of traditional (often non-generative) recommender models as well as generative models. While some techniques discussed here will be detailed in subsequent chapters, the primary objective is to provide an overview of various techniques and methods to accommodate audiences with diverse backgrounds. Another goal is to serve as a resource for teaching purposes.

1The chapter begins by introducing and motivating the use of generative models in the context of recommendation systems, explaining their fundamental concepts and distinguishing them from discriminative models. It highlights the evolving scope of generative models in generating rich, complex, or structured outputs, such as personalized reviews, natural language explanations, and new item designs.

Overall the chapter aims to equip readers with an understanding of the foundational techniques in recommender systems, providing the space for exploring more advanced generative models and their applications in subsequent chapters.

---

Strictly speaking, *generative models* are models that estimate the distribution of objects in a dataset $X$ ('observables') given some label $y$ ('target'); *discriminative models,* in contrast, estimate (the probability of) a label given an observation:

$$\text{Generative:} \quad P(X|Y = y) \tag{2.1}$$

$$\text{Discriminative:} \quad P(Y|X = x). \tag{2.2}$$

More straightforwardly, discriminative models estimate a label $y$ given (the features associated with) an object $X$, while generative models estimate the distribution of $X$ given an observed label. For example, a discriminative model might be used to estimate the probability that an image contains a cat given its pixels, whereas a generative model would be used to generate (by sampling from the distribution of) pictures of cats.

Throughout this book, we will use the term 'generative model' a little more colloquially, to describe *methods that generate rich, complex, or structured outputs.* Whereas traditional ('discriminative') recommender systems might estimate the next item a user will purchase, or the rating they'll give, generative recommender systems might instead estimate the text of a user's review (Ni *et al.*, 2017), explain recommendations to a user in natural language (Ni *et al.*, 2019a), or suggest new item designs (Kang *et al.*, 2017).

Broadly speaking, generative recommender systems may be useful in applications such as the following:

1. Estimating more complex outputs than simple regressors or classifiers. This may range from models that generate multiple outputs simultaneously (see e.g. methods that recommend trades (Rappaz *et al.*, 2017), heteroscedastic regression (Ng *et al.*, 2017), etc.), to models that generate sets of items, such as clothing outfits

(Hsiao and Grauman, 2018) or bundle recommenders (Pathak *et al.*, 2017).[1]

2. Models that solve 'traditional' generative tasks with a personalization component. These include systems for personalized question answering (Wan and McAuley, 2016), or personalized image generation (Kang *et al.*, 2017).

3. Systems whose recommendation outputs are themselves complex, structured, outputs, e.g. an estimated heart-rate profile or a cooking recipe (Ni *et al.*, 2019b; Majumder *et al.*, 2019).

4. Methods that generate text for the sake of explaining or justifying recommendations (Ni *et al.*, 2019a).

5. Methods that use text (or otherwise) to facilitate more complex interactions with users, including conversational recommendation (Li *et al.*, 2018a; Kang *et al.*, 2019).

6. Methods that use generative models in the context of product design, i.e., to suggest (the characteristics of) potential new items (Kang *et al.*, 2017).

Below we attempt to outline approaches from these various categories, all of which will be presented in more depth later in the book. Our goal here is to gently introduce the common methodology and ideas involved to build generative recommender systems.

In Chapter 3 we will also briefly discuss approaches that use generative modeling techniques within discriminative frameworks, for example approaches that use Large Language Models as data processing and feature engineering tools to produce better item representations (which are then used for 'traditional' recommendation tasks downstream).

## 2.1 Recommendation Modalities

Before diving in to more complex recommendation methodologies, we will briefly revise the notation and methodology involved in traditional

---

[1]Though strictly speaking, these models may not use generative frameworks as in Equation (2.2) to generate sets.

**Name:** Pomberrytini;
**Ingredients:**
pomegranate-blueberry juice,
cranberry juice, vodka;

Combine all ingredients except for the ice in a blender or food processor. Process to make a smooth paste and then add the remaining vodka and blend until smooth. Pour into a chilled glass and garnish with a little lemon and fresh mint.

**A:** *Pours a very dark brown* with a *nice finger of tan head that produces a small bubble and leaves decent lacing* on the glass. **S:** Smells like a nut brown ale. It has *a slight sweetness* and a bit of a woody note and a little cocoa. The nose is *rather malty* with some chocolate and coffee. The taste is strong but not overwhelmingly sweet. The sweetness is overpowering, but not overwhelming and is a pretty strong *bitter* finish. **M:** Medium bodied with a slightly thin feel. **D:** *A good tasting beer. Not bad.*

**Figure 2.1:** A (personalized) generated recipe (left) and a product review (right). From (Majumder *et al.*, 2019; Ni and McAuley, 2018).

recommendation approaches. While we assume some passing familiarity with recommender systems, our goal here is to establish common notation and terminology; for a more in-depth treatment of recommendation fundamentals, see e.g. McAuley (2022).

### 2.1.1 Traditional Representations for Recommendation

At their core, recommender systems are models that describe *interactions between users and items;* the goal of modeling *interactions* is the main distinguishing feature between recommendation approaches and other types of machine learning.

A fundamental (discriminative) task of a recommender might be to estimate an outcome associated with a user/item interaction (for user $u \in U$ and item $i \in I$) such as predicting a rating or a click:

$$f : U \times I \to \mathbb{R}. \tag{2.3}$$

Learning such a model is generally cast as a *supervised learning* task; that is, we have a *training set* of historical user/item interactions from which we would like to estimate the above function.

**Representing data as matrices, sets, and graphs**

Most straightforwardly, training data for Equation (2.3) might simply be described using a set of tuples $(u, i, r)$ describing historical user interactions:

$$
\begin{array}{llll}
(\text{Scott}, & \textit{John Wick}, & 3) \\
(\text{Julian}, & \textit{John Wick}, & 5) \\
(\text{Julian}, & \textit{Harry Potter}, & 4) \quad \text{or} \\
(\text{Scott}, & \textit{Signs}, & 2) \\
& \vdots
\end{array}
\qquad
\begin{array}{llll}
(113, & 12, & 3) \\
(899, & 12, & 5) \\
(899, & 172, & 4) \\
(113, & 988, & 2) \\
& \vdots
\end{array}
\tag{2.4}
$$

where the latter consists of anonymized user and item IDs; such a dataset may be further augmented with side information such as a timestamp associated with each interaction, the text of a review, etc.

**Activities as matrices**   While the representation in Equation (2.4) may be a complete record of historical interactions, it does little to express the relationships among entries in the data. A more expressive way to represent the data may be as a matrix whose rows and columns are users and items, and whose entries are observed interactions (e.g. ratings):

$$
R = \left.
\begin{bmatrix}
\cdot & 4 & 3 & \cdot & \cdot \\
5 & \cdot & \cdot & 2 & 3 \\
\cdot & \cdot & 2 & \cdot & 4 \\
\cdot & 2 & \cdot & 3 & \cdot \\
\cdot & \cdot & 4 & \cdot & 1
\end{bmatrix}
\right\} \text{users.}
\tag{2.5}
$$

$$\underbrace{\phantom{xxxxxxxxxxx}}_{\text{items}}$$

Note that such a representation is largely conceptual: in a real dataset such a matrix would have millions of rows and columns, the vast majority of which would be empty (i.e., the user has never interacted with the corresponding item). However such a representation is useful for several models, e.g. those based on the principle of matrix factorization (Koren *et al.*, 2009).

**Activities as sets and graphs**   We will also frequently describe users and items as *sets*, i.e., a user may be described as a set of items they interacted with, or an item may be described as a set of users who interacted with it. Such sets can be extracted from the representation in Equation (2.5) as:

$$I_u = \{i \mid R_{u,i} \neq 0\}; \quad U_i = \{u \mid R_{u,i} \neq 0\}. \tag{2.6}$$

Occasionally, it may also be convenient to describe interactions as a (bipartite) graph $\mathcal{G}$, where users and items are nodes, and edges exist between nodes $u$ and $i$ whenever a user interacts with an item.[2]

**Activities as sequences**   A critical feature of users' interactions with items is the *order* in which they occur, as future activities may be heavily guided by recent interactions. Thus it may be useful to have a representation that naturally describes the *ordering* among interactions. Here, we might describe each user (or user ID) in terms of an ordered list of items they've interacted with:

$$u_{113} = [12, 988, 15, 141, \ldots]; \quad u_{899} \quad = [12, 172, 411, 311, \ldots]. \tag{2.7}$$

Note that the subscript in $u_{113}$ (for example) refers to a specific user (ID), while the corresponding list corresponds to an ordered list of item IDs.

In Chapter 4 we'll explore methods that adopt language models to recommendation tasks. In the context of language modeling, these models generally operate on documents, each of which is an ordered sequences of words; to adapt them for recommendation, documents become *users*, each of which is an ordered sequence of *items*. In this way, next word prediction, a commonly language modeling task, naturally maps to next item prediction.

## 2.2   Introduction to modeling approaches

Throughout out this book, we'll explore representations for many types of data, including interactions, sequences, text, images, etc. Here, we show some of the most common data modalities used in recommender systems, and explore some standard techniques for representing users and items that will reappear throughout the book.

---

[2]Though in terms of notation, the matrix representation in Equation (2.5) and the set representation in Equation (2.6) are already equivalent to an adjacency matrix or edge list, respectively.

### 2.2.1 Set-based recommendation

Before exploring recommendation approaches based on machine learning, it is worth briefly exploring 'heuristic' approaches based on item-to-item (or user-to-user) similarity. Such systems remain widely deployed, e.g. recommender systems of the form 'people who bought X also bought Y' are typically based on an appropriate definition of similarity between X and Y based on historical purchases.

Essentially our goal here is to define a similarity function

$$sim(i, j) \tag{2.8}$$

between items $i$ and $j$ (or between users $u$ and $v$) that measures whether those items are similar based on historical interactions.

In terms of our set-based definition, possibly the simplest example of such a similarity measure might be the Jaccard similarity:

$$\text{Jaccard}(i, j) = \frac{|U_i \cap U_j|}{|U_i \cup U_j|}; \tag{2.9}$$

this measure takes a value between 0 and 1: 1 indicates that the two items have been purchased by identical sets of users (denoted by $U_i$ and $U_j$); 0 indicates the two items have not been purchased by any common users. The purpose of the denominator in the above expression is to ensure that the measure is appropriately normalized such that it does not favor items with a large or small number of historical interactions.

While Equation (2.9) is designed to operate over sets (e.g. the set of users who have interacted with an item), interaction histories may include additional signals, such as star ratings, which might be helpful to assess similarity. Ideally, a similarity function should only regard items as being similar if they were consumed by similar users, *and those users agreed in their sentiment toward them.* To capture such a notion one might compute the angle between interaction vectors via the Cosine similarity:

$$\text{Cosine Similarity}(u, v) = \frac{R_u \cdot R_v}{|R_u| \cdot |R_v|}. \tag{2.10}$$

Note here that we are operating on vector-based (rather than set-based) interaction histories.

Such a definition may be further improved by appropriately *normalizing* scores based on user/item averages; roughly speaking, ratings of 3 and 5 should be oriented in opposite directions if a user's average rating is 4 stars. This type of normalization is achieved by the Pearson similarity:

Pearson Similarity$(u, v) =$

$$\frac{\sum_{i \in I_u \cap I_v}(R_{u,i} - \bar{R}_u)(R_{v,i} - \bar{R}_v)}{\sqrt{\sum_{i \in I_u \cap I_v}(R_{u,i} - \bar{R}_u)^2}\sqrt{\sum_{i \in I_u \cap I_v}(R_{v,i} - \bar{R}_v)^2}}. \quad (2.11)$$

**Similarity estimation versus prediction**   Below we'll discuss machine learning-based approaches to recommendation. In particular these approaches will focus on *prediction* (e.g. of a star rating) rather than similarity estimation. Generally, building such predictive models is the realm of machine learning; that is, a predictor is built by training a model to fit historical groundtruth observations as accurately as possible.

However, the distinction between similarity estimation and prediction is not so clear, and indeed, a good similarity function can also be used to build an accurate predictor. For example, a user's future rating of item may be predicted by taking a weighted average of their previous ratings:

$$r(u, i) = \frac{\sum_{j \in I_u \setminus \{i\}} R_{u,j} \cdot \text{Sim}(i, j)}{\sum_{j \in I_u \setminus \{i\}} \text{Sim}(i, j)}. \quad (2.12)$$

Here the weight is determined by the item similarity function $\text{Sim}(i, j)$ such that items $j$ are given higher weight if they are more similar to item $i$. Many such predictors could be defined, e.g. by weighting the average according to recency, interchanging users and items, choosing different similarity functions etc. In practice such models often serve as a simpler alternative to the complex machine learning models we'll explore throughout the remainder of this book.

### 2.2.2 Matrix factorization and latent-factor recommendation

The approaches described above, which operate by estimating similarity functions between users and items, are ultimately *heuristics*: while they may provide effective recommendations, at no point did we discuss a specific objective, i.e., a way to measure whether a set of recommendations is 'good'. Such approaches are examples of *memory-based* methods, i.e., they make predictions on the basis of sets (or lists, etc.) of items in users' histories.

*Model-based* methods, in contrast, attempt to distill users and items into (low-dimensional) representations, i.e., a collection of *model parameters*.

Perhaps the simplest instance of such a model is based on the principle of matrix factorization, which is typically referred to as a *latent factor* model. Here, each user is represented by a $K$-dimensional vector $\gamma_u$ and each item is represented by a $K$-dimensional vector $\gamma_i$. A user is assumed to have a high probability of interacting with an item (or giving it a high rating) if the user vector $\gamma_u$ is 'similar to' the item vector $\gamma_i$.

If the similarity operation between user and item vectors is captured by an inner product relationship ($\gamma_u \cdot \gamma_i$, then the interaction matrix $R$ can be written as the product of two low-rank[3] matrices:

$$\underbrace{\left[ \begin{array}{c} \\ \text{R} \\ \\ \end{array} \right]}_{|U| \times |I|} = \underbrace{\left[ \begin{array}{c} \\ \gamma_U \\ \\ \end{array} \right]}_{|U| \times K} \times \underbrace{\left[ \begin{array}{cc} \gamma_I^T & \end{array} \right]}_{K \times |I|} . \qquad (2.13)$$

Here $\gamma_U$ is a matrix describing *all* users; each row $\gamma_u$ describes a single user. Likewise each row $\gamma_i$ of $\gamma_I$ describes a single item. Thus the prediction for a single entry $R_{u,i}$ is given by the inner product:

$$R_{u,i} \simeq \gamma_u \cdot \gamma_i, \qquad (2.14)$$

i.e., a user is assumed to be compatible with an item if the item vector (roughly speaking, the item's 'properties') points in the same direction as the user vector (roughly speaking, the user's 'preferences' toward those properties).

---

[3]Specifically, rank $K$.

**Relationship to the Singular Value Decomposition**   The form of factorization in Equation (2.14) is reminiscent of the SVD, in which a matrix $M$ is factorized according to $M = U\Sigma V^T$, where $U$ and $V$ contain the left and right singular vectors and $\Sigma$ contains the singular values. Critically, the best rank-$K$ approximation to $M$ (in terms of the MSE) is given by taking the top-$K$ singular vectors from $U$ and $V$.

Thus, in principle, we might seek to estimate the matrices $\gamma_U$ and $\gamma_I$ by taking the SVD of $R$; however this turns out to be infeasible as (a) the matrix $R$ is prohibitively high-dimensional (e.g. millions of users and items); and (b) the matrix $R$ is *partially-observed* (i.e., most users don't interact with most items), such that the SVD isn't well-defined. Instead, as we discuss below, we'll *approximate* the SVD using gradient-based methods.

**Optimization**   Fitting the parameters in Equation (2.14) is generally accomplished by casting the problem as a supervised learning task. That is, the model parameters should be chosen such that the model's predictions are as close as possible to some groundtruth observations. One such objective is the *Mean Squared Error* (MSE), which says that $\gamma$ should be chosen to minimize average (or sum) of squared differences between predictions $f(u, i)$ and observations $R_{u,i}$:

$$\arg\min_{\gamma} \frac{1}{|R|} \sum_{(u,i) \in R} (f(u, i) - R_{u,i})^2. \qquad (2.15)$$

Though we leave discussion of the choice of the MSE (or other objectives) in the context of recommendation, learning strategies (e.g. gradient descent), and other issues such as model regularization, for more introductory texts (see e.g. McAuley (2022)).

**Bias terms**   The model in Equation (2.14) captures the basic features of latent-factor recommendation: the interaction matrix from Equation (2.13) is modeled by a low-rank approximation, in which each user and item is represented via a low dimensional vectors. While the model has the expressive power we need, such a model is often slightly expanded through the use of additional offset and bias terms, e.g.:

$$r(u, i) = \alpha + \beta_u + \beta_i + \gamma_u \cdot \gamma_i. \qquad (2.16)$$

Roughly speaking, user biases $\beta_u$ and item biases $\beta_i$ capture whether a user gives high ratings and whether an item receives high ratings, while $\alpha$ is a global offset.[4] These additional offset terms are typically added for the sake of model regularization: assuming that the basic function of a regularizer is to 'push' parameters toward zero, the above expression puts additional weight in the bias terms such that $\gamma$ can have zero mean.

**What happened to features?**

A remarkable property of the models presented above, including the latent factor model in Equation (2.16), or even traditional models based on item-to-item similarity, is that they make predictions without the use of item *features*. That is, the system stores no knowledge of whether a user is a female in their 30s, or whether the item they're considering is an action movie. Such attributes may exist within the latent representations (e.g. $\gamma_u$ and $\gamma_i$ in Equation (2.16)) of the model, but they are never considered explicitly.

The above property is surprising to learners first exposed to recommender systems, and stands in contrast to other types of machine learning (and especially 'deep learning') where success generally depends on learning representations from features. Recommender systems, in contrast, learn representations from user histories and patterns of user/item interaction. To the extent that a user's gender, or a movie's genre, is useful to make a prediction, it will be captured via a latent representation and is not needed as an explicit feature.[5]

**Deep learning approaches**

The latent factor model in Equation (2.16) is in essence a simple bilinear model. Given the wide success of deep learning for tasks in computer vision and natural language processing, one might ask whether deep learning approaches could also be effective for recommendation. The answer is not immediately obvious: deep learning has been successful in the context of computer vision and natural language processing tasks because of its effectiveness in learning *feature representations*; but as we argued above, recommender systems generally eschew features in favor of latent representations of users and items.

---

[4] Though such interpretations should be made carefully due to interactions among terms.

[5] Though we'll discuss various exceptions to this throughout the book.

Indeed, given that the latent factors of models like those in Equation (2.15) have the freedom to learn *any* representations of users and items, one might wonder what additional value a 'deep' model might provide.

However, there are some choices in models like that of Equation (2.15) which seem arbitrary. In particular, the choice of the inner product to compare user and item terms. Although it was motivated via a connection to the Singular Value Decomposition as in Equation (2.13), it could just as easily have been replaced by a Euclidean distance, or the maximum product between latent dimensions, etc. Ideally, we might let the data tell us what the ideal aggregation function should be.

Several works have explored deep learning approaches to learn better recommenders, and we'll discuss many examples throughout this book. An early attempt can be found in He *et al.* (2017b), which tried to learn an aggregation function via a multilayer perceptron. Zhang *et al.* (2019a) broadly discusses the benefits and pitfalls of deep learning within the context of recommender systems.

### 2.2.3 Sequential recommendation

Temporal dynamics play a major role in a variety of recommendation settings. It is critical to know both what items a user has interacted with as well as *when* those interactions occur. Such dynamics can account for a variety of critical factors such as evolving movie preferences (Koren *et al.*, 2009), fashion trends (He and McAuley, 2016), or user sessions (Hu *et al.*, 2018).

Although temporal dynamics are not a major focus of this book, it is worth discussing a specific form of temporal dynamics, namely *sequential* dynamics. Rather than dealing with interaction timestamps explicitly, such models make a simpler assumption that the most important temporal factor is simply the *order* in which items are consumed. In particular the *most recent* item a user interacted with is a powerful predictor of their next interaction.

Such models are particularly relevant to us as they form the basis of adapting language models to recommendation, as we'll see below in Section 2.3.1. That is, they treat interaction histories as a sequence of discrete tokens, such that recommendation is simply a form of next-token prediction.

**Markov chain methods**

At their essence, *Markov Chain* methods assume that in a sequence of events, the probability of the next event in a sequence depends only on the previous event, i.e., that the next event is conditionally independent of all other events given the previous event.[6] Formally:

$$p(i^{(t+1)} = i \mid i^{(t)} \dots i^{(1)}) = p(i^{(t+1)} = i \mid i^{(t)}) \tag{2.17}$$

In the context of recommender systems, the Markovian assumption corresponds to the notion that a user's next interaction can be predicted as a function of their most recent interaction. A *personalized* Markov chain (Rendle *et al.*, 2010) further assumes that the next action should be a function of both the previous action and a model of user preferences. Dropping the mathematical formalism from Equation (2.17), we might simply state that the user's next action should be related to their previous action, as well as our model of the user:

$$f(i|u,j) = \overbrace{\gamma_u^{(ui)} \cdot \gamma_i^{(iu)}}^{\text{user's compatibility with the next item}} + \underbrace{\gamma_i^{(ij)} \cdot \gamma_j^{(ji)}}_{\text{next item's compatibility with the previous item}}. \tag{2.18}$$

This type of model is known as a *Factorized Personalized Markov Chain*, or FPMC (Rendle *et al.*, 2010).

**2.2.4  "User-free" recommendation**

A challenge in implementing and deploying large scale recommender systems is the need to store and update representations associated with every user and every item, the vectors $\gamma_u$ and $\gamma_i$ in Equation (2.15).

---

[6]This definition corresponds to a *first order* Markov Chain.

**Table 2.1:** Sequential recommendation approaches. Markov-chain based models (top) and deep-learning approaches (bottom).

| Name | Method | Description |
| --- | --- | --- |
| FPMC | Factorized Personalized Markov Chains (Rendle *et al.*, 2010) | Sequential items should be related to each other, and also relevant to the user. |
| SPMC | Socially-Aware Personalized Markov Chains (Cai *et al.*, 2017) | The next item should be related to friends' recent interactions. |
| PRME | Personalized Ranking Metric Embedding (Feng *et al.*, 2015) | Replaces the compatibility function in FPMC by a metric embedding. |
| FME | Factorized Markov Embeddings (Chen *et al.*, 2012) | Models sequential dynamics, but without a personalization term. |
| TransRec | Translation-based Recommendation (He *et al.*, 2017a) | Uses a knowledge-graph formulation to model sequential relations. |
| item2vec | Distributed item representations (Barkan and Koenigstein, 2016) | Adapts word2vec (Mikolov *et al.*, 2013) to item representations |
|  | Session-based Recommendation with RNNs (Hidasi *et al.*, 2016) | Uses an RNN to model sequential interactions |
| NARM | Neural Attentive Recommendation (Li *et al.*, 2017) | Uses an attention mechanism to 'focus' on recent interactions. |
| SASRec | Self-Attentive Sequential Recommendation (Kang and McAuley, 2018) | Uses self-attention (a Transformer model) rather than an RNN. |
| BERT4Rec | (Devlin *et al.*, 2019) | Uses a bidirectional Transformer (BERT). |

Especially problematic are *user* representations, as the number of users can be very large, and ideally user representations should be updated as users continue to interact with the system.

'User free' models attempt to address this via modeling approaches that store only item representations, while operating directly on the user history. A representative example of such a model is the *Factored Item Similarity Model* (FISM) (Kabbur *et al.*, 2013), in which the user term $\gamma_u$ in Equation (2.15) is replaced by an average of *item* terms from the user's history:

$$f(u, i) = \alpha + \beta_u + \beta_i + \frac{1}{|I_u \setminus \{i\}|} \sum_{j \in I_u \setminus \{i\}} \gamma'_j \cdot \gamma_i. \qquad (2.19)$$

Note a few features of such a model:

- The user representation is replaced by an item representation; the model now estimates whether the target item ($i$) is similar to the *average* of the items from the user's history.

- There are now *two* item representations ($\gamma$ and $\gamma'$), i.e., we have doubled the number of item parameters, in order to discard user parameters.

- The user's history (represented by the set of items $j \in I_u$) is ingested at inference time.

The final point above is critical, as it means that the model doesn't need to be updated as the user performs further actions: as new items are added to the user history, the model's predictions will automatically change.

Although we present FISM just as an example, the above benefits mean that user-free approaches are adopted by a wide variety of models, ranging from traditional approaches (e.g. Sparse Linear Methods, or SLIM (Ning and Karypis, 2011)), to approaches based on deep learning. This paradigm is especially ubiquitous when we explore approaches based on language models below and in Chapter 4: here a 'user' is simply regarded as a sequence of discrete tokens (items), and predictions change at inference times as new actions are appended to the sequence.

## 2.3 Representations for generative models

### 2.3.1 Language models

**Language models as sequence encoders**

Before discussing recommender systems that make use of language *per se*, it is worth discussing recommendation approaches that use language modeling techniques as a means of representing and predicting user actions, i.e., based on interaction data but *without* any text.

In Section 2.1.1 we discussed various ways to represent user interaction data, including the possibility of representing each user's interaction history as an (ordered) sequence of events. In essence, such a representation is just a sequence of discreet tokens for each user; thus a 'user' is roughly analogous to a 'document' whose 'words' are the items the user interacted with.

In principle, this means that any model used to capture patterns in language (sequences of words) might be used to capture patterns in sequences of items, e.g. for next-item prediction.

Following the item-to-item recommendation approaches we explored in Section 2.2, we might develop new item-to-item recommenders by borrowing models that assess word-to-word similarity. One such model is *word2vec* (Mikolov *et al.*, 2013), which learns latent word representations that are capable of predicting which words will appear in similar contexts. Specifically, the model tries to predict whether a word ($w_t$, indexed by position $t$) will appear in the 'context' of another word within a range of $c$ tokens:

$$\frac{1}{T}\sum_{t=1}^{T}\sum_{-c\leq j\leq c, j\neq 0}\log p(w_{t+j}|w_t).\qquad(2.20)$$

The probability is parameterized by a vector $\gamma_w$ associated with each word, and words are related by an inner product:

$$p(w_o|w_i)=\frac{e^{\gamma'_{w_o}\cdot\gamma_{w_i}}}{\sum_{w\in\mathcal{W}}e^{\gamma'_w\cdot\gamma_{w_i}}}.\qquad(2.21)$$

*item2vec* (Barkan and Koenigstein, 2016) adopts this technique to sequences of item representations:

$$\log p(i|j) \simeq \log \sigma(\gamma_i' \cdot \gamma_j) + \sum_{i' \in \mathcal{N}} \log \sigma(-\gamma_{i'}' \cdot \gamma_j); \qquad (2.22)$$

Note the similarity between Equations (2.19) and (2.22), both of which use latent representations to capture sequential dynamics.

**Recurrent networks and transformers** Following the success of adapting word representations to item representations via methods like item2vec (Barkan and Koenigstein, 2016), state-of-the-art language models have repeatedly been adapted for recommendation, replacing next-word prediction with next-item prediction.

Hidasi *et al.* (2016) does so by adapting recurrent neural networks (which had proven effective for generating language) to predict next items in user sequences. Later models adopted similar settings to predict next items using attention mechanisms (Li *et al.*, 2017; Kang and McAuley, 2018), which arguably represent the current state-of-the-art. We'll revisit these models in Chapter 3.

**Interactivity**

Before discussing conversational recommendation below, it is worth briefly mentioning that early models for conversational recommendation treat the idea of 'conversational' quite broadly. Historically, any recommender system involving back-and-forth iterative interactions between a user and a system might be referred to as 'conversational,' even if the model does not make use of text.

Examples of models in this paradigm include Thompson *et al.* (2004), which is essentially is a form of interactive query refinement, in which users' preferences and constraints are gathered across multiple turns.

Mahmood and Ricci (2009) and later Christakopoulou *et al.* (2016) explore strategies to develop policies to efficiently query users about their preferences over item attributes.

**Review Generation**

One of the most important paradigms of generative recommenders that we'll explore in Chapter 4 is that of systems that generate natural language outputs. The genesis of such models is arguably approaches that focus on generating product *reviews.*

Here, a prediction function $f(u, i) \rightarrow y$ that might previously have predicted a rating or interaction is now replaced by one that estimates a review for the user $u$ about item $i$. While perhaps not an obviously useful task in and of itself, learning how users write about items (and express their preferences, etc.) is a step toward developing more sophisticated approaches based on natural language.

Early approaches generated reviews using recurrent neural networks (Radford *et al.*, 2017), though this method lacks a user or item model and simply learns the 'background' distribution of what reviews look like.

Later approaches generate personalized reviews conditionally based on specific attributes associated with the user or item (Dong *et al.*, 2017; Li *et al.*, 2017; Ni *et al.*, 2017). We showed an example of a review generated using the approach from (Ni and McAuley, 2018) in Figure 2.1.

**Interpretability**    Generated reviews can also be used to make recommender systems more *interpretable* or *explainable*, e.g. by surfacing language (from reviews or otherwise) that explain why a user selected an item. For example Li *et al.* (2017) learns to generate personalized 'tips' that explain why a user might select a restaurant; or Ni *et al.* (2017) learns a model of which sentences in reviews are explanatory in nature, so that a generative model can be trained to generate explanatory sentences.

## 2.4    Conclusion

In summary, this chapter has provided a foundational overview of both traditional and generative models used in recommender systems. We explored various methodologies and concepts, ranging from discriminative models that predict user interactions to advanced generative models capable of producing rich, structured outputs such as personalized reviews and item designs. This comprehensive examination equips readers with the necessary understanding to address more specialized applications of generative models in subsequent chapters. As we move forward to Chapter 3, we will shift our focus to the use of generative models in traditional user-item recommender systems, exploring how these models can enhance recommendation accuracy and address common challenges such as data sparsity and cold-start problems.

# 3

## Generative Models for ID-Based Recommender Systems

### 3.1 Introduction

In this chapter, we will examine the use of generative models in traditional recommender systems for various purposes. This chapter differs from the following two chapters in that it considers only traditional user-item "ID" data, such as user-item interactions (for user ID $u \in U$ and item ID $i \in I$) and the corresponding features like timestamps and ratings, but without any content information like textual descriptions or item images.

The reasons that we have taken this standalone chapter to introduce generative models in traditional ID-based recommender systems are multi-fold:

First, the ID-based recommendation setting has been a focal point in recommender system research and industry. On the one hand, it is because ID-based recommendation setting reprsents the simplest and most generic recommendation setting, requiring minimal data sources – only user-item interactions, without any additional content information. On the other hand, user-item interactions, represented by IDs, are usually strong signals for recommendation tasks (i.e., *collaborative* signals). Recommendation models based on these signals often show impressive performance (e.g., Netflix Prize (Bennett, Lanning, *et al.*, 2007)). Meanwhile, the ID-based recommender systems are extendable, for example, models relying on user-item ID features can be effectively combined with content-based models for further improvements (e.g., Hybrid Recommender Systems (Burke, 2002)). Therefore, a discussion of the role of generative models in this influential ID-based recommendation setting is warranted.

Second, the ID-based recommendation setting demonstrates different applications of generative models compared to generative models for content data (e.g., textual or visual content). For instance, generative models for image generation, such as Generative Adversarial Networks (GANs) (Goodfellow *et al.*, 2014) or Variational Autoencoders (VAEs) (Kingma and Welling, 2013), are designed to model the underlying distribution of image data and then sample (i.e., generate) new images from this learned distribution. However, in ID-based recommendation, beyond the interest of modeling data distribution for generation purposes, generative models often play an indirect role as well, serving for e.g., model regularization, hard example mining, or sequence encoding. Therefore, it is necessary to dedicate a separate chapter to these applications of generative models in the ID-based recommendation setting.

With such considerations, below we attempt to categorize different ID-based recommendation tasks by their output structures Then, we explain the role of various generative models used in such recommendation tasks, along with an overview of different generative model paradigms, designs, and applications. The organization of this chapter is illustrated in Figure 3.1.

## 3.2 Formulations in ID-based Recommender Systems

The core data source of ID-based recommender systems is a set of user-item interactions $(u, i, r, t)$. Here, user $u \in U$ and item $i \in I$ interact at timestamp $t$, and $r$ is the feedback from the user to the item. This feedback can be explicit (e.g., star ratings $r = \{1, \ldots, 5\}$) or implicit (e.g., clicks $r = \{0, 1\}$). These historical user interactions are used to train a recommender to output $k$ items of interest to the user $u$.

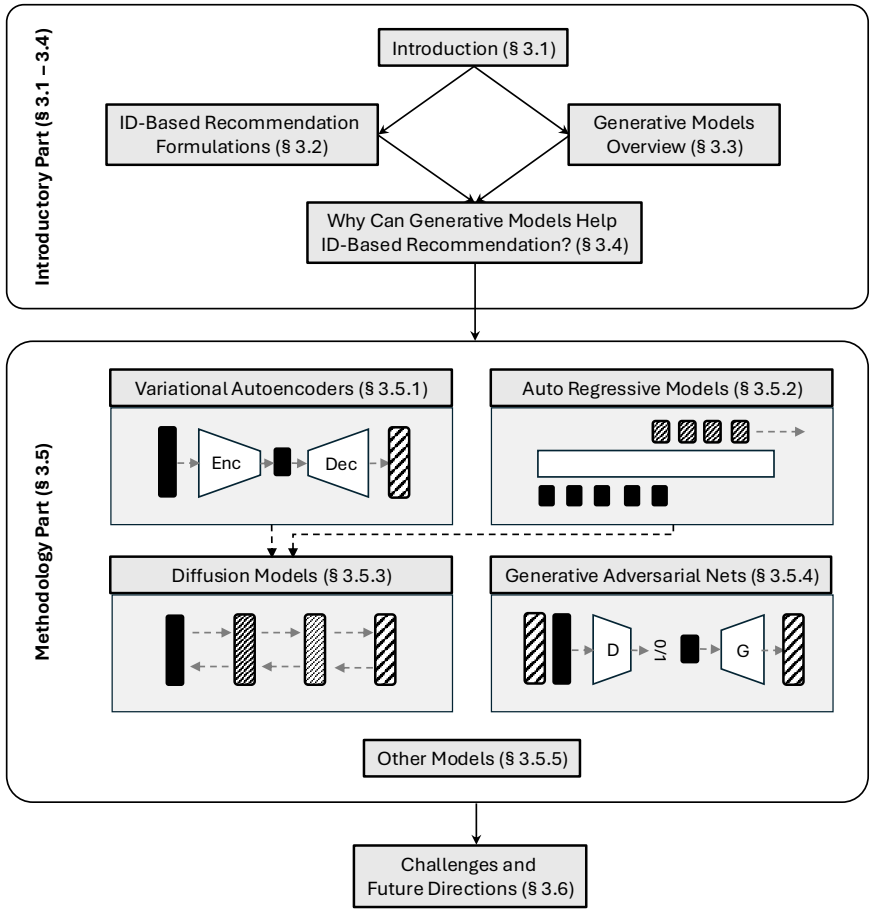

**Figure 3.1:** Generative Models for ID-Based Recommendation covered in Chapter 3. The illustrations are partially adapted from (Weng, 2021).

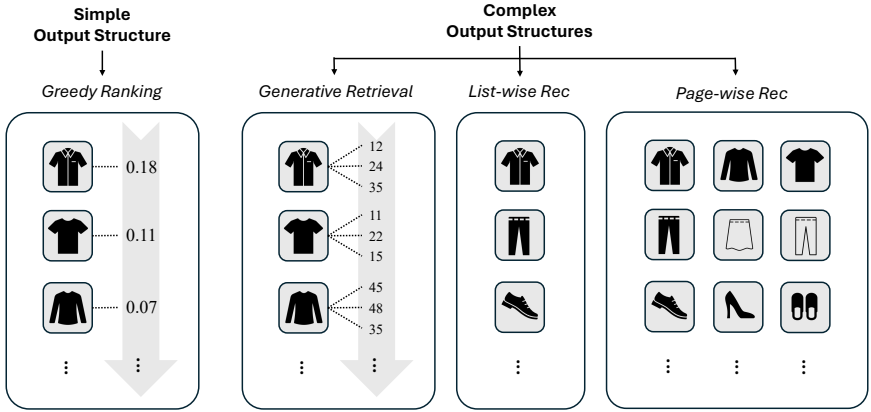

**Figure 3.2:** ID-Based Recommendation Formulations.

Based on how those $k$ items are presented and optimized, we categorize recommendation output structures into two types and illustrate them in Figure 3.2.

### Recommendation with Greedy Ranking

The output structure of greedy ranking is simple. Recommendation with greedy ranking aims to learn a scoring function for a given user-item interaction, i.e., $(u, i)$, with some ground truth such as user ratings or clicks. It then assigns scores to all the interaction candidates and sorts the rated interactions as a recommendation list. This is a typical discriminative task, as it assigns scores to each interaction candidate independently. Even in this simple output structure, generative models can be used in, for example, the negative sampling step of model training (Wang *et al.*, 2017) or introducing stronger model assumptions for regularization (Liang *et al.*, 2018).

### Recommendation with Complex Structures

The output structures for recommendation results can be complex. We consider three examples:

In *recommendation with generative retrieval*, single items are repre-sented by multiple sub-tokens (Rajput *et al.*, 2023). In detail, generative retrieval reformulates the recommendation task in Section 2 as a generative task. Items are represented by multiple sub-tokens and decoded auto-regressively as the predicted users' future interactions. The output structure is complex, as models are required to generate one (or more) token sequences as the identifier of the recommended items, instead of greedy ranking.

In *list-wise recommendation*, models output a list of recommended items to improve overall user engagement beyond greedy ranking (Jiang *et al.*, 2018). It optimizes a recommendation list as a whole to capture more signals than independent user-item interactions. For example, in greedy ranking, many similar items may achieve high predicted scores and hurt the diversity of the final recommendation list. In contrast, list-wise recommendation may capture item diversity (and more factors) when recommending a list of items. The output structure of the list-wise recommendation is complex, and it is usually not treated as a standard discriminative task.

In *page-wise recommendation*, models output a grid of items to users, considering individual items, item-item relationships, layout, and even personalized scanning orders in many services such as e-commerce and streaming (Gomez-Uribe and Hunt, 2015). This can be viewed as a two-dimensional extension of *list-wise* recommendation and is the most complex output structure, which discriminative models struggle to handle trivially.

In these cases, generative models have shown promising results in modeling the distributions of the complex output structures of personalized recommendations.

## 3.3   Paradigms of Generative Models

We briefly go through representative generative model paradigms.

### Variational Autoencoders (VAEs)

Variational Autoencoders (Kingma and Welling, 2013) are models that learn stochastic mappings from an input $\mathbf{x}$ from often a complicated probability distribution $p$ to a typically simpler probability distribution $q$ (e.g., a normal distribution). This enables the use of a decoder to generate output $\hat{\mathbf{x}}$ by sampling from $q$.

### Auto-Regressive Models (AR)

Auto-Regressive Models (AR) (Hochreiter and Schmidhuber, 1997) learn the conditional probability distribution $p(\mathbf{x}_i|\mathbf{x}_{<i})$ given an input sequence $\mathbf{x}$, at step $i$, where $\mathbf{x}_{<i}$ represents the subsequence before step $i$. Auto-regressive models are primarily used for sequence modeling (Bengio *et al.*, 2000; Vaswani *et al.*, 2017; Van Den Oord *et al.*, 2016).

### Diffusion Models

By extending VAEs to an autoregressive latent space, diffusion models (Sohl-Dickstein *et al.*, 2015; Ho *et al.*, 2020) show impressive generation quality in image and video domains. Diffusion models generate outputs through a two-step process: (1) corrupt inputs into noise by adding different levels of white noises through a forward process, and (2) learn to recover the original inputs from the noise iteratively in a reverse process.

### Generative Adversarial Networks (GANs)

Generative Adversarial Networks (GANs) (Goodfellow *et al.*, 2014) include two components: a generator network $G$ and a discriminator network $D$. These networks engage in adversarial training, in which $G$ learns to generate fake samples to fool $D$. These two components reach the equilibrium when $G$ has successfully learned the underlying data distribution (Goodfellow *et al.*, 2014).

## 3.4 Why Generative Models for ID-based Recommendations?

Generative models are usually used to model data (conditional or unconditional) distribution $p(\mathbf{x}|\cdot)$[1]. In the ID-based recommendation setting, there are multiple reasons to use generative models, either **directly** or **indirectly**.

---

[1] For simplicity, we omit the user identifier $u$ in conditional distributions below.

## Direct Usage: Generative Models as Recommenders

Generative models are typically used to generate new sample $\hat{\mathbf{x}}$ from learned distribution $p_{\text{model}}$. Therefore, the direct reason for using generative models in the ID-based recommendation setting is that those models can learn the data distribution, and the new samples from this learned distribution can serve as recommendation results. For example, generative models can model the distribution of *user interaction vectors* (Liang *et al.*, 2018), *item lists* (Jiang *et al.*, 2018) and *item pages* (Bai *et al.*, 2019) and generate recommendation results for greedy ranking, list-wise recommendations and page-wise recommendations, respectively.

## Indirect Usage: Generative Models Help Recommenders

For many cases in ID-based recommendation settings, generative models are often not the final recommender systems or the sole component. In such cases, rather than focusing solely on the quality of the learned distribution $p_{\text{model}}$, people typically require additional "byproducts" from generative models to enhance the performance of their recommender systems. For instance, researchers are interested in *disentangled representations* obtained through a VAE framework (Ma *et al.*, 2019), *hard negative mining* for discriminative models using the GAN models (Wang *et al.*, 2017), or *data augmentation* techniques leveraging diffusion models (Lin *et al.*, 2023b; Wu *et al.*, 2023d), where generative models improve the overall quality of a recommender system indirectly.

## 3.5 Generative Models for ID-based Recommendation Setting

We explain in more detail about using different generative model paradigms in the ID-based recommendation setting below.

### 3.5.1 Variational Autoencoders (VAEs)

**Normal Priors as Model Regularizations**

In the context of recommendations, Variational Autoencoders (VAEs) emerged from Autoencoders (AE) or Denoising Autoencoders (DAEs) with a latent space representation and the training objective is to reconstruct missing or incomplete user data. For example, we typically treat a user's feedback (clicks or ratings) over all items as a user interaction vector, where each element represents an item, and the binary value or like-rt value indicates the user's preference. AutoRec (Sedhain *et al.*, 2015) is a DAE model that recommends items by reconstructing this partially observed input vector.

Unlike DAEs, the first VAE model for recommendation, Mult-VAE (Liang *et al.*, 2018), assumed that the latent variable $\mathbf{z}$ follows a simple distribution (e.g., normal distribution). As a result, this model optimizes a traditional reconstruction objective term with an additional regularization term:

$$\mathcal{L}_\beta^{\text{ELBO}}(\mathbf{x};\theta,\phi) = \underbrace{\mathbb{E}_{q_\phi(\mathbf{z}|\mathbf{x})}\left[\log p_\theta(x|\mathbf{z})\right]}_{\text{Reconstruction}} - \beta \cdot \underbrace{\text{KL}(q_\phi(\mathbf{z}|\mathbf{x})||p(\mathbf{z}))}_{\text{Regularization}}. \quad (3.1)$$

Generally speaking, the first "reconstruction" term in Eq 3.1 is used for traditional AE-based model training. By introducing the additional "regularization" term, MultVAE is under the $\beta$-VAE framework, aiming to learn $p(x)$. In particular, when $\beta = 1$, optimizing the objective function in Eq 3.1 is equivalent to maximizing the lower bound of $\log p(\mathbf{x};\theta,\phi)$, which is a well-known result in variational inference as Evidence Lower Bound (Kingma and Welling, 2013) (i.e., ELBO). Therefore, MultVAE is considered a generative model for collaborative filtering tasks.

MultVAE's normal-prior assumption effectively regularizes recommendation models, leading to improved performance, especially with sparsely rated items (Liang *et al.*, 2018). This demonstrates the effectiveness of incorporating VAE-based models as additional regularization techniques for recommendation systems. Beyond normal priors in VAE for collaborative filtering, the follow-up works extend similar priors to *sequential recommendation* settings (Sachdeva *et al.*, 2019), include more flexible priors such as the *composite prior* in RecVAE (Shenbin *et al.*, 2020), and combine VAE with other generative model paradigms such as *Adversarial Variation Bayes* (Mescheder *et al.*, 2017) in ACVAE (Xie *et al.*, 2021).

## Disentanglemented or Casual Representations

Beyond using normal priors for model regularization, researchers are exploring how VAEs can introduce structural assumptions into the latent representation $\mathbf{z}$ within a probabilistic framework. One approach involves incorporating additional disentangled or causal structures into the VAE framework to further refine and improve the user representations $\mathbf{z}$.

Disentangled representations are representations where each dimension of an embedding is independent and semantically meaningful. (Ma *et al.*, 2019) defines a two-level disentanglement: macro-disentanglement (category level) and micro-disentanglement (aspect level) to model a user's preference for various aspects of an item category. To encourage statistical independence across different dimensions of a user representation under a VAE framework, this work introduces a regularization using prior distributions with independent factors, resulting in more disentangled representations. Beyond disentanglement, researchers have explored incorporating causal structures into user representations within a VAE framework. For example, causal representations (Wang *et al.*, 2022b) assume a causal structure for user clicks, driven by observed features like user income, and unobserved features like conformity. Generative models, such as VAEs, are more likely to capture causal relationships due to their focus on modeling the generation process (Kingma and Welling, 2013), leading to better generalization ability. Thus, causal representations show the improved Out-Of-Distribution recommendation performance (Wang *et al.*, 2022b).

## Personalized List or Page Generation

As a generative model, VAEs can directly model the high-dimensional distributions of personalized recommendation lists or pages. The core idea is to treat recommendation as a generation task. We can sample new recommendation lists or pages from a distribution conditioned on "positive user feedback" (e.g., one user's satisfaction). This approach offers the advantage that VAE models can directly learn the conditional distributions of "good lists or pages" without additional assumptions like "good lists should put better items at the top" as required in traditional greedy ranking methods. Instead, additional user preferences like scanning orders, and item diversity can be naturally incorporated into such a generative modeling framework. In literature, ListCAVE (Jiang *et al.*, 2018) and PivotCVAE (Liu *et al.*, 2021b) are following this research line to model and generate item lists or bundles directly.

Furthermore, the representations of item lists or bundles from such a VAE model can be further used to represent latent "actions". GeMS (Deffayet *et al.*, 2023) encodes an item list to a latent representation $\mathbf{z}$, and then does reinforcement learning with this latent representation (action) space to find an optimal $\mathbf{z}^*$. Then, the VAE decoder decodes a new item list from this representation $\mathbf{z}^*$ as recommendations, avoiding the untractable discrete action space over all the possible item lists.

### 3.5.2 Auto-Regressive Models (AR)

### 3.5.3 Next-Item Prediction / Sequential Recommendation

Auto-regressive (AR) models factorize the high-dimensional distribution $p(\mathbf{x})$ to $\prod_i^N p(\mathbf{x}_i \mid \mathbf{x}_{<i})$ given the input sequence $\mathbf{x}$. Thus, modeling $p(\mathbf{x})$ with AR models can produce powerful models $p_\theta(\mathbf{x}_i \mid \mathbf{x}_{<i})$, which can be naturally used to predict the next item (e.g., item at step $i$) for each user, given the user history (e.g., interacted items before step $i$). This is called next-item prediction, or sequential recommendation.

Under the AR framework, two typical model architectures are RNNs (Hidasi *et al.*, 2016) and Self-Attentive models (Kang and McAuley, 2018). There are some differences in RNNs and Self-Attentive models for recommendations. First, regarding training parallelism, RNNs, due to their recurrent structure, have a sequential dependency. The output at each time step depends on the current input and the hidden state of the previous time step, which prevents parallelization. In contrast, Self-Attentive Models (e.g., Transformers (Vaswani *et al.*, 2017)) use the self-attention mechanism, allowing the model to focus on all positions in the input sequence simultaneously, enabling parallel training. This sequential independency makes Self-Attentive models more efficient to train. Second, regarding the model effectiveness, the *multi-head attention*, *positional encoding* (Vaswani *et al.*, 2017) components make Self-Attentive models more expressive and achieve better accuracy metrics empirically. In addition to those components, some inductive biases (e.g., locality bias (He *et al.*, 2021)) can be incorporated into Self-Attentive models to improve the accuracy further.

However, due to the design of the vanilla self-attention mechanism, the time and space complexity grows quadratically with the length of the input sequence, $\mathbf{x}_{<i}$. This poses challenges for sequential recommendations with long user sequences (e.g., lifelong) or low-latency requirements for incrementally changing user sequences, such as updating recommendations from $\mathbf{x}_{<i}$ to $\mathbf{x}_{<i+1}$. Many works focus on developing efficient self-attentive models or non-self-attentive AR models to address these challenges while maintaining model effectiveness and training parallelism (Yue *et al.*, 2024).

## Generative Retrieval for Recommendation

Recently, an intriguing paradigm in AR for sequential recommendation is called generative retrieval. Generative retrieval represents single items with multiple subtokens. The benefits are multi-fold: The first is vocabulary efficiency. By combining existing subtokens, models with generative retrieval can represent different items more efficiently. The second is about item similarity. Overlapping subtokens can reflect similarities among items in terms of collaborative information (Petrov and Macdonald, 2023)[2]. Moreover, this method shows the potential of structured representations. Subtokens can be arranged to represent coarse-to-fine descriptions.

For these reasons, generative retrieval has become a promising direction in recent years. It also shows potential for integration with Large Language Models and improved scalability. However, replacing single item IDs with multiple subtokens lengthens user sequences, leading

---

[2]Due to the scope of this chapter, we only mention collaborative information in ID-based RecSys. However, generative retrieval is more commonly combined with content similarity (Rajput *et al.*, 2024)

to challenges in training and serving efficiency as well.

## Attacks to Sequential Recommendation

It is noted that since AR models are generative models for $p(\mathbf{x})$, we can sample new user interaction sequences from a trained AR model, such as SASRec (Kang and McAuley, 2018). One finding from Yue *et al.*, 2021 is that by using user sequences generated by AR-based sequential recommenders, we can create a local surrogate model of this sequential recommender by retraining a similar model on those sequences. Attacks on the surrogate model (e.g., adversarial examples (Goodfellow *et al.*, 2015), or poison attacks (O'Mahony *et al.*, 2002)) can be transferred to the black-box original AR-based sequential recommender. This finding highlights the vulnerability concerns of AR-based sequential recommenders.

## Multi-Item Autoregressive Generation

The major applications of AR models in recommendations beyond standard next-item prediction are generating multiple items that satisfy specific complex structures, such as item bundles (or sets), lists of item bundles, and user long-term item interactions. Item sets aim to serve multiple items to a user simultaneously, with the promise that multi-item combinations (e.g., playlists, outfits) offer mutual benefits compared to recommending those items individually. These items often have complementary or substitutive relationships, allowing AR models to model these relationships and generate the items one by one.

Personalized Outfit Generation Model (Chen *et al.*, 2019b) is an example of item-set or bundle generation. It's a self-attentive model with encoder-decoder architectures that "translates" a user's historical behaviors into a personalized outfit with the following loss function:

$$\mathcal{L}_{(U,F)} = -\frac{1}{N} \sum_{i=1}^{N} \log p\left(\mathbf{x}_{i+1} | \mathbf{x}_1, \ldots, \mathbf{x}_i, U; \Theta_{(U,F)}\right), \qquad (3.2)$$

where $\mathbf{x}$ represents the vector of items to be generated. While $U$ and $F$ represent some additional context information, specifically the user's general contexts and outfit preference history, respectively.

For lists of item sets or item bundles, the approach extends general bundle or set generation models to a bi-level generation pipeline where multiple sets of items are generated to form a list. It also can be treated as generating a personalized "page" for users. Bundle Generation Network (Bai *et al.*, 2019) demonstrates that AR models can be used to generate multiple sets or bundles sequentially, with additional objectives on item diversity (e.g., combining with Determinantal Point Processes (Zhou *et al.*, 2010)).

### 3.5.4  Diffusion Models

**Generate User Interactions via Denoising**

One direction of using diffusion models for recommendation is learning the distribution of user interaction vectors (i.e., user feedback over all items) via denoising. For example, representative works like DiffRec (Wang *et al.*, 2023d) aim to denoise corrupted a user interaction vector through diffusion models and learn to reconstruct the user interaction vector, where the user feedback for unrated items can be used for recommendations. The core idea of DiffRec is to optimize the objective function below:

$$
\mathcal{L}^{\text{ELBO}}(\mathbf{x}; \theta) = \underbrace{\mathbb{E}_{q(\mathbf{x}_1|\mathbf{x})}[\log p_\theta(\mathbf{x}|\mathbf{x}_1)]}_{\text{Reconstruction}}
$$
$$
- \sum_{t=2}^{T} \underbrace{\mathbb{E}_{q(\mathbf{x}_t|\mathbf{x})}[\text{KL}(q(\mathbf{x}_{t-1}|\mathbf{x}_t, \mathbf{x}) \| p_\theta(\mathbf{x}_{t-1}|\mathbf{x}_t)))]}_{\text{Denoising Matching}}, \quad (3.3)
$$

where $\mathbf{x}$ represents the original user interaction vector, and $\mathbf{x}_i$ represents the intermediate use representations with different levels of white noise. $\mathbf{x}_T$ follows a normal distribution obtained after the iterative forward pass in the diffusion model. We can have several observations from this objective function: (1) Diffusion models are a special variant of VAEs derived from maximizing ELBO (Kingma and Welling, 2013; Ho *et al.*, 2020); (2) This objective function, like VAEs in Section 3.5.1, can be treated as a "reconstruction" term and a "regularization" term, but here the regularization term matches autoregressive white noises. (3) Compared to traditional VAEs, the beauty of diffusion models is that we do not need to learn a parameterized encoder. Instead, diffusion models explicitly define the white noise per step (i.e., using $q(\mathbf{x}_t|\mathbf{x})$ rather than $q_\phi(\mathbf{x}_t|\mathbf{x})$) and only have a neural-network-based decoder. It empirically shows a better trade-off between model tractability and flexibility than standard VAEs with better recommendation accuracy.

**Training Data Augmentation**

Another direction is using diffusion models for data synthesis and augmentation. Recent works (Liu *et al.*, 2023c; Wu *et al.*, 2023d) have demonstrated the effectiveness of augmenting training user interaction sequences with diffusion models, addressing data sparsity and long-tail issues in sequential recommendation. For instance, DiffuASR (Liu *et al.*, 2023c) leverages diffusion models to learn the distribution of user interaction sequences. A specialized sequential U-Net architecture (Liu *et al.*, 2023c) enables the generation of augmented interaction sequences, which can then be used to enhance model training and improve recommendation accuracy.

### 3.5.5   Generative Adversarial Networks (GANs)

**Training Example Selection**

In recommender systems, GANs were first introduced by IRGAN (Wang *et al.*, 2017) to select informative training samples and improve recommendation performance. Specifically, a generator $G$ learns to select items by fitting the user-related conditional data distribution over all items, while a discriminator $D$ learns to distinguish between generated and ground-truth items. Through adversarial training, $G$ and $D$ learn to leverage informative unobserved items as training samples during model training, leading to better model performance. IRGAN is based on the framework of Conditional GAN (CGAN) (Mirza and Osindero, 2014b), with the key difference being that $G$ in IRGAN is used to select existing items rather than generate new ones.

One way to interpret the IRGAN-like approach is from the *hard negative mining* perspective. In recommendation model (i.e., the discriminative model $D$) training, loss functions are typically designed to maximize the "distance" between positive items and negative items given a user (Rendle *et al.*, 2009). In this paradigm, negative examples are usually sampled uniformly (Rendle *et al.*, 2009) or by some heuristic rules (Zhang *et al.*, 2013). Yet in the GAN framework, generator $G$ plays the role of a learning-based hard negative sampler to select more useful negatives for recommendation model (i.e., discriminator $D$) training. Therefore, IRGAN-like methods improve recommendation accuracy against the traditional negative sampling methods.

Variants after IRGAN have explored multiple directions. One direction is to improve the capacity of G and D to better capture sequential patterns in recommendation data. For example, RecGAN (Bharadhwaj *et al.*, 2018) and PLASTIC (Zhao *et al.*, 2018) proposed using AR models (e.g., RNNs) to replace the traditional $G$ and $D$ in IRGAN, while ACVAE (Xie *et al.*, 2021) introduced the sequential VAE model SVAE (Sachdeva *et al.*, 2019) to model user historical sequences. Additionally, another group of variants focuses on proposing multiple generators or discriminators (Fan *et al.*, 2019; Ren *et al.*, 2020).

**Training Data Augmentation**

Another direction of GAN models for recommender systems is to augment the user preference or interaction data via models' generation abilities. For instance, AugCF (Wang *et al.*, 2019a) introduces a two-phase training procedure via a GAN framework. In Phase I, a GAN model is trained to model the distribution of user-item interactions and generate user-item interaction triplets (i.e., <user, item, feedback>) as augmented training data. In Phase II, a standard collaborative filtering model is trained on the augmented dataset to improve the recommendation performance. Similar data augmentation techniques via GAN models can also be found in CFGAN (**chae2018cfgan**), UGAN (Wang *et al.*, 2019b), RSGAN (Yu *et al.*, 2019) methods to alleviate data sparsity issues or long-tail problems.

**Attacks to ID-based Recommender Systems**

However, not all training data generation is beneficial. (Christakopoulou and Banerjee, 2019) demonstrated that GANs can generate fake user profiles, potentially harming oblivious recommenders through poisoning attacks (O'Mahony *et al.*, 2002). In this work, the authors show that a GAN framework can be used to model the distribution of a set of user interactions. By sampling from this learned distribution, people can create a set of hard-to-detect fake user profiles. With some adversarial attack techniques, those augmented user interactions can hurt model re-training performance by manipulating the recommendation results (i.e., decrease the recommendation scores of a target item). Related adversarial attacks can be found in AUSH (Lin *et al.*, 2020) as well.

### 3.5.6 Other Generative Models

In addition to the aforementioned generative models, other types have been explored in ID-based recommendations. VASSER (Zhong *et al.*, 2020) leverages normalizing flows (Rezende and Mohamed, 2015) and VAE (Kingma and Welling, 2013) for session-based recommendation to improve the generalibility of the representations. GFN4Rec (Liu *et al.*, 2023d) adapts generative flow networks (Bengio *et al.*, 2021; Pan *et al.*, 2023) for generating multiple items for list-wise recommendation. Furthermore, IDNP (Du *et al.*, 2023) uses generative neural processes (Garnelo *et al.*, 2018b; Garnelo *et al.*, 2018a) for sequential recommendation.

### 3.6 Challenges and Future Directions

Deep generative models were originally proposed for generating unstructured data, such as text or visual content. However, they are also playing important roles in improving the quality of recommender systems with structured data in different ways. They are widely used in areas such as model regularization, representation learnings, data augmentation, and complex structured recommendation generation such as bundles, lists and pages, with different emphases according to the characteristics of the different generative model paradigms.

However, we are still facing several challenges in using generative models for traditional recommendations, we discuss them and list some future research directions below.

- **Training Stability.** Some generative models, such as GANs, can be unstable during training due to their adversarial nature. Generally speaking, adapting generative models that were previously used in continuous space to discrete space (i.e., related to items) is usually challenging. More effort may be needed to overcome training instabilities in these recommendation models.

- **Generation Efficiency.** Using generative models for generative retrieval or complex structured recommendation generation is intriguing but often suffers from efficiency issues. For example, generative retrieval with auto-regressive models requires multiple decoding steps, and many decoding issues in Large Language Models are also present in this area.

- **Datasets and Benchmarks.** For recommendation tasks with complex output structures, such as bundle recommendation or page-wise recommendation, large-scale public datasets are needed. Corresponding benchmarks on such datasets are also urgently needed to accelerate the development of generative models for the recommendation community.

- **Recommendation-tailored Model Design.** Most generative models were originally motivated by text or visual content generation, with little consideration of the characteristics of recommendation tasks, such as the sparsity of user-item interactions and low-latency serving requirements. There have been some efforts to simplify generative models into recommendation-tailored models, but more novel model designs are still valuable in this direction.

# 4

## Large Language Model Driven Recommendation

ABSTRACT

While previous chapters focused on recommendation systems (RSs) based on standardized, non-verbal user feedback such as purchases, views, and clicks – the advent of LLMs has unlocked the use of natural language (NL) interactions for recommendation. This chapter discusses how LLMs' abilities for general NL reasoning present novel opportunities to build highly personalized RSs – which can effectively connect nuanced and diverse user preferences to items, potentially via interactive dialogues. To begin this discussion, we first present a taxonomy of the key data sources for language-driven recommendation, covering item descriptions, user-system interactions, and user profiles. We then proceed to fundamental techniques for LLM recommendation, reviewing the use of encoder-only and autoregressive LLM recommendation in both tuned and untuned settings. Afterwards, we move to multi-module recommendation architectures in which LLMs interact with components such as retrievers and RSs in multi-stage pipelines. This brings us to architectures for conversational recommender systems (CRSs), in which LLMs facilitate multi-turn dialogues where each turn presents an opportunity not only to make recommendations, but also to engage with the user in interactive preference elicitation, critiquing, and question-answering.

---

## 4.1   Introduction

The advent of LLMs has enabled conversational, natural language (NL) interactions with users – while also unlocking the rich NL data sources within recommendation systems (RSs) such as item descriptions, reviews, and queries. These advances create the opportunity for highly personalized RSs which harness the general reasoning abilities of LLMs to accommodate diverse and nuanced user preferences through customized recommendations and interactions. Such NL-based personalization contrasts starkly to the ID-based RSs described in Chapters 2 and 3 which are highly specialized for standard recommendation tasks (e.g., rating prediction, sequential recommendation) and require large volumes of non-textual interaction data – though many synergies between these two paradigms are possible, as discussed below.

### 4.1.1   Natural Language vs. Non-Textual Interaction Data

Both textual and non-textual data are important in this chapter – Figure 4.1 illustrates how such data can represent key RS information, covering items, users, and system interactions. This figure is discussed in detail in Section 4.2.

**Non-textual Interaction Data:**

On one hand, non-textual user-item interactions such as purchases, views, clicks, or ratings facilitate the collection of large volumes of data in fixed formats (e.g., rating matrices, item ID sequences) and enable the large-scale training of conventional RSs – namely, collaborative filtering (CF) and content-based filtering (CBF) systems (c.f. Ch 2). On the other hand, these non-textual data forms only provide a narrow, highly standardized view of user-system interactions – limiting the degree of personalization that can be achieved.

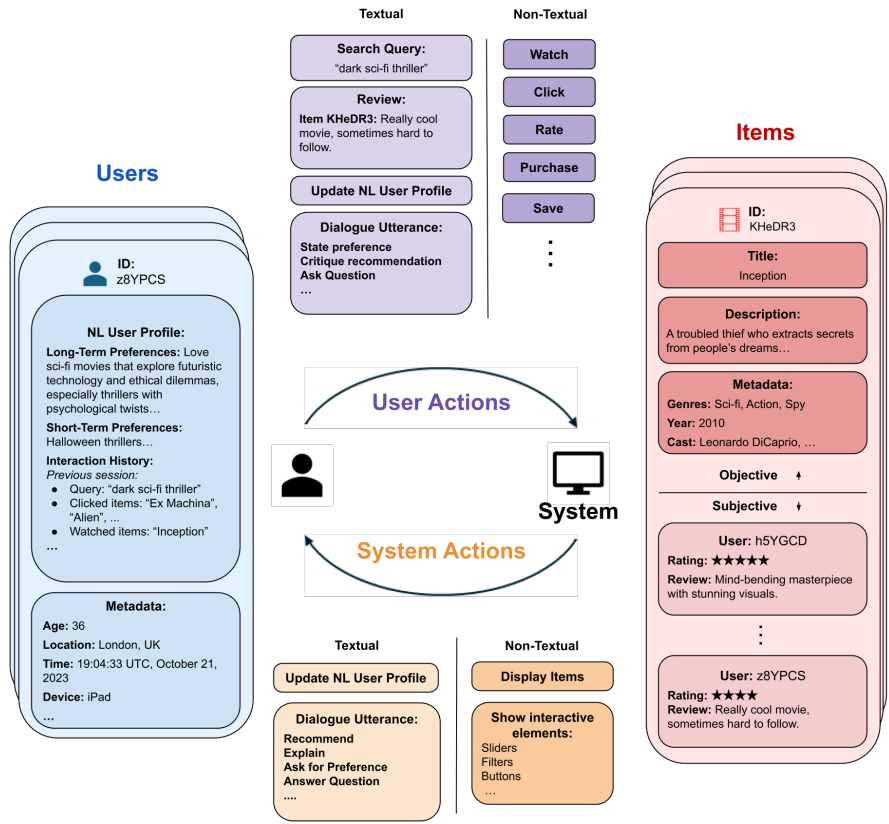

**Figure 4.1:** Sources of data in LLM-driven RSs, including item descriptions (right), user-system interactions (top/bottom), and user profiles (left).

**Natural Language in Recommendation:**

In contrast, while NL is a much more complex medium, it is also far more expressive and constitutes a unified format to represent nuanced information about items, preferences, and user-system interactions (Geng *et al.*, 2022). Many RSs already contain an abundance of NL data in the form of item descriptions, reviews, and user query histories, and systems can also generate new NL representations from interaction histories by using LLMs or templates (Radlinski *et al.*, 2022a; Sanner *et al.*, 2023; Zhou *et al.*, 2024). NL conversational recommendation dialogues are also emerging as a key source of data, capturing a variety of user and system intents in multi-turn interactions (c.f. Sec. 4.7).

### 4.1.2   General vs. Specialized Recommendation Reasoning

#### Task-specific Reasoning with Conventional RSs

Conventional RSs must be trained on large amounts of task-specific interaction data – making them highly-specialized tools that are typically optimized for either rating prediction or top-$k$, sequential, or page-wise recommendation. While their performance on *offline* benchmarks for these tasks is generally very strong (c.f. Ch. 6), these specialized systems are limited by the inability of standardized interaction data (e.g., purchases, views, clicks, etc.) to capture nuanced and diverse user preferences. These systems also often require considerable data engineering efforts and design time to deploy.

However, despite these limitations, conventional RS remain powerful and scalable methods for predicting future interactions, able to work with millions of users and items. Therefore, as will be discussed in Sections 4.5-4.7, many researchers are actively studying the integration of conventional RSs as specialized sub-components within larger, LLM-powered system architectures (e.g., Friedman *et al.*, 2023; Hou *et al.*, 2023; Zhang *et al.*, 2023d).

#### General Recommendation Reasoning with LLMs

In contrast to conventional RSs, the pretraining of LLMs on large text corpora provides them with emergent abilities for *general* reasoning – with LLMs achieving impressive performance on many diverse and previously unseen tasks (Bubeck *et al.*, 2023). Through pretraining, LLMs have internalized fine-grained knowledge about a wide range of entities, human preferences, and interaction patterns – knowledge which could enhance RS personalization while reducing data and design time requirements.

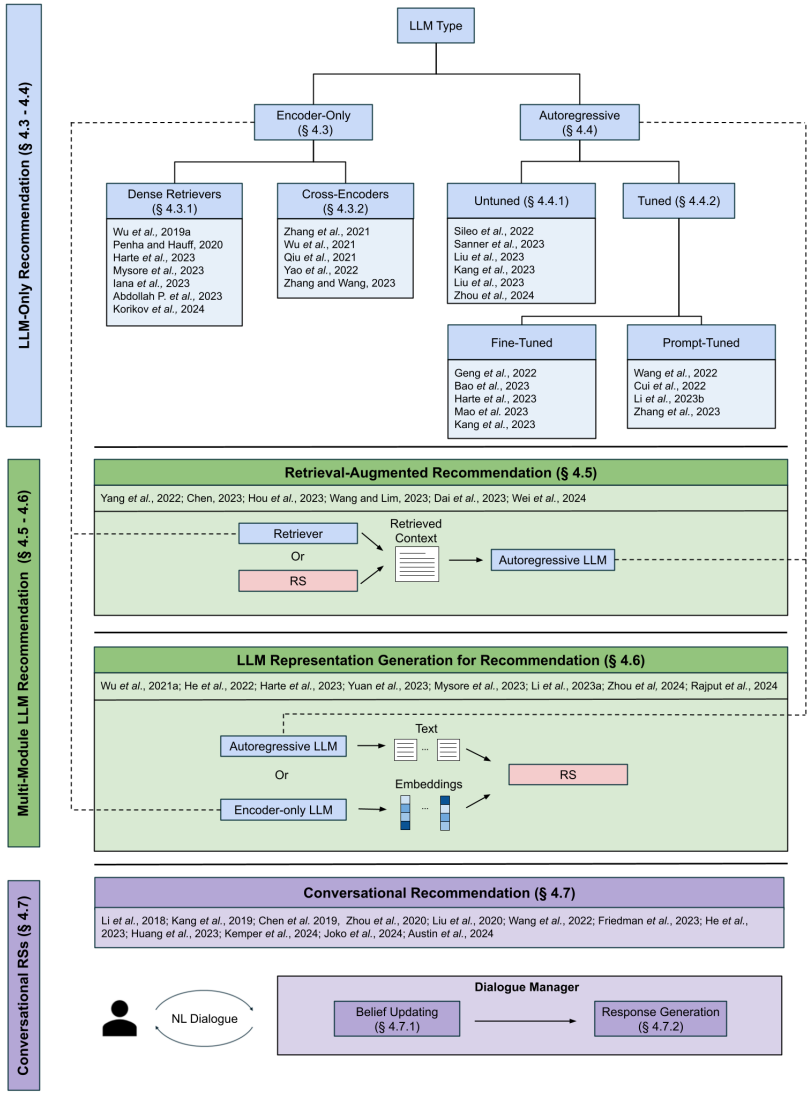

**Figure 4.2:** Outline of key techniques and publications in LLM-driven RSs.

**Recommendation and Explanation** Firstly, LLMs can use NL data (e.g., descriptions of items and a user's preferences) to make recommendations – either by generating text (c.f. Sec 4.4) or via embedding-based item scoring (c.f. Sec 4.3). Further, LLMs can generate textual explanations (c.f. 4.4.3) to help users understand why recommendations were made and enable better user feedback, such as NL critiques or follow-up questions.

**Conversational Recommendation**   LLMs can also drive NL recommendation dialogues where users *interactively* convey various intents, including: stating and refining preferences, critiquing recommendations, asking questions, or engaging in trade-off negotiations (c.f. Sec 4.7). Conversational recommendation systems (CRSs) can facilitate such complex dialogues by leveraging LLMs to generate a variety of personalized responses, including recommendations and explanations, answers to questions, and requests for more information. Thus, the general reasoning abilities of LLMs provide opportunities to better personalize not only recommendations, but also user-system interaction sessions more broadly.

**Limitations of LLM-Driven Recommendation**   Unfortunately, these LLM-driven opportunities for RSs also come with new risks, biases, and limitations, as discussed further in Chapter 7. Firstly, LLMs may hallucinate, generating outputs which are incorrect or misleading (Ji *et al.*, 2023) which creates significant risks in settings where reliability is key. More broadly, our ability to control LLM behaviour is limited: while prompt engineering and fine-tuning influence outputs, neither approach achieves total control (Mialon *et al.*, 2023). More optimistically however, this chapter also discusses approaches to mitigate some of these limitations, including through retrieval-augmented generation (RAG) and external tool calls to improve system control and reliability (c.f. Sec 4.5-4.7).

### 4.1.3   Chapter Outline

Before diving into recommendation methodologies, we first present a structured overview of NL data sources for describing items, users, and interactions in Section 4.2. Then, as summarized in Figure 4.2, the subsequent sections cover key techniques and research in LLM-driven recommendation. First, we describe single turn LLM recommendation, covering the use of both encoder-only and autoregressive models (c.f. Sec. 4.3-4.4). The next two sections focus on synergies between LLMs, conventional RSs, and information retrieval in a discussion of RAG (c.f. Sec. 4.5) and LLM-based representation generation (c.f. Sec. 4.6). Finally, we look at architectures for conversational recommendation, surveying various approaches for managing multi-turn and multi-intent dialogues (c.f. Sec 4.7).

## 4.2 Data Sources in LLM-Driven RSs

The use of language in recommendation is not new. For instance, text has long been leveraged for content-based recommendation (Lops *et al.*, 2011), key-phrase explanations (McAuley and Leskovec, 2013; Wu *et al.*, 2019c), and metadata-driven Dialogue State Tracking (DST, Yan *et al.*, 2017). However, LLM's have enabled far more advanced NL reasoning about items, users, and their interactions (Geng *et al.*, 2022), creating opportunities for more nuanced and interactive RS personalization. This section thus outlines the primary data sources which could be used by LLM-era RS, covering item data, interaction data, and user profiles, summarized in Figure 4.1.

### 4.2.1 Item Text

The right side of Figure 4.1 illustrates data sources for textual item representations, including titles, descriptions, metadata, and reviews.

### Titles

In most settings, items are assigned a *title*: a short, descriptive text span which aims to distinguish the item. A title's capacity to represent an item can vary greatly with domain and item popularity: for example while some hit movies and books can be summarized in one or two words (e.g., *"Titanic (1999)"*, *"Dune (1965)"*), less unique and less popular items (e.g., articles of clothing) do not have such information-dense titles. Items may even entirely lack titles in certain domains, such as user-generated social media video recommendation, though in these cases, it may be possible to generate titles from visual content and metadata with a multimodal LLM (c.f. Ch. 5).

**Descriptions**

Items may also be associated with longer and more detailed NL content which we call *descriptions*, which may be human-written or LLM generated (e.g. Acharya *et al.*, 2023; Wei *et al.*, 2024; Li *et al.*, 2023e). The length and format of descriptions can vary greatly across domains, ranging from non-existent (e.g., social media videos), to short summaries (e.g., eShopping products), to many pages long (e.g., news articles and books).

**Metadata**

Item metadata often contains *structured* information about item attributes such as product categories, brands, technical specifications, price, release date, and so on. It can contain various types of information including numerical, categorical, temporal, geographical, and visual data.

**Reviews**

A plentiful source of user-generated text are often reviews – which express nuanced opinions across a diverse range of item attributes and user experiences. However, reviews are often highly subjective, especially when describing "soft" attributes such as "inexpensive", "funny", or "safe" (Radlinski *et al.*, 2022b; Balog *et al.*, 2021).

### 4.2.2   Interaction Data

The top and bottom of Figure 4.1 illustrate sources of user-system interaction data, including both non-textual interactions (e.g., clicks, purchases, etc.) and NL interactions such as queries, dialogue utterances, and reviews (discussed in the previous section).

**Verbalizing Non-Textual Interactions**

Conventional non-textual user-item interaction history – such as views, clicks, likes, purchases, and ratings – can easily be represented as text. For this, a simple template may be sufficient – for instance, Hou *et al.* (2023) represent a user's movie viewing history with a template such as: *"I've watched the following movies in the past in order: 1. Multiplicity, 2. Jurassic Park, … "*. Alternatively, LLMs can be prompted to summarize such interaction histories (Yin *et al.*, 2023a; Zhou *et al.*, 2024; Wei *et al.*, 2024). In either case, such verbalized histories offer an alternative to the sometimes arbitrary mapping of distinct interactions types to numerical or categorical formats, and can also cover pointwise, sequential, and bundle interactions.

**NL Interactions**

Other user-system interactions are inherently text-based – for instance, reviews were already discussed in Section 4.2.1.

**Queries**   While RSs have traditionally focused on query-less personalization, *queries* – short text spans expressing a user's real-time information need – are becoming a key part of language-driven RSs (Reddy *et al.*, 2022; He *et al.*, 2022b). In fact, this integration of NL queries into RSs is contributing to the convergence of the recommendation and information retrieval (IR) fields.

**User-System Dialogue**   As conversational recommendation systems (CRS) develop, user-system *dialogues* are emerging as a primary source of textual interaction data (Li *et al.*, 2018a). Each user and system utterance can reflect diverse intents, including conveying preferences, recommendations, and explanations, as well as asking and answering questions about items or preferences (Lyu *et al.*, 2021), as discussed further in Section 4.7.

### 4.2.3    NL User Profiles

The left side of Figure 4.1 illustrates various textual representations of user preferences, including metadata and NL user profiles. Recently, Radlinski *et al.* (2022) proposed that language-driven recommendation could be centered on *scrutable* NL user profiles: textual descriptions of user preferences that are editable and understandable by humans. Such user profiles could succinctly summarize user interests through both specific examples of preferred items and generic preference descriptions.

While research on effectively generating, editing, and leveraging these interpretable NL preference representations is still nascent, several initial studies have emerged. For instance, Sanner *et al.* (2023) have users directly express generic NL item preferences in personal NL profiles, while other authors (Yin *et al.*, 2023a; Zhou *et al.*, 2024) use and LLM to generate NL profiles based on item rating histories – both works find that recommendation performance is competitive with or better than conventional baselines in cold-start settings.

**User Agency via Editable NL Profiles**    In principle, an editable NL profile has the potential to empower users to correct system errors, safeguard their privacy, and exert natural control over their preference representations (Radlinski *et al.*, 2022a). It may also be well-suited to handle *preference shifts* (Hosseinzadeh Aghdam *et al.*, 2015; Pereira *et al.*, 2018) by allowing users to delete obsolete interests or describe temporary contexts. In addition, users can express *aspirations* - desires that do not necessarily align with past behavior, but should influence future recommendations (Ekstrand and Willemsen, 2016). Finally, user edits that result in improved recommendations can incentivize further feedback and increase user satisfaction (Bostandjiev *et al.*, 2012; Harper *et al.*, 2015; Knijnenburg *et al.*, 2012).

### 4.3    Encoder-only LLM Recommendation

We now begin our discussion of how textual data can be used in LLM-driven RSs – starting with recommendation with encoder-only LLMs. As shown in Figure 4.3, encoder-only LLMs can be used in two main architectures: as dense retrievers (c.f. Sec 4.3.1) or as cross-encoders (c.f. Sec 4.3.2). The key difference is that dense retrievers encode preference and item descriptions *separately* while cross-encoders encode them *jointly*, which generally makes cross-encoders slower but more accurate.

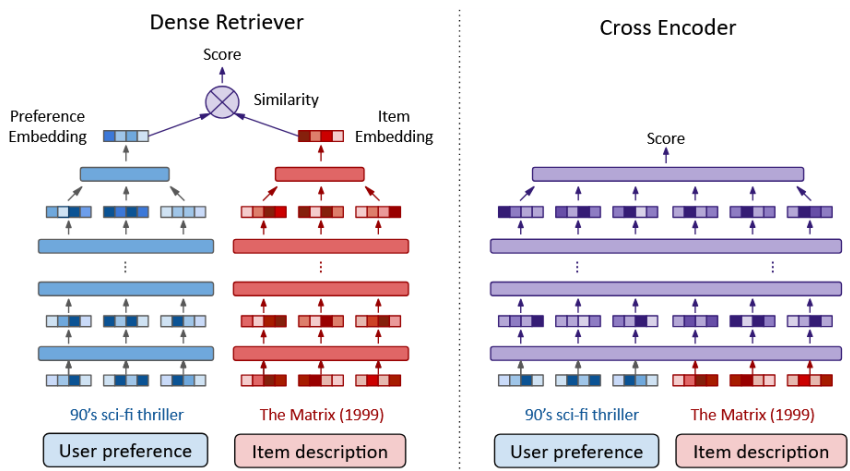

**Figure 4.3:** Two main architectures for encoder-only LLM recommendation: a) dense retrievers (left) which encode item and preference descriptions separately and then compute a preference-item embedding similarity score and b) cross-encoders (right) which jointly encode user preference and item descriptions to predict a score.

### 4.3.1 Dense Retrievers

First introduced in the field of information retrieval (IR), dense retrievers produce a ranked list of documents given a query by evaluating the similarity (e.g., dot product or cosine similarity) between a document embedding and query embedding (Fan *et al.*, 2022). Dense retrieval is highly scalable (especially with approximate search libraries like FAISS[1]) since document embeddings can be pre-computed and only the query embedding needs to calculated at query time.

To use dense retrieval for recommendation, first, a component of each item's text content, such as its title, description, reviews, etc., is treated as a document. Then, a query is formed by some NL user preference description – for instance: an actual search query, the user's recently liked item titles, or text generated based on a user utterance in a dialogue (Penha and Hauff, 2020).

---

[1]https://github.com/facebookresearch/faiss

Several recent works explore recommendation as standard dense retrieval, including with off-the-shelf (Penha and Hauff, 2020; Harte *et al.*, 2023; Zhang *et al.*, 2023a) and fine-tuned (Mysore *et al.*, 2023; Li *et al.*, 2023b; Hou *et al.*, 2023) retrievers. Dense retrieval is especially common in news recommendation since news articles are very rich in text – here, a common approach is to aggregate embeddings of a user's recently liked articles into a query embedding, either via mean pooling (Iana *et al.*, 2023; Iana *et al.*, 2024) or a multi-level encoder (e.g., Wu *et al.*, 2019b; Wu *et al.*, 2019a; Li *et al.*, 2022a), before scoring with news candidate embeddings. Another line of work focuses on retrieval using item reviews, and includes studies of contrastive retriever tuning (Abdollah Pour *et al.*, 2023) and multi-aspect retrieval (Korikov *et al.*, 2024).

### 4.3.2  Cross-Encoders

In contrast to dense retrievers, cross-encoders embed a query and document *jointly*, allowing cross-attention between query and document tokens (Fan *et al.*, 2022). Several works approach rating prediction by jointly embedding NL item and preference descriptions in LLM cross-encoder architectures with a rating prediction head (Zhang *et al.*, 2021b; Yao *et al.*, 2022; Wu *et al.*, 2021; Qiu *et al.*, 2021; Zhang and Wang, 2023). Such fusion-in-encoder methods often exhibit strong performance because they allow interaction between user and item representations, but are much more computationally expensive than dense retrieval and thus may be best used for small item sets or as rerankers.

### 4.4  Generative Recommendation and Explanation

We next move beyond item-preference scoring to generative recommendation (c.f. Sec 4.4.1-4.4.2) and explanation (c.f. Sec 4.4.3) with autoregressive LLMs. Autoregressive LLM inputs are called *prompts*, which are sequences of tokens expressing a task such as top-k recommendation, rating prediction, or explanation generation (Geng *et al.*, 2022). As illustrated in Figure 4.4, prompts typically consist of word tokens (Figure 4.4 a-b), but may also feature non-word tokens such as user and item IDs (Figure 4.4 c) that can be learned through tuning (c.f. Sec. 4.4.2). While the space of possible LLM inputs and outputs is extremely large, common tasks include generating:

- a recommended list of item titles or ids (e.g., Mao *et al.*, 2023; Harte *et al.*, 2023; Sanner *et al.*, 2023; Sileo *et al.*, 2022)

- item ratings (e.g. Bao *et al.*, 2023; Kang *et al.*, 2023; Zhou *et al.*, 2024)

- explanations of recommendations (e.g., Ni *et al.*, 2019a; Li *et al.*, 2020; Hada and Shevade, 2021; Geng *et al.*, 2022; Li *et al.*, 2023e)

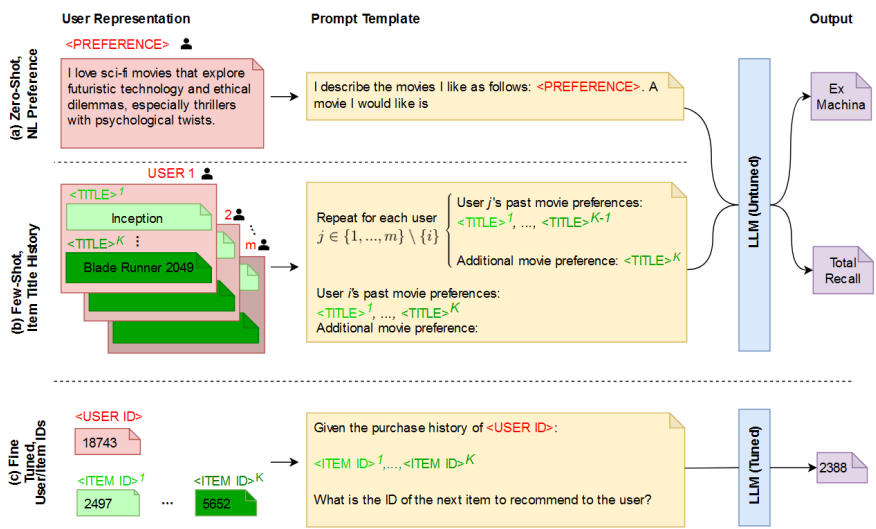

**Figure 4.4:** Three examples of generative recommendation with autoregressive LLMs, showing alternative user preference representations and prompting approaches. a) A ZS prompt with a user-generated NL preference description. b) A FS prompt based on sequences of liked item titles from multiple users. c) An ID-based prompt to a tuned LLM which has learned to use item and user ID tokens.

This section further discuses such single-turn (i.e., non-conversational), autoregressive LLM recommendation and explanation, covering both untuned and tuned approaches.

### 4.4.1   Zero- and Few-Shot Recommendation

The simplest way an autoregressive LLMs can be used is in the *untuned* (i.e., "off-the-shelf") setting (e.g. Sileo *et al.*, 2022; Sanner *et al.*, 2023). As illustrated in Figure 4.4, this includes:

- zero-shot (ZS) approaches, which rely solely on the LLM's pre-trained knowledge without any additional training data (Figure 4.4 a)

- few-shot (FS) approaches, which provide a small number of input-output examples in the prompt (Figure 4.4 b), and thus are also referred to as in-context-learning (ICL) methods.

**Zero- and Few-Shot Prompt Engineering**   Clearly, there is a large design space for prompting approaches, including choices for representing user preferences, specifying task instructions, and selecting few-shot examples. Figure 4.4 illustrates two specific examples from this design space (Sanner *et al.*, 2023): a) a ZS prompt with a user-generated NL preference description, and b) a FS prompt based on sequences of liked movie titles from multiple users. Many other variants are possible, for instance: constructing FS examples based on interactions from the *same* user instead of from different users, including user *dis*preferences, or using LLM-generated user profiles (Zhou *et al.*, 2024).

**Initial Experimental Findings**   Several recent publications have evaluated off-the-shelf LLMs for movie and book recommendation – domains where relevant knowledge is likely to be internalized during pretraining. Specifically, these methods construct prompts using NL representations of user preferences and instructions to recommend item titles (Sanner *et al.*, 2023; Sileo *et al.*, 2022; Liu *et al.*, 2023b) or predict ratings (Kang *et al.*, 2023; Liu *et al.*, 2023b; Zhou *et al.*, 2024). These initial studies find that while untuned LLMs generally underperform supervised CF methods given sufficient training data (Kang *et al.*, 2023; Sileo *et al.*, 2022), they are competitive in near cold-start settings (Sileo *et al.*, 2022; Sanner *et al.*, 2023; Zhou *et al.*, 2024). They also suggest that FS typi-cally outperforms ZS prompting and that LLMs struggle with negated reasoning in recommendation (e.g. reasoning about *dis*preferences).

### 4.4.2  LLM Tuning for Generative Recommendation

To improve an LLM's generative recommendation performance, multiple works study LLM tuning on historical RS data. First, historical data is converted into textual input-output training pairs for generative recommendation tasks (Geng *et al.*, 2022), which may include:

- top-*k* recommendation (e.g., Geng *et al.*, 2022; Li *et al.*, 2023f), such as in Figure 4.4 a-b

- sequential recommendation (e.g., Harte *et al.*, 2023; Li *et al.*, 2023f; Mao *et al.*, 2023), such as in Figure 4.4 c

- rating prediction (e.g.,Bao *et al.*, 2023; Kang *et al.*, 2023; Zhang *et al.*, 2023d)

Notably, Geng *et al. et al.* (2022) point out that text is a unifying format for training data, enabling multi-task LLM tuning on all of the above tasks. They also show that LLMs can learn to use user and item ID tokens for these tasks (e.g., Figure 4.4 c).

Given textual RS training data, the two main approaches to LLM tuning are:

- *fine-tuning*, where training set performance is optimized by adjusting LLM weights (Geng *et al.*, 2022; Bao *et al.*, 2023; Harte *et al.*, 2023; Mao *et al.*, 2023; Kang *et al.*, 2023)

- *prompt-tuning* where training set performance is optimized by learning prompt tokens (hard prompt-tuning) or embeddings (soft prompt-tuning) (Li *et al.*, 2023e; Cui *et al.*, 2022; Zhang *et al.*, 2023d)

### 4.4.3  Generative Explanation

Autoregressive LLMs can also generate explanations that aim to help users comprehend why recommendations were made. Similarly to recommendation generation, explanations can be generated by prompting

LLMs that are either untuned (e.g., Rahdari *et al.*, 2024) or tuned (e.g., Li *et al.*, 2023e) – though the latter is more common for academic publications. To tune LLMs for RS explanation, the typical data source for constructing textual training examples are user reviews, as these are assumed to contain justifications for users' opinions about the reviewed items (Ni *et al.*, 2019a).

Broadly, recently studied techniques for RS explanation generation include:

- ZS and FS prompting (Rahdari *et al.*, 2024)

- fine-tuning (Geng *et al.*, 2022; Li *et al.*, 2023e; Wang *et al.*, 2024c)

- prompt-tuning (Li *et al.*, 2023e; Li *et al.*, 2020)

- controllable decoding – where predicted parameters such as ratings steer LLM decoding (Ni and McAuley, 2018; Ni *et al.*, 2019a; Hada and Shevade, 2021; Xie *et al.*, 2023)

## 4.5   Retrieval Augmented Recommendation

While the previous two sections explore the use of LLM knowledge internalized through pretraining or tuning to generate recommendations and explanations, relying solely on internal LLM knowledge has several limitations (Lewis *et al.*, 2020):

- a relatively large number of LLM weights are needed to store knowledge

- retraining is needed for each knowledge update

- there is no inherent source attribution mechanism

To address these limitations, recent work (e.g., Lewis *et al.*, 2020; Izacard and Grave, 2020; Borgeaud *et al.*, 2022; Mialon *et al.*, 2023) explores a framework called retrieval-augmented generation (RAG) in which: 1) relevant content is retrieved from an external knowledge source and 2) the retrieved content is used to prompt an autoregressive LLM to generate textual output. These studies provide evidence that, in contrast to approaches relying solely on internal LLM memory, RAG can: reduce the number of LLM parameters (since knowledge can be externalized); provide a convenient mechanism for knowledge updates; improve factuality through better source attribution and by reducing hallucinations.

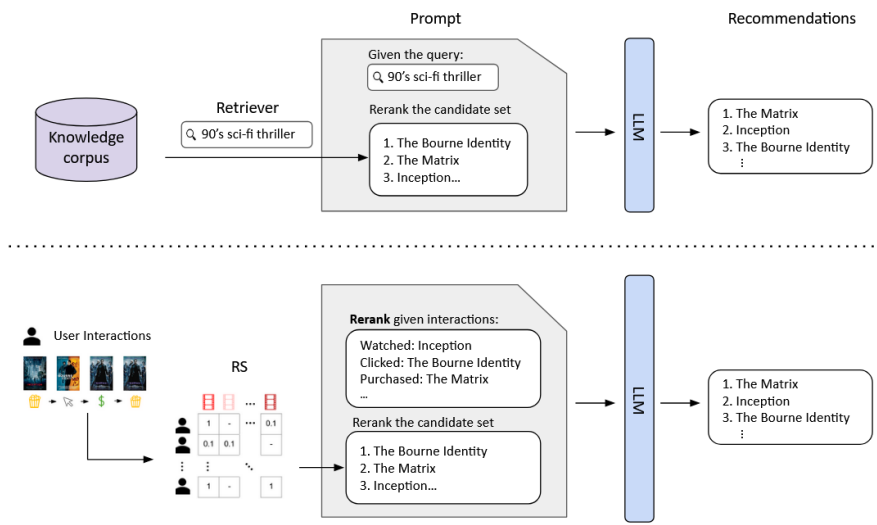

**Figure 4.5:** Two examples of RAG for top-$k$ recommendation, where an external tool produces a candidate item set which an LLM is prompted to rerank given a textual description of user preferences. Top: A query is used by a retriever to search for a candidate item set, which an LLM is prompted to rerank given the query. Bottom: A user's interaction history is used by an RS to select a candidate item set, which an LLM is prompted to rerank given the interaction history.

## 4.5.1   RAG in RSs

There are many opportunities to use RAG in RSs, including to generate recommendations, explanations, and question answers. More broadly, the RAG framework introduces us to modular architectures for LLM-driven RSs – a concept we'll explore further when discussing LLM representation generation, and conversational RS architectures (c.f. Sec. 4.7).

**RAG for Recommendation**    As illustrated in Figure 4.5, the most commonly studied RAG recommendation method has been LLM candidate item set reranking. Specifically, this method involves: 1) selecting a candidate item set based on RS interaction data and 2) prompting an LLM to rerank the candidate set given some information about user preferences (Yang *et al.*, 2022a; Hou *et al.*, 2023; Chen, 2023; Wang and Lim, 2023; Dai *et al.*, 2023; Wei *et al.*, 2024). Recall from Section 4.2 that user preferences can be represented in diverse forms, leading to many alternatives for candidate selection and reranking methods. For instance, Figure 4.5 a) illustrates candidate selection with a retriever driven by a user query, and LLM candidate reranking given this query. As another example, Figure 4.5 b) shows candidate selection with an RS based on item interaction history, and LLM reranking given this history.

**RAG for Explanation and Conversational Recommendation**    Other examples of how RAG can be used in RSs include explanation generation and as a tool for conversational recommendation. For RAG-based explanation generation, Xie *et al.* (2023) generate queries based on interaction history to retrieve an item's reviews, which are then used as context to generate an explanation of the recommendation. Examples of RAG in conversational recommendation, discussed further in Section 4.7, include work by Friedman *et al.* (2023) to retrieve relevant user preference descriptions from a user "memory" module, and by Kemper *et al.* (2024) to retrieve information from an items reviews to answer user questions.

## 4.6   LLMs Representation Generation

While Section 4.5 explored using an external module such as retrievers or RSs to select LLM inputs, this section discusses the converse setting in which LLMs generate inputs to downstream modules, as illustrated in 4.6. Specifically, LLMs can be used to transform text such as item and preference descriptions into representations (e.g., embeddings, text, item ratings) that serve as inputs to modules such as RSs, retrievers, or autoregressive LLMs. Such text transformations can be interpreted as an LLM encoding step (Zhou *et al.*, 2024), including for the case of text-to-text transformations – such as the generation of user preference profiles based on user history.

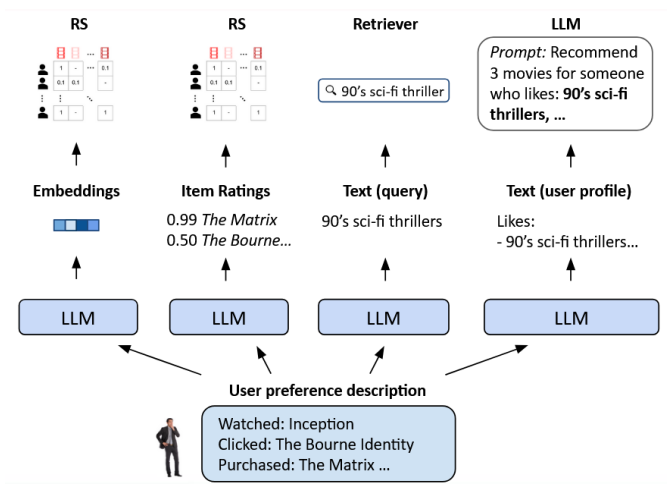

**Figure 4.6:** Examples of LLM generated-inputs for downstream modules, including embeddings and predicted item ratings for downstream RSs, and text for downstream search or LLM prompting.

### 4.6.1 Text to Text

Two common settings where an LLM generates text that serves as input to a downstream module are search queries and LLM prompt elements, with several example studies for both approaches discussed below.

**Search Queries** Given a user's item interaction history (and item text), MINT (Mysore *et al.*, 2023) and GPT4Rec (Li *et al.*, 2023b) prompt an LLM to generate an search query for a dense retriever with the goal of finding new items to recommend. Similarly, recent work on conversational recommendation uses dialogue history to generate search queries to find items to recommend (Friedman *et al.*, 2023; Kemper *et al.*, 2024).

**Prompt Elements**    As mentioned in Section 4.2.3, recent work (Yin *et al.*, 2023a; Zhou *et al.*, 2024) first uses an LLM to generate a NL user profile based on their interaction history, then prompts an LLM to generate recommendations given this profile. In conversational recommendation, LLMs are often employed to generate prompt elements (or entire prompts) for downstream tasks based on dialogue history (e.g., Wang *et al.*, 2022c; Friedman *et al.*, 2023; Gao *et al.*, 2023b; Kemper *et al.*, 2024).

### 4.6.2    Text to Embeddings

As discussed in Chapters 2 and 3, many RS techniques rely on latent representations of items and user preferences. Thus, a recent line of research uses LLMs to encode semantic information from text into latent embeddings which are then used for recommendation. While encoder-only LLM recommendation was already discussed in Section 4.3, this section covers multi-stage pipelines where LLM embeddings are used for downstream recommendation modules, discussed next.

**Sequential Recommendation with LLM Embeddings**    Recall that Chapter 3 discussed several attention-based sequential recommendation models (e.g., BERT4Rec **REF**, SASRec **REF**) which predict the next item ID to recommend given a sequence of item IDs that a user has interacted with. Recent work uses LLMs to initialize item ID embeddings based on item text (Harte *et al.*, 2023; Yuan *et al.*, 2023; Rajput *et al.*, 2024), reporting significant performance gains. Similarly, Query-SeqRec (He *et al.*, 2022b) incorporates user query information into sequential recommendation by using an LLM to encode queries alongside item embeddings.

**Rating Prediction with LLM Embeddings**    LLMs have also been used to generate latent representations of items (based on item text) and users (based on item interactions) that serve as inputs to a neural network which predicts a user-item rating (Wu *et al.*, 2021; Yuan *et al.*, 2023). These approaches essentially augment the well-known neural collaborative filtering (He *et al.*, 2017b) framework with LLM embeddings.

### 4.6.3  Text to Item Ratings

As discussed further in Section 4.7, LLMs have been used to map dialogue history to item rating predictions, with these ratings then treated as observations by an RS to make recommendations. Examples include sentiment analysis towards items mentioned in a dialogue (Li *et al.*, 2018b) and natural language inference between user-stated aspect preferences and item text (Austin *et al.*, 2024).

## 4.7  Conversational Recommendation

The previous sections mostly focused on single-turn LLM recommendation personalized based on *pre-existing* user history information, such as non-verbal interactions, queries, reviews, and so on (c.f. Figure 4.1), and item text. However, LLMs provide new opportunities for multi-turn conversational recommender systems (CRSs) where each turn presents a chance for the user to clarify or revise their preferences, critique and ask questions about recommended items, or convey a variety of other real-time intents such as those shown in Table 4.1 (Lyu *et al.*, 2021). Correspondingly, CRSs should facilitate a wide range of responses including revising recommendations, responding to questions, and personalizing explanations. Further, each turn is also an opportunity for the CRS to generate proactive utterances such as clarifying questions or explanations focused on key topics to help elicit user preferences. As with user intents, various possible CRS actions are summarized in Table 4.2 (Lyu *et al.*, 2021). LLM-driven CRSs thus present opportunities not only to personalize recommendations, but also to personalize system interactions more broadly.

**Pre-LLM CRS Architectures**   Historically, most pre-LLM CRSs followed a two-step process in each turn: 1) *belief tracking* to maintain a dialogue state, and 2) *response generation* to produce a system utterance (Jannach *et al.*, 2020). Belief tracking was often implemented with slot-filling techniques, where the dialogue history would be used to update a *dialogue state* consisting of variables, or slots (e.g., "*preferred_cuisine: _*"), that were filled from a set of predefined values (e.g., "*Mexican*", "*French*", ...) (Williams *et al.*, 2014; Budzianowski *et al.*, 2018). These slot-based states would then be used to determine the system response, which often used NL templates and item metadata. However, these slot-based architectures exhibited limited NL reasoning capabilities, constraining their capacity to understand and represent complex user intents and generate deeply personalized responses and recommendations.

**Table 4.1:** The user intent conversational recommendation taxonomy of Lyu *et al.* (2021), with † indicating a additional category to the taxonomy of Cai and Chen (2020).

| Category | Description | Example |
|---|---|---|
| **Ask for Recommendation** | | |
| Initial Query | User asks for a recommendation in the first query | "Hi I am looking for a place to have a family brunch..." |
| Continue | User asks for another recommendation in a subsequent query | "Maybe you can give me one more choice so I can pick one..." |
| **Provide Preference** | | |
| Provide Context † | User provides background information for the restaurant search | "I am looking for a restaurant for my Valentine's day dinner." |
| Provide Preference | User provides specific preference for the desired item | "I would prefer a place that has a very good scenic view." |
| Refine Preference † | User improves over-constrained/under-constrained preferences | "It does not have to be chicken fingers." |
| **Answer** | User answers the question issued by the recommender | "Yes that's correct." |
| **Acknowledgement †** | User shows understanding towards a previous recommender utterance | " I see." |
| **Recommendation Rating** | | |
| Been to (modified) | User has been to the restaurant before | "Oh I have been there before." |
| Accept | User accepts the recommended item, either explicitly or implicitly | "Ok our final choice will be Eggspectation." |
| Reject | User rejects the recommended item, either explicitly or implicitly | "Maybe there is a private room in the other three restaurants?" |
| Neutral Response | User does not indicate a decision with the current recommendations | "I will take a look in the menu and compare and maybe ask my partner." |
| Inquire | User requires additional information regarding the recommendation | "So what about the interior design, the decorations and environment?" |
| **Critiquing** | | |
| Critique - Feature | User critiques on a specific feature of the recommended item | "I am pretty sure it will be expensive so what is the price range?" |
| Critique - Add | User adds further constraints on top of the current recommendation | "I want sushi." |
| Critique - Compare | User requests comparison between recommended item with another item | "How about the price compared with Miku??" |
| **Others** | Greetings, gratitude expression, chit-chat utterance | "Thank you so much for your recommendation." |

**Table 4.2:** The recommender intent conversational recommendation taxonomy of Lyu *et al.* (2021), with † indicating a additional category to the taxonomy of Cai and Chen (2020).

| Category | Description | Example |
|---|---|---|
| **Request** | | |
| Request Information | Recommender requests the user's preference | "What kind of food do you like?" |
| Clarify Question | Recommender asks for clarification on a previous requirement | "So you would like to reserve a private room?" |
| Ask Opinion † | Recommender requests the user's opinion to a choice question (e.g., yes/no) | "So it is just open space but separated from others, is that ok?" |
| Ensure fulfilment † | Recommender confirms task fulfillment during the conversation | "Anything else I can do for you today?" |
| **Inform progress †** | Recommender discloses the current item being processed | "So let me just check the closest nearby parking." |
| **Acknowledgement †** | Recommender shows understanding towards a previous user utterance | "...you mentioned that one of the attendees is vegetarian..." |
| **Answer** | Recommender answers the question issued by the user | "So for the Michael's on Simcoe, the price varies a lot..." |
| **Recommend** | | |
| Recommend - Show | Recommender provides recommendation by showing it directly | "So I found a restaurant called paramount." |
| Recommend - Explore | Recommender provides recommendation and asks if the user has prior knowledge | "The first one that comes to mind is Miku, have you heard of it before?" |
| **Explain** | | |
| Preference | Recommender explains recommendations based on the user's said preference | "Because it has vegetarian options, it has a full bar and a good view..." |
| Additional Information † | Recommender explains recommendations with features not previously discussed | "There are a couple different varieties (of food) that your guests might enjoy." |
| **Personal Opinion** | | |
| Comparison † | Recommender compares recommended item with another item | "I would say the price for this place is a bit higher than HY steakhouse but..." |
| Persuasion † | Recommender provides positive context towards the recommended item | "It is on the pricier side but it is worth the experience..." |
| Prior Experience † | Recommender refers to past experience with the recommended item | "I have been there before during the summerlicious." |
| Context † | Recommender provides opinion considering the given context or current reality | "Since it's summer I don't think the weather will be that much of an issue..." |
| **Others** | Greetings, gratitude expression, chit-chat utterance | "Yeah a lot of people recommended me to go there." |

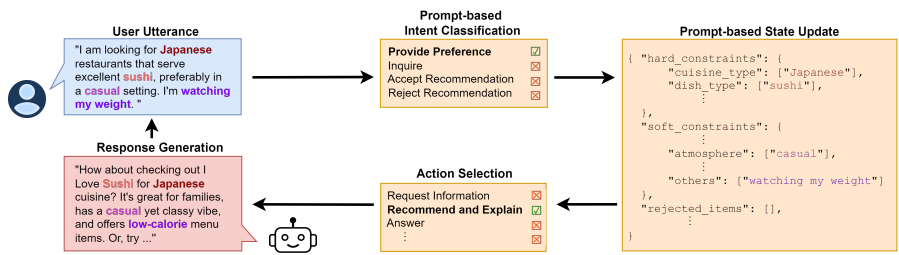

**Figure 4.7:** The *RA-Rec* prompt-driven dialogue management approach (Kemper *et al.*, 2024). LLM prompting is used to maintain a JSON semi-structured NL state which tracks user preferences and intents. The predefined state keys (e.g., "*cuisine_type*") provide a configurable structure, while the LLM-generated state values can dynamically express nuanced concepts in NL (e.g., " *"watching my weight"*).

**LLM-Driven CRSs**   In contrast to highly constrained slot-based dialogue states described above, LLMs enable rich textual representations of the dialogue state. For instance, if we consider the basic case of using a monolithic LLM *as* a CRS (e.g., He *et al.*, 2023), the dialogue state simply becomes the conversation history, which preserves the full dialogue context. As discussed further in Section 4.7.1, recent CRS research often explores more complex text-augmented dialogue states – expanding or replacing the raw conversation history with elements such as JSON-like dialogue states (e.g., Figure 4.7), user preference summaries (Friedman *et al.*, 2023; Joko *et al.*, 2024), or numerical variables (e.g., inferred item ratings). These text-augmented dialogue states are then used to form inputs to response generation modules, which may include not only LLMs but also tools such as retrievers, RSs, and KGs, as discussed in Section 4.7.2.

### 4.7.1   Belief Tracking

LLMs enable a CRS to track its beliefs about the conversation through text-augmented dialogue states. This section covers the use of purely textual components in dialogue states as well as the incorporation of non-textual elements alongside text. Both textual and non-textual dialogue state elements can form inputs to response generation modules, discussed further in Section 4.7.2.

**Textual Dialogue State Components**

The most basic form of a textual dialogue state is simply the conversation history – the default state representation when a monolithic LLM is used as a CRS (e.g., He *et al.*, 2023). However, the pure conversation history can also be extended or substituted with other textual state components. For instance, Kemper *et al.* (2024) prompt an LLM to convert the conversation history into a JSON dialogue state, as shown in Figure 4.7, which provides a semi-structured format to represent beliefs about the user's item preferences (e.g., preferred cuisine type) and the dialogue flow (e.g., whether the user is expecting an answer to a question). Other authors (Friedman *et al.*, 2023; Joko *et al.*, 2024) prompt an LLM to generate textual user preference *memories,* and store these memories as documents in a long-term memory corpus.

**Semi-textual Dialogue States**

In addition to textual components such as those described above, other works also feature categorical or numerical variables in the dialogue state. Specifically, the conversation history can be used to assign values to variables such as:

- liked/disliked item titles (Liu *et al.*, 2020) or item categories (Huang *et al.*, 2023)

- mentioned item-related KG entities (Chen *et al.*, 2019a; Zhou *et al.*, 2020; Liu *et al.*, 2020; Ma *et al.*, 2020; Wang *et al.*, 2022c)

- user constraints in a constraint reasoning-based CRSs (Zeng *et al.*, 2024)

Dialogue states can also include numerical variables – for instance Li *et al.* (2018) use a sentiment classification module to infer a user rating towards mentioned items, while Austin *et al.* (2024) maintain Bayesian beliefs over user-item preferences.

### 4.7.2 System Response Generation

The primary purpose of the dialogue state elements discussed above is to guide the system response by forming inputs to response generation modules. For instance, textual state components may be used for LLM prompts and/or retriever queries, while non-textual state elements such as item ratings may function as arguments to tools such as RSs. Broadly, LLM-driven CRS response generation can be guided through prompting, tuning, and interfacing with external tools.

**Prompting**

There are many diverse ways that prompting can be used to guide CRS utterance generation. Several works study *single-intent* systems, where the prompt is constructed using the dialogue history and an instruction to generate an utterance with a fixed intent, such as to recommend (He *et al.*, 2023) or ask a preference elicitation query (Handa *et al.*, 2024; Austin *et al.*, 2024). Other methods (Joko *et al.*, 2024; Kemper *et al.*, 2024) express hand-crafted dialogue rules through prompts, asking an LLM to select the best system action given the dialogue state and some rules (e.g., "*Ask a user for their location if they have not provided it, otherwise, recommend a restaurant.*"). In addition, prompts can be partially or fully system generated, for instance using the output of another LLM module, a retriever, or a reasoning tool such as an RS or KG (e.g., Zhou *et al.*, 2020; Wang *et al.*, 2022c; Friedman *et al.*, 2023; Gao *et al.*, 2023b).

**Tuning**

CRS dialogue policies can also be controlled by tuning LLMs on human-human or synthesized conversation data (Li *et al.*, 2018a; Kang *et al.*, 2019; Zhou *et al.*, 2020; Liu *et al.*, 2020; Ma *et al.*, 2020; Li *et al.*, 2022c; Wang *et al.*, 2022c; Friedman *et al.*, 2023; Joko *et al.*, 2024). While these methods cover many tuning approaches, they all lead to an LLM internalizing some knowledge about how to respond to the user in NL or execute some reasoning step such as a search or recommendation.

**Tool Use**

LLM-driven CRS response generation can also be augmented by the use of external tools such as retrievers or RSs. The techniques for interfacing

LLMs with such tools – discussed in Sections 4.5-4.6 – remain applicable in multi-turn settings, with the dialogue now state serving as the basis for representing tool inputs and outputs. In the context of retrieval-augmented CRSs, recent work (Friedman *et al.*, 2023; Kemper *et al.*, 2024) generates a search query based on preferences represented in the dialogue state to retrieve relevant items based on item text. Similarly, RS modules are used to make recommendation based on inferred item ratings (Li *et al.*, 2018a; Austin *et al.*, 2024) and recognized KG entities (e.g., Chen *et al.*, 2019a; Wang *et al.*, 2022c).

# 5

## Multi-modal Generative Models in Recommendation System

ABSTRACT

The recommendation systems discussed so far typically limit user inputs to text strings or behavior signals such as clicks and purchases, and system outputs to a list of products sorted by relevance. With the advent of generative AI, users have come to expect richer levels of interactions. In visual search, for example, a user may provide a picture of their desired product along with a natural language modification of the content of the picture (e.g., a dress like the one shown in the picture but in red color). Moreover, users may want to better understand the recommendations they receive by visualizing how the product fits their use case, e.g., with a representation of how a garment might look on them, or how a furniture item might look in their room. Such advanced levels of interaction require recommendation systems that are able to discover both shared and complementary information about the product across modalities, and visualize the product in a realistic and informative way. However, ex-isting systems often treat multiple modalities independently: text search is usually done by comparing the user query to product titles and descriptions, while visual search is typically done by comparing an image provided by the customer to product images. We argue that future recommendation systems will benefit from a multi-modal understanding of the products that leverages the rich information retailers have about both customers and products to come up with the best recommendations.

In this chapter we discuss recommendation systems that use multiple data modalities simultaneously. As we shall see, a key challenge in developing multimodal generative models is to ensure that the features extracted from each modality are adequately *aligned* across modalities, i.e., mapped to nearby points in the embedding space. Since the problem of jointly learning a generative model for each modality and their alignment is extremely difficult (Chen *et al.*, 2020a), a common approach is to use contrastive learning methods to approximately align the modalities before learning a multimodal generative model. Therefore, in this chapter we will review both contrastive and generative approaches to multimodal recommendation. More specifically, in Section 5.1 we will provide a brief introduction to multimodal recommendation systems, in Section 5.2 we will review contrastive approaches to multimodal recommendation, and in Section 5.3 we will discuss generative approaches. Finally, in Section 5.4 we will overview various applications of multimodal recommendation systems. Throughout the chapter, we will center the discussion around vision and language models due to the larger volume of work for these two modalities, but we note there is a growing literature of generative recommendation systems that combine other modalities such as audio and text (Vyas *et al.*, 2023), video and audio (Ruan *et al.*, 2023), or even more than two modalities (Wu *et al.*, 2023b).

## 5.1 Introduction to Multimodal Recommendation Systems

### 5.1.1 Why do we need multimodal recommendation systems?

Retailers have a lot of information about their customers and the items they sell, including purchase history, customer interactions, product descriptions, product images and videos, and customer reviews. However, existing recommendation systems typically process each data source independently and then combine the recommendation results. For example, text search is typically done by comparing a short user query to product title, descriptions and reviews, while visual search is typically done by comparing an image provided by the customer to product images. Both search approaches produce a list of products sorted by relevance, and current "multimodal" systems simply fuse unimodal relevance scores to produce a single list of products from both modalities. In practice, there are many use cases in which such a "late fusion" approach may be insufficient for satisfying the needs of the user.

One such use case, known as the *cold start problem*, occurs when new users start using the system, or new products are added to the catalog, hence user behavioral data cannot be leveraged to recommend new products to existing users or existing products to new users. To alleviate this problem, it is useful to gather diverse information about the items so that preference information can be transferred from existing products or users to new ones. To this end, models that combine information from multiple modalities offer a unique advantage. For example, if a store receives a new product (e.g., a dress), but no purchases have been made yet, we can use the visual similarities between the new dress and existing ones in the store to determine which customers could be interested in it.

Another use case occurs when different modalities are necessary to understand the user request. For example, to answer the request "best metal and glass black coffee table under $300 for my living room", the system would need not only the text query but also an image of the customer's living room in order to find a table that best matches the room. Moreover, answering this customer's question requires reasoning about the appearance and shape of the item in context with the shape

and appearance of many other objects, as well as limiting the search by price, which cannot be achieved by searching with either the text or image independently. Other examples of multimodal requests include an image or audio of the desired item together with modification instructions in text (e.g., a dress like the one in the picture but in red, a song like the sound clip provided but in acoustic), or a complementary related product (e.g., a kickstand for the bicycle in the picture, or other movies from the actress talking in the video clip).

A third use case where multimodal understanding becomes crucial is when considering more complex recommendation systems, like those featuring virtual try-on capabilities, or intelligent conversational shopping assistants (Mehta, 2024; Templeton, 2024). To be effective, AI shopping assistants will need to be able to understand the context of previous interactions in the conversation history. Let's consider the example of a customer looking for a complete outfit he is planning to wear during the summer in Cairo, to attend the wedding of a friend with traditional tastes. An AI shopping assistant interacting with the customer will have to resort to visual cues to recommend products compatible as an outfit, as well as other customer preferences expressed earlier, the climate in Cairo during the summer, and cultural or dress code norms.

### 5.1.2   Key challenges in designing multimodal recommendation systems

The development of multimodal recommendation systems faces several challenges.

- First, combining different data modalities to improve recommendation results is not simple. Existing systems learn joint representations that capture information that is shared across modalities (e.g., the text query refers to a visual attribute of the product that is visible in the image), but they ignore complementary aspects that could benefit recommendations (Guo *et al.*, 2019); e.g., the text mentions inside pockets not visible in the picture, or the image contains texture patterns that are hard to describe precisely in text. Therefore, when learning multimodal representations it is important to ensure adequate alignment of the aspects that need to be aligned, while leaving some flexibility to capture complementary information across modalities as well. In general we will want the modalities to compensate for one another and result in a more complete joint representation.

- Second, collecting aligned data from multiple modalities to train multimodal recommender systems is significantly more difficult than collecting data for individual data modalities. For example, in the unimodal case one can define positive pairs for contrastive learning via data augmentation, but in the multimodal case such positive pairs often need to be annotated (see Section 5.2). In practice, existing annotations may be incomplete for some modalities (Rahate *et al.*, 2022). For example, visual search with text modification would require examples of an input image, the textual modification, and the modified image, but typically only two of the three are available, e.g., image-caption pairs.

- Third, learning a latent space that can be used for generative tasks is often harder than for discriminative tasks, as it typically requires larger datasets and computational resources to be able to adequately learn the data distribution by using more complex losses (Chen *et al.*, 2020a). This challenge is further exacerbated in the case of multimodal data because we need to not only learn a latent representation for each modality but also ensure that these latent representations are adequately aligned.

Despite these challenges, multimodal generative models are a promising technology for improving recommendation systems. Indeed, recent literature shows tremendous advances on the necessary components to achieve effective multimodal generative models for recommender systems, including 1) the use of LLMs and diffusion models to generate synthetic data for labeling purposes (Brooks *et al.*, 2023; Rosenbaum *et al.*, 2022; Nguyen *et al.*, 2024), 2) high quality unimodal encoders and decoders (He *et al.*, 2022a; Kirillov *et al.*, 2023), 3) better techniques for aligning the latent spaces from multiple modalities into a shared one (Radford *et al.*, 2021; Li *et al.*, 2022b; Girdhar *et al.*, 2023), 4) efficient re-parametrizations and training algorithms (Jang *et al.*, 2016), and 5) techniques to inject structure to the learned latent space to make the problem tractable (Croitoru *et al.*, 2023; Yang *et al.*, 2023b). Once trained, generative recommender systems are more versatile, and can produce better recommendations in more general, open ended tasks.

### 5.1.3   Multimodal recommendation systems covered in this chapter

In the remainder of this chapter, we will review both contrastive and generative approaches to multimodal recommendation. In Section 5.2 we will review contrastive approaches, such as CLIP, which learns to map each modality to a common latent space in which the modalities are approximately aligned. In Section 5.3 we will discuss generative approaches, such as ContrastVAE, which learns a probabilistic embedding from each modality to a common latent space where modalities are approximately aligned, and DALL-E 2, Stable Diffusion, LLAVA and multimodal LLMs, which learn to generate image recommendations given an input text prompt.

## 5.2   Contrastive Multimodal Recommendation Systems

As discussed in Chapter 4.3.1, many recommendation approaches like Du *et al.*, 2022 rely on learning an embedding of the data such that similar items are close to each other in the embedded space. In the case of multimodal data, a natural approach to learning a *multimodal embedding* would be to learn one embedding per modality, as done by (He *et al.*, 2020; Grill *et al.*, 2020; Chen *et al.*, 2020b; Caron *et al.*, 2021) for images, or (Saeed *et al.*, 2021; Won *et al.*, 2020; Wang *et al.*, 2022a) for audio, and then concatenate such *unimodal embeddings*. Such an approach is adequate when different modalities capture complementary aspects of an item. However, as discussed in Section 5.1.2, when different modalities capture related aspects of an item, unimodal embeddings need to be adequately aligned to ensure that similar items are close to each other in the multimodal embedded space. For example, to ensure that the embedding of a textual description of a product is close to the embedding of an image of the same product we need to learn both embeddings with that constraint in mind, which often requires large amounts of aligned training data (e.g., text-image pairs). One way to address this challenge is to first learn an alignment between data modalities and then learn a generative model on *aligned* representations. Hence, in this section we will focus on the problem of learning aligned representations across multiple modalities.

A popular approach to learning aligned representations is contrastive learning (Gutmann and Hyvärinen, 2010) which, for a pair of data points from different modalities, minimizes a loss that encourages their embeddings to be close when the points are similar (positive pairs), and far when the points are very different (negative pairs). In the single modality setting, positive pairs are generated by simply altering one sample (e.g., slightly shifting the image, flipping the image, transforming it to grayscale). In the multimodal setting, however, it is hard to generate a corresponding positive pair in the other modality via simple augmentation strategies. Instead, positive pairs are typically obtained by labeling similar pairs in a coarse-grained or fine-grained manner. Coarse-grained labels (e.g., a pair of an image and a caption) are easier to obtain, but they may not be sufficiently discriminative. Fine-grained labels (e.g., a bounding box for each object in the image and the corresponding word in the caption) are harder to obtain, but they provide more detailed correspondences between image regions and words in the caption.

### 5.2.1 Contrastive Language-Image Pre-training (CLIP)

Contrastive Language-Image Pre-training (CLIP) (Radford *et al.*, 2021) is one of the most popular contrastive learning approaches to multimodal pre-training. The main idea behind CLIP is that coarse labels in natural language have a sufficient degree of supervision to enable the learning of general concepts, while being much easier to scale using internet data. Indeed, the authors of CLIP found that trying to predict the exact words, as previous works had done, led to very hard training objectives that converged very slowly, due to the variety of ways in which the same information can be conveyed. Therefore, they proposed to use coarse labels, i.e., to pair an entire image with a caption.

Figure 5.1 shows CLIP's model architecture, which consists of two towers, an image encoder and a text encoder, that project an input image-text pair to a shared embedding space. Semantically equivalent image-text pairs should be projected to the same point in the embedding space, and unrelated image-text pairs should be projected to far apart points. This is achieved by computing the cosine similarity for all possible image-text pairs in a training minibatch, and applying a symmetric cross-entropy loss over the rows and columns of the similarity matrix.

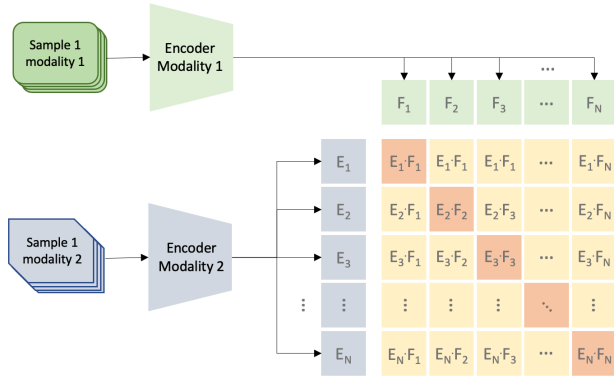

**Figure 5.1:** Contrastive pre-training used to train models such as CLIP. For each minibatch, the positive (diagonal) and negative (off-diagonal) pairs are used to compute the loss.

To be effective, CLIP was trained on a large dataset of 400 million image and text pairs, obtained from downloading images and their associated *alt-text* (text to be displayed in place of an image that fails to load) from the Internet. The dataset was curated to guarantee coverage of concepts by balancing the word occurrences, and filtering with methods like making sure each text included at least one word from a pre-defined list obtained from Wikipedia data to remove noisy or irrelevant image-text pairs. While aligning crawled Internet images and their alt-text is bound to find many irrelevant or misleading examples, dataset curation techniques to improve the quality of training samples (Cao *et al.*, 2023; Fan *et al.*, 2023a), or to reduce harmful or undesired examples (Bansal *et al.*, 2023; Yu *et al.*, 2023) have proven useful to improve results. Furthermore, scaling up the datasets to billion-scale (Schuhmann *et al.*, 2022) has shown that noisy examples not previously removed can be cancelled by overwhelming numbers of positive ones, resulting in better overall performance Jia *et al.* (2021).

The simple idea behind CLIP demonstrated to scale very well and achieved state-of-the-art in many zero-shot benchmarks. For example, it obtained impressive zero-shot classification and retrieval results (Novack *et al.*, 2023; Baldrati *et al.*, 2023; Hendriksen *et al.*, 2022), and has been successfully fine-tuned to a multitude of tasks, such as object detection (Gu *et al.*, 2021), semantic segmentation (Zhou *et al.*, 2023b) or action recognition (Huang *et al.*, 2024b). The same contrastive alignment objective has also been used between other modalities, including audio and images (Cheng *et al.*, 2020), tables and images (Hager *et al.*, 2023), tables and medical images (Huang, 2023), and with multiple modalities at the same time (Girdhar *et al.*, 2023). The datasets used for pre-training these models are typically composed of data scrapped from the Internet (e.g., pairs of images and alt-text), generated as a byproduct of another process (e.g., e-commerce purchases (Chen *et al.*, 2023c), robot sensor logs (Huang *et al.*, 2021)), or automatically generated by an existing ML model (e.g. speech in audio or video (Zhang *et al.*, 2021a), human poses in images computed with OpenPose (Cao *et al.*, 2019)).

The generalization ability of CLIP and similar Vision-Language Models (VLM) greatly benefited from scaling the training in model size, batch size, and dataset size (Pham *et al.*, 2023; Cherti *et al.*, 2023). Researchers have also studied how preferring adaptation of the text branch over the language branch affected results (Zhai *et al.*, 2022). Furthermore, many approaches have been proposed to improve the semantic accuracy of the resulting models (Li *et al.*, 2023a), such as loss functions to improve the image and text encoders (He *et al.*, 2022a; Shen *et al.*, 2022), or to encourage desirable properties such as multilanguage understanding, interpretability and fairness in the embedding space (Chen *et al.*, 2023a; Carlsson *et al.*, 2022; Dehdashtian *et al.*, 2024). Other interesting improvements include training better encoders with additional loses like image masking (He *et al.*, 2022a) and Triple Contrastive Learning (Yang *et al.*, 2022b), or enhancing the text with Wikipedia definitions of entities (Shen *et al.*, 2022).

### 5.2.2  Other Contrastive Pre-training Approaches

Other approaches have looked into novel architecture designs and novel losses to further improve results. Align BEfore Fuse (ALBEF) (Li *et al.*, 2021), for example, uses a multimodal encoder to combine the text and image embeddings generated from the unimodal encoders, and propose two additional objectives to pre-train a model in addition to the Image-text contrastive (ITC) learning: masked language modeling (MLM) to predict masked words on the unimodal text encoder, and image-text matching (ITM) to classify if a pair of image and text match or not. The authors also introduce *momentum distillation*, where a moving average version of the model weights provides pseudo-labels in order to compensate for the potentially incomplete, or wrong, text descriptions in the noisy web training data. Using their proposed architecture and training objectives, ALBEF obtains better results than CLIP in several zero-shot and fine-tuned multimodal benchmarks, despite using orders of magnitude less images for pre-training. In a subsequent work, Li *et al.* (2022b) replace the multimodal encoder by cross-attention layers to the text tower to model vision-language interactions, and replace the MLM loss by a Language Modelling (LM) loss that trains the model to maximize the likelihood of a generated caption given an image.

Finally, other works explore how to bring more modalities into alignment. Girdhar *et al.* (2023) propose ImageBind, an approach to learn an aligned embedding across six different modalities, including text, audio, image, depth, thermal and Inertial Measurement Unit (IMU) data. Instead of requiring paired data for all modalities, they only rely on readily available paired data between image and other modalities (e.g., web scale text-image data, audio for a video clip or depth in RGBD images). All modality encoders use transformer networks and the joint model is learned using the InfoNCE loss.

These contrastive multimodal models can then be used in multimodal recommendation systems such as (Sevegnani *et al.*, 2022; Alpay *et al.*, 2023; Wu *et al.*, 2023b). They are also used to initialize the weights of generative multimodal systems, that will make the generative training much more tractable.

## 5.3 Generative Multimodal Recommendation Systems

Despite their advantages, purely contrastive recommendation systems often suffer from data sparsity and data uncertainty (Wang *et al.*, 2022e; Lin *et al.*, 2023c). For example, users may provide reviews for very few items and some of them may have errors. Generative models address these issues by imposing structure on the data generation process, e.g., by using latent variable models, and by adequately modeling uncertainty. Moreover, generative models allow for more complex recommendations, e.g., those involving image generation.

In this section, we will survey generative recommendation systems that utilize multiple modalities in order to better understand the user, or provide the recommendations. Depending on how the generative models are designed and learned, we will distinguish between three types of models: Generative Adversarial Networks (see Section 5.3.1), Variational AutoEncoders (see Section 5.3.2), and Diffusion Models (see Section 5.3.3). All these three types of models posit the existence of a latent variable $Z$ (continuous or discrete) such that the distribution of the data $X$ (e.g., image and text) can be written as

$$p(X) = \int p(X \mid Z)p(Z)dZ. \tag{5.1}$$

The main differences among these models are how the prior $p(Z)$ and posterior $p(X \mid Z)$ are defined and parametrized with deep networks, and what losses are used to learn the network weights from data. The following subsections describe each one of these models in more detail, how network architectures are modified to accomodate multimodal data, and how these models are used for building recommendation systems.

### 5.3.1 Generative Adversarial Networks for Multimodal Recommendation

Proposed by Goodfellow *et al.* (2014), Generative Adversarial Networks (GANs) are an innovative approach to learning a distribution from multimodal data. GANs have been used in various recommender systems, including collaborative filtering (Wei *et al.*, 2023) and content-based retrieval (Tautkute and Trzcinski, 2021). In this subsection, we will briefly summarize the basic formulation of GANs for unimodal data, show how it can be extended to multimodal data, and discuss adaptations of GANs for collaborative filtering and content-based retrieval.

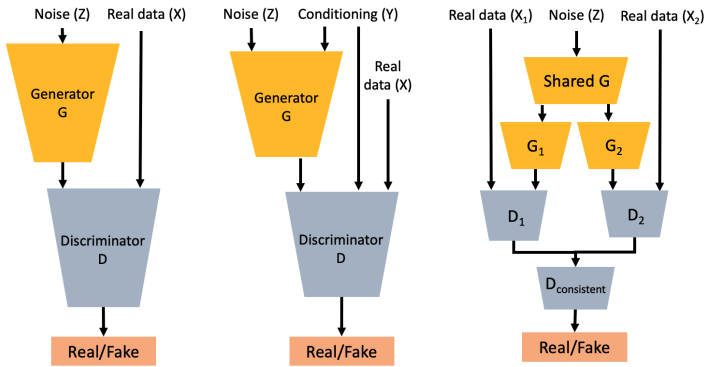

**Figure 5.2:** Generative Adversarial Networks (GANs) are composed of a generator $G$ that generates a data point (e.g. an image) from a latent variable $Z$, and a discriminator $D$ which tries to determine if a data point is fake (synthesized by the generator) or real. Left: Standard GAN architecture for unconditional generation. Center: Conditional GAN architecture. Right: Multimodal GAN architecture.

**Unimodal GANs**   As discussed in Section 3.5.5, GANs are a class of latent variable models in which a generator maps a latent variable $Z$ to a sample data point $X$, while a discriminator $D$ decides whether its input is real or generated (see Fig. 5.2 left). More specifically, GANs learn a probability distribution as in Eq. (5.1), where $p(Z)$ denotes the prior which is typically assumed to be a standard Gaussian (continuous) or categorical (discrete) and $p(X \mid Z)$ denotes the posterior. What is unique in GANs is that the posterior $p(X \mid Z)$ is not modeled explicitly (e.g., as a Gaussian). Instead, the posterior is represented by a generator $G : \mathcal{Z} \to \mathcal{X}$ that produces samples $G(Z)$ from $p(X \mid Z)$ without having to represent $p(X \mid Z)$. Then, a second component of a GAN is the discriminator $D : \mathcal{X} \to \{0, 1\}$, which is designed to discriminate real from generated images, i.e., $D(X) = 1$ if $X$ is real and $D(X) = 0$ if $X$ is generated. The generator and discriminator are then jointly learned from samples of $p(X)$ by optimizing a min-max objective

$$\min_G \max_D \mathbb{E}_{X \sim p(X)}[\log D(X)] + \mathbb{E}_{Z \sim p(Z)}[\log(1 - D(G(Z)))] \quad (5.2)$$

in which the generator $G$ tries to generate samples that fool the discriminator $D$, while $D$ tries to discriminate between real samples $X \sim P(X)$ and generated samples $G(Z)$, $Z \sim P(Z)$.

One advantage of GANs is that sampling is straightforward: all we need to do is to sample $Z$ (e.g., categorical or standard Gaussian) and pass it through the generator to produce $X$. Another advantage is that, in the ideal case in which $G$ and $D$ have infinite capacity, one can show that the optimal discriminator $D^*$ can't tell true from generated (i.e., $D^*(X) = 1/2$) and the optimal generator $G^*$ is such that the distribution of the generated data $G^*(Z)$ matches the distribution of the true data $X$. In practice, $D$ and $G$ are parametrized with neural networks, and the expectation in the min-max objective is computed as the average over samples. As a consequence, while there is no guarantee that GANs learn the true distribution of the data, in the case of images it has been empirically shown that GANs produce high quality generations.

Despite these advantages, GANs also suffer from some limitations. One of them is the issue of mode collapse, which happens when the generator produces samples that are not representative of the full data distribution, such as generating only the most likely outputs, or a specific output that fools the discriminator (Zhang *et al.*, 2019b). GANs also suffer from training instabilities due to the nature of the min-max objective optimized by the generator and discriminator networks. For example, (Arjovsky *et al.*, 2017) show that a small change in one network leads to major adjustments in the other, which can result in destabilizing the learning process and failing to converge. Moreover, gradient vanishing problems happen when one network dominates the other, e.g., when the discriminator becomes very accurate and produces a loss with little gradient information for the generator (Su, 2018; Chakraborty *et al.*, 2024).

**Multimodal GANs**  The vanilla GAN formulation discussed so far assumes that $X$ is generic, i.e., $X$ can be unimodal or multimodal. In principle, we could use such a vanilla formulation to learn generative models for multimodal data. However, doing so may require collecting, annotating and aligning very large datasets and using them to train a very complex multimodal generator. In practice, it may be preferable to design specialized models that leverage existing unimodal generators, such as models that can generate one modality conditioned on another, or models that can ensure adequate alignment across modalities.

Conditional GANs (see Fig. 5.2 center) generate data for data modality $X$ conditioned on another modality $Y$, such as the product type, a textual description of a product, an image mask, etc. In this case, the goal is to learn a conditional model of the form

$$p(X \mid Y) = \int p(X \mid Z, Y)p(Z)dZ. \qquad (5.3)$$

To model $p(X \mid Z, Y)$, the generator must take both $Z$ and $Y$ as inputs to generate samples $G(Z, Y)$. Likewise, the discriminator $D(X, Y)$ must also depend on the conditioning variable $Y$. Different modalities can be used for the condition; examples are class-conditioning (Mirza and Osindero, 2014a), conditioning on an input image (Isola *et al.*, 2017), or using a latent code vector (Chen *et al.*, 2016). Huang *et al.* (2022) proposed an approach to allow conditioning on multiple input modalities (e.g., text, sketch, segmentation mask) to generate new images. This allowed very fine-grained control of the generated image layout and content. Ziegler *et al.* (2022) also use conditioning on multi-modal clinical tabular data for the generation of realistic 3D medical images.

Alternatively, we may want to generate multiple data modalities. For the sake of simplicity, assume that the data is composed of two modalities $X = (X^1, X^2)$. We can design a multimodal generator that leverages unimodal generators by assuming that $X^1$ and $X^2$ are conditionally independent given $Z$. Under these assumptions, the model in Eq. (5.1) factorizes as the product of two unimodal models because

$$p(X^1, X^2) = \int p(X^1, X^2 \mid Z)p(Z)dZ = \int p(X^1 \mid Z)p(X^2 \mid Z)p(Z)dZ. \qquad (5.4)$$

Therefore, we can use one generator per modality, $X^1 = G^1(Z)$ and $X^2 = G^2(Z)$, to represent $p(X^1 \mid Z)$ and $p(X^2 \mid Z)$, respectively. Note, however, that the latent representation $Z$ must be shared to ensure alignment across modalities. Alternatively, we may want the generators $G_1$ and $G_2$ to have a shared backbone that then splits into separate branches for each modality (see Fig. 5.2 top right). For example, Zhu *et al.* (2024) use a StyleGAN backbone with three modality specific branches.

Regarding the design of multimodal discriminators, we note that the discriminator should take generated data for both modalities and compare it with the true data for both modalities. To leverage pre-trained discriminators for each specific modality, say $D_1$ and $D_2$, we could simply fuse the predictions of unimodal discriminators. Alternatively, we could fuse intermediate features from unimodal discriminators and have a simple discriminator $D_{consistent}$ predict whether the data is real or fake from the fused representation of both modalities (see Fig. 5.2 bottom right). Zhu *et al.* (2024) also use two types of discriminators: *fidelity discriminators* are unimodal discriminators that assess the quality of an individual data modality, while *consistency discriminators* judge whether two modalities are consistent with each other.

**Multimodal GANs for collaborative filtering**  As discussed before, a natural approach to building multimodal recommendation systems is to incorporate multiple modalities when learning a latent representation of items and/or users. However, existing multimodal representation learning methods lack robustness to scarce labels for user-item interactions. Self-supervised learning methods address this problem by exploiting supervisory signals in unlabeled data, e.g., by using data augmentation. However, a key challenge is generating augmentations that are consistent across multiple modalities. Recent work (Wei *et al.*, 2023) proposes an adversarial multi-modal self-supervised learning paradigm in which a generator proposes collaborative relations which are then vetted by a discriminator. In addition, Wei *et al.* (2023) propose a cross-modal contrastive learning framework for preserving inter-modal semantic commonality and user preference diversity. On the other hand, GANs have been used to model and improve user-item interaction data. For example, Gao *et al.* (2021) review several works that use GANs to mitigate noise and perform informative sample selection in user preferences data, and to synthesize new samples through data augmentation.

**Multimodal GANs for fashion recommendation** Due to its visual nature, GANs have found an application area in fashion-related tasks such as compatible outfit generation, virtual try-on, or product search.

Liu *et al.* (2021a) tackle clothing compatibility learning. Given an image of a clothing product, and a target product category name, the proposed method generates an image of a compatible item with a GAN. A compatibility matrix representing a style space is used to condition the GAN and make sure the generated item is compatible with the input one. The style space is learnt using triplets of anchor with compatible and incompatible clothing items, and additional losses for feature matching, reconstruction, and a discriminator loss. Zhou *et al.* (2023a) propose a method to generate multiple options for compatible clothing simultaneously, with attention to diversity. They use a style embedding discriminator to provide supervision to the generator through a binary real/fake classification loss, and a compatibility discriminator that uses a contrastive loss. They also include a diversity loss to ensure variety in the generated items.

For virtual try-on, an input garment image, and a target image with a person onto which the garment has to be placed are used. Generators often use several modalities derived from the target image, such as the person mask and a pose image, to condition the generation. For example, Liu *et al.* (2019) take conditional and reference images and transfer the clothing from the person in the conditional image to the one in the reference image. For that, they use a pose map, segmentation map, mask map and head map, derived from the input images. They combine three generators and a discriminator to have a single system for clothing transfer. Similarly, Pandey and Savakis (2020) use segmentation masks, pose estimation and clothing parsing (i.e., detecting all clothing in a picture) to transfer a reference garment in a white background image to a person in a model image. Their proposed system combines multiple tasks previously done by different networks into a single architecture.

Tautkute and Trzcinski (2021) use GANs for query expansion (i.e., augmenting or reformulating the user query to improve retrieval results) on a multimodal fashion product retrieval scenario. Instead of combining or fusing the text and image representations, as is commonly done in multimodal search, they generate an image of the desired product, and then use it for visual search. The authors trained their image generation network using both a discriminator and a triplet loss to make sure the generated image is not too close to the original query image.

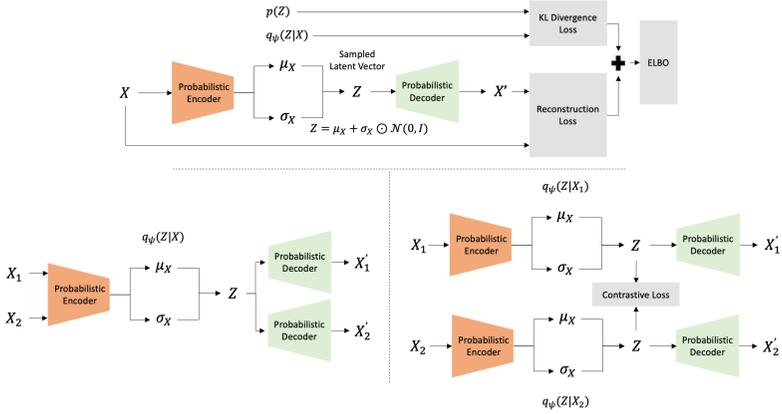

**Figure 5.3:** A Variational AutoEncoder (VAE) is composed of a probabilistic encoder that maps an input data point to a latent distribution from which a latent vector is sampled, and decoded with a probabilistic decoder with the objective of reconstructing the original input. Top: Standard VAE architecture for unconditional generation. Bottom left: Multimodal VAE architecture with a shared encoder and two unimodal decoders. Bottom right: Contrastive VAE architecture with two unimodal encoders and decoders and a contrastive loss for aligning the latent spaces.

### 5.3.2 Variational AutoEncoders for Multimodal Recommendation

Proposed by Kingma and Welling (2014), Variational AutoEncoders (VAEs) have been used as a key component of various recommendation systems, including collaborative filtering with implicit feedback (Liang *et al.*, 2018), collaborative filtering with side information (Karamanolakis *et al.*, 2018), and content-based retrieval (Yi *et al.*, 2021). In this subsection, we briefly summarize the formulation of VAEs for unimodal data, and show how it can be extended to multimodal data, including adaptations of VAEs for collaborative filtering and content-based retrieval.

**Unimodal VAEs** As discussed in Section 3.5.1, VAEs are a class of latent variable models in which a *probabilistic encoder* maps the input data $X$ to a latent variable $Z$, and a *probabilistic decoder* maps $Z$ back to $X$ (see Fig. 5.3 top). More specifically, VAEs learn a probability distribution $p(X)$ for data $X$ (e.g., image or text) by positing the existence of a latent variable $Z$ (continuous or discrete) such that $p(X) = \int p_\theta(X \mid Z)p(Z)dZ$. The prior $p(Z)$ is typically assumed to be a standard Gaussian (continuous) or uniform (discrete). The posterior $p_\theta(X \mid Z)$ is typically assumed to be Gaussian or categorical and is implemented by a neural network (encoder) that maps $Z$ to the parameters of $p_\theta(X \mid Z)$, e.g., its mean $\mu_\theta(Z)$. The posterior $p_\theta(Z \mid X)$ is typically intractable and thus approximated by a simpler distribution $q_\psi(Z \mid X)$ (e.g., Gaussian or categorical) implemented by another neural network (encoder), which maps $X$ to, e.g., the mean $\mu_\psi(X)$. The weights of the encoder-decoder pair are learned by maximizing a lower bound for the log likelihood $\log(p(X))$, known as the Evidence Lower Bound (ELBO),

$$\mathcal{L} = \mathbb{E}_{Z \sim q_\psi(Z|X))} \Big[ \log p_\theta(X \mid Z) - KL(q_\psi(Z \mid X) \mid\mid p_\theta(Z \mid X)) \Big] \quad (5.5)$$

which is the sum of a reconstruction term $\log p_\theta(X \mid Z)$ and a regularization term $KL(q_\psi(Z \mid X) \mid\mid p(Z))$. The variable $Z$ is then used for downstream recommendation tasks.

**Multimodal VAEs**   In the case of multimodal data, say $X = (X^1, X^2)$ consists of both image and text, we can still use the VAE model described so far. However, as we argued in the case of GANs, doing so may require designing a very complex decoder. A better approach is to design multimodal VAEs that leverage unimodal VAEs. For example, as we did in (5.4), we can assume that $X^1$ and $X^2$ are conditionally independent given $Z$, i.e., $p_\theta(X^1, X^2 \mid Z) = p_{\theta_1}(X^1 \mid Z)p_{\theta_2}(X^2 \mid Z)$, so that we can use one decoder per modality. However, since the latent space $Z$ is shared, this requires the design of a shared encoder $q_\psi(Z \mid X^1, X^2)$. Fig. 5.3 (bottom left) shows the design of such a multimodal VAE with a single probabilistic encoder and two modality specific decoders.

To leverage modality specific encoders and decoders pretrained on large datasets, two families of approaches have been proposed. The first family approximates $q_\psi(Z \mid X^1, X^2)$ with a product of experts (Wu and Goodman, 2018), a mixture of experts (Shi *et al.*, 2019) or a mixture of products of experts (Sutter *et al.*, 2020), allowing one to fuse multiple unimodal encoders into a multimodal one. The second family, partitions the latent space per modality, i.e., $Z = (Z^1, Z^2)$, and assume that $q_\psi(Z \mid X) = q_\psi(Z^1 \mid X^1)q_\psi(Z^2 \mid X^2)$ and $p_\theta(X \mid Z) = p_\theta(X^1 \mid Z^1)p_\theta(X^2 \mid Z^2)$. However, doing so reduces the entire model to two independent VAEs, one per modality, which defeats the purpose of having a multimodal model. ContrastVAE (Wang *et al.*, 2022e) addresses this issue by adding a contrastive loss to the ELBO objective, the InfoNCE loss (Oord *et al.*, 2018), which aligns the latent spaces of the two modalities. Experiments in Wang *et al.* (2022e) show that ContrastVAE improves upon purely contrastive models by adequately modeling data uncertainty and data sparsity, and being robust to perturbations in the latent space.

**Multimodal VAEs for collaborative filtering** Traditional VAEs for recommendation systems are unimodal in nature as they aim to model user ratings. For example, Liang *et al.* (2018) extends VAEs to collaborative filtering for implicit feedback by using a multinomial likelihood conditional likelihood. However, such models often use the standard Gaussian as a prior, which has been shown to give poor latent representations [16]. Karamanolakis *et al.* (2018) extend VAEs to collaborative filtering with side information. Their key contribution is to replace the standard Gaussian prior in the latent space of the VAE (which is user-agnostic) by a prior that incorporates multimodal user preferences (e.g., user reviews and ratings). The resulting VAE achieves around 30% relative improvement in ranking metric with respect to standard VAEs for collaborative filtering.

**Multimodal VAEs for content-based retrieval** Yi *et al.* (2021) proposes a multimodal VAE for content-based retrieval. The proposed approach takes three modalities (music, video, and text), maps each modality to a separate latent space using modality-specific encoders, and then aligns these latent spaces via cross-modal generation. More specifically, the video and text modalities are first fused via a product-of-experts model and the fused representation is passed through a cross-modal decoder that generates music. Conversely, the encoding of music is passed through another crossmodal decoder that generates the visual representation. The resulting representation is trained in 150000 video clips of 3000 different music backgrounds and used to build a music recommendation system.

**Graph VAE for Multimodal Recommendation**    In many applications, multimodal data are better represented by a graph. For example, the graph nodes can be items with hand-crafted or learned features from all modalities, while the graph edges can represent item-item similarities. If we want to learn a VAE for the graph and its features, the encoder needs to be able to process a graph as an input and the decoder needs to generate a graph as an output. Graph neural networks (GNNs) are specialized architectures for processing graphs and can be used as both encoders and decoders. The latent space can be a Gaussian vector, as before, or a graph with one Gaussian vector per node. The resulting model is known as a Graph VAE or GVAE for short (Kipf and Welling, 2016), and has been used in various recommendation systems.

One example is the work if Zhou and Miao (2024), which proposes a Disentangled Graph Variational AutoEncoder (DGVAE) for interpretable multimodal recommendation. DGVAE harnesses contrastive pretraining approaches to map multimodal data to a common space in which user-to-item and user-to-word similarities are used to build an item-to-item graph, which is processed by a GNN. Mutual information maximization is used to regularize the learning objective. Experiments show significant improvements in retrieval performance, especially in terms of the interpretability of the recommendations.

Another example is the work of Chattopadhyay *et al.* (2023), which uses a conditional GVAE to generate decoration recommendations for a room given its type (e.g., bedroom) and its layout (e.g., room elements such as floor and walls). A graph is used to represent both room and furniture layouts, e.g., the nodes capture attributes such as the location, orientation and shape of room and furniture elements, while the edges capture geometric relationships such as relative orientation. Their GVAE then generates a furniture graph, e.g., a collection of furniture items such as bed and night stand that is consistent with the room type and layout, which is then rendered to obtain images of the decorated room. Experiments on the 3D-FRONT dataset show that their method produces scenes that are diverse and adapted to the room layout.

### 5.3.3   Diffusion Models for Multimodal Recommendation

Diffusion models (Sohl-Dickstein *et al.*, 2015) have recently emerged as the state-of-the-art approaches for generation of images and other data modalities. They take inspiration from stochastic differential equations (Feller, 1949), dynamical systems (Anderson, 1982) and non-equilibrium thermodynamics (Sohl-Dickstein *et al.*, 2015) to learn highly complex data distributions. Several works have used multimodal diffusion models for recommendation. For example, Li *et al.* (2023j) use diffusion models for sequential recommendation, and Seyfioglu *et al.* (2024) propose a fast diffusion model for virtual try-on that takes a picture with a human model and a white background catalog picture of a garment as input, and places the garment on the model. In this subsection we will discuss the basic diffusion model architecture, its extensions to multimodal data, and applications in content generation.

**Unimodal diffusion models**   The main idea behind a diffusion model is to generate a new image by sampling a random Gaussian vector and transforming it via multiple denoising steps. This is done by defining a forward diffusion process[1] $X \rightarrow Z_1 \rightarrow Z_2 \rightarrow \cdots \rightarrow Z_T$ that iteratively maps a data sample $X$ to Gaussian noise $Z_T$, and a reverse diffusion process $Z_T \rightarrow Z_{T-1} \rightarrow \cdots \rightarrow Z_1 \rightarrow X$ that recovers the original data from noise. More specifically, diffusion models are VAEs with a sequential latent space $Z = (Z_1, \ldots, Z_T)$. The VAE encoder $q(Z \mid X) = q(Z_1 \mid X) \prod_{t=2}^{T} q(Z_t \mid Z_{t-1})$ assumes that $Z \mid X$ is Markovian with Gaussian transition probabilities $q(Z_t \mid Z_{t-1})$. The VAE decoder $p(X \mid Z) = p(Z_T)p(X \mid Z_1) \prod_{t=2}^{T} p(Z_{t-1} \mid Z_t)$ assumes that $Z$ is Markovian with Gaussian transition probabilities $p(Z_{t-1} \mid Z_t)$ and $p(Z_T)$ a standard Gaussian. The transition probabilities are parametrized with deep networks whose parameters are learned by maximizing the ELBO objective in (5.5). Once trained, the model can generate high-quality original data examples by sampling from the noise distribution and simulating the reverse diffusion process.

---

[1]The forward process is also called corruption or noising process, while the reverse process is also called restoration or denoising process.

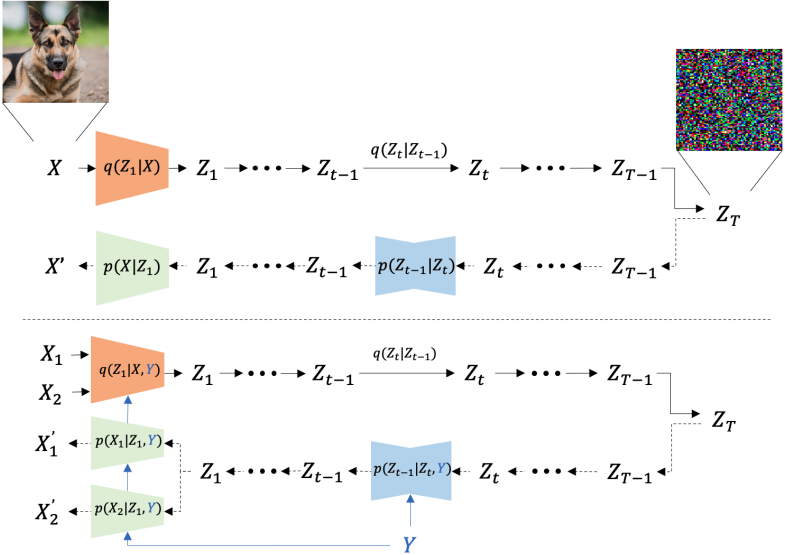

**Figure 5.4:** A diffusion models consists of a forward process, which iteratively corrupts an input data sample until it becomes Gaussian noise, and a reverse process, which reconstruct the original data from white noise. Top: Latent diffusion model architecture for unconditional generation. The standard diffusion model architecture is obtained by removing the encoder $p$ and the decoder $q$. Bottom: Conditional multimodal diffusion model with a shared encoder and two unimodal decoders. An unconditional multimodal model is obtained by simply removing the condition on $Y$.

The original diffusion model for images (Sohl-Dickstein *et al.*, 2015) operates directly in the image space. That is, $X$ is an image and $Z_1, \ldots, Z_T$ are noisy images of the same dimensions as $X$. Therefore, the encoder does not need to be trained because $q(Z_1 \mid X)$ simply adds noise to the image $X$. This makes the model simpler, since only the decoder needs to be learned. However, generating high-quality samples requires dividing both the forward and backward processes into small steps, which can be computationally costly when $T$ is large.

To address this issue, stable diffusion (Rombach *et al.*, 2022) uses latent variables $Z_1, \ldots, Z_T$ of smaller dimensions, which makes inference faster, but adds the cost of learning an encoder $q(Z_1 \mid X)$ and decoder $p(X \mid Z_1)$. In practice, pre-trained models are often used for $p$ and $q$ to avoid the training cost. Figure 5.4 (top) shows the architecture of a diffusion model for image generation. The standard diffusion model operated of Sohl-Dickstein *et al.* (2015) operated in pixel space, and thus did not have the $p$ and $q$ decoder and encoder models.

Even with the addition of latent variables, at inference time diffusion models still require many evaluation steps (Yang *et al.*, 2023b). Recent research has focused on reducing the time spent in the inference process by reducing the number of steps required (Song *et al.*, 2020b; Karras *et al.*, 2022; Dockhorn *et al.*, 2021; Song *et al.*, 2020a; Lu *et al.*, 2022), or training a better sampler to directly select the best possible steps (Watson *et al.*, 2021; Salimans and Ho, 2022; Meng *et al.*, 2023).

**Multimodal diffusion models**   In the case of multimodal data, we could also build a multimodal diffusion model as above, say with $X = (X^1, X^2)$ being images and text. However, two challenges emerge. First, the same challenges of building multimodal encoders and decoders as in VAEs. Second, even if we can build models with separate encoders and decoders per modality, the issue is that diffusion models are not as suitable for text generation as they are for image generation. Specifically, while diffusion models for text generation have been developed, e.g., by using a discrete latent space $Z$ with categorical transition probabilities (Austin *et al.*, 2021), text encoders based on transformers or other sequence-to-sequence models are preferred in practice. As a consequence, multimodal models for both text and images, such as text-to-image generation models, combine text encoders with diffusion models for images. Figure 5.4 (bottom) shows one possible architecture for such a model. As with the other generative approaches, an additional input $Y$ (e.g., a text description) can be used to condition the generation.[2]

Recently, many conditional diffusion models for image generation have been proposed using text and other modalities as the conditioning variables. For example, *DALL-E* (Ramesh *et al.*, 2022; Betker *et al.*, 2023) uses the CLIP (Radford *et al.*, 2021) embedding space as a starting point to generate novel images. To this objective, the authors train a decoder to invert the CLIP representation back to images. Working on a space that jointly represents text and images allows one to apply language-guided image manipulations. Betker *et al.* (2023) improve the quality of the generated images by performing an automated cleaning and improvement of the training image captions with a dedicated captioning model. *Stable Diffusion* (Rombach *et al.*, 2022) is able to generate images from an input text. Since directly training in the pixel space is very computationally demanding, the generative part of Stable Diffusion is trained on a lower-dimension feature space, and relies on a UNet (Ronneberger *et al.*, 2015) autoencoder separately pre-trained on a perceptual loss and a patch-based adversarial objective. To condition the generation based on other modality inputs, such as texts or semantic maps, they train a cross-attention layer to project the new modality inputs to the intermediate layers of the UNet. *Imagen* (Saharia *et al.*, 2022) train a diffusion model for image generation based on a U-Net image model and a T5 text encoder pre-trained only with text. To condition the image generation based on text, the authors find that using cross-attention significantly outperforms other pooling strategies, and achieves high image-text alignment as well as photo-realistic results.

---

[2]Note that, unlike the case of VAEs where the conditioning affects both the encoder $q(Z \mid X, Y)$ and the decoder $p(X \mid Z, Y)$, in the case of diffusion models the conditioning does not affect $q(Z_t \mid Z_{t-1})$ because it a simple noising process.

Other works expanded diffusion models in different directions. For example, Zhang *et al.* (2023c) increase the controllability of the generated results, Brooks *et al.* (2023) add instruction-following capabilities for image modification, Ruiz *et al.* (2023) improve the consistency of the generated subject's identity by fine-tuning the model with a few images, and Chen *et al.* (2024) propose a multi-modal, multi-task, diffusion model, where multiple input modalities are fused and fed to various decoders to accomplish multiple tasks simultaneously.

Diffusion models have also been used for sound generation (Yang *et al.*, 2023a), video generation (Jeong *et al.*, 2023; Brooks *et al.*, 2024), and other modalities (Kotelnikov *et al.*, 2023; Lin *et al.*, 2023b), or multiple modalities simultaneously (Tang *et al.*, 2024; Ruan *et al.*, 2023). See (Cao *et al.*, 2024a) for a recent survey on diffusion models and its applications.

This rapid progress in diffusion model research shows great potential in their usefulness for recommendation applications. Zhu *et al.* (2023b) propose a virtual try-on system based on a diffusion model, which outperforms earlier ones based on GANs. Ma *et al.* (2024) and Jiang *et al.* (2024) use diffusion models to combine multimodal item information with user-item interaction data.

### 5.3.4 Interactive Multimodal Recommendation Models

As seen in Chapter 4, Large Language Models (LLMs) have been widely used in recommender systems, to do tasks like top-k recommendation, rating prediction or explanation generation (Geng *et al.*, 2022; Wu *et al.*, 2023a; He *et al.*, 2023). In this section we discuss interactive multimodal recommendation models based on LLMs, which have demonstrated impressive generalization capabilities and apparent emergent properties to solve tasks not directly targeted during training (Brown *et al.*, 2020).

**Multimodal Large Language Models** One approach to designing interactive multimodal recommendation systems is to train or adapt specialized *X-to-text* encoders that allow LLMs to accept multimodal input, such as images (Liu *et al.*, 2024a) or other modalities (Wu *et al.*, 2023b; Tang *et al.*, 2023). These new models are called Multimodal Large Language Models (MLLM), and greatly extend the capabilities of LLMs not only at the input side, by accepting information expressed in different modalities, but also at the output side, where appropriate decoders can be used to allow the model to generate content in various modalities, replacing or complementing the textual answer. Figure 5.5 shows a high-level diagram of an MLLM architecture.

With the addition of multiple modalities, MLLMs can become versatile task solvers for recommendation problems. They provide a natural language interface for users to express their queries in multiple modalities, they can tackle complex zero-shot recommendation tasks thanks to their emergent properties, or orchestrate several sub-systems to obtain the best recommendation, they can also generate fluent natural language explanations for a multi-modal recommendation, or even generate documents in different modalities to help the user visualize the products.

As discussed earlier, given the complexity of training large generative models end-to-end, researchers typically assemble systems composed of discriminatively pre-trained components (encoders, decoders and LLM "reasoning" models), usually connected by adaptation layers. These adaptation layers are usually pre-trained using paired data for the different modalities, sometimes together with or followed by some form of parameter-efficient fine-tuning of the base models. For example, Low Rank Adaptation (LoRA) (Hu *et al.*, 2021) freezes the pre-trained parameters and introduces low-rank decomposable trainable adaptation matrices to each transformer layer. This fine-tuning step ensures that representations from different modalities are aligned. In some cases, an existing expert model able to produce textual descriptions of the multi-media content is used in place of adaptation layers (Li *et al.*, 2023d; Yang *et al.*, 2024; Gao *et al.*, 2023a; Wang *et al.*, 2024b). This may lead to lower data needs for adaptation, but could also result in information lost in translation.

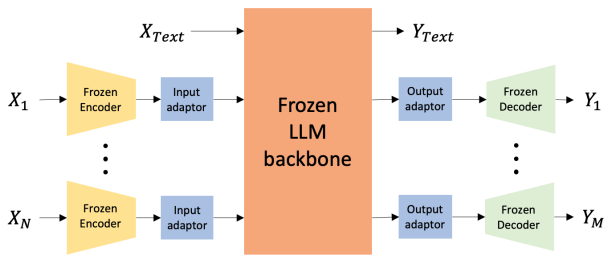

**Figure 5.5:** High-level architecture of an MLLM model. Each input is processed by a specialized encoder to obtain modality specific features which are then projected to a representation adequate for an LLM backbone via suitable adaptors. Similarly, the output of the LLM is then projected to serve as input for specific generators for each modality via a similar adaptor. The input and output modalities are independent.

**Controller LLMs** Similar to the "dialogue controller" described in Chapter 4.7.2, instead of training adaptation layers, another approach is to allow a "controller" LLM to use external tools (e.g., foundation models, classical recommendation systems, arbitrary functions), to deal with the multi-modal input and output (Yin *et al.*, 2023b). This approach has the advantage that it usually involves little or no training. With a carefully constructed prompt explaining all the available tool capabilities and providing usage examples, the controller LLM can create an execution plan with the multi-modal input and generate the desired result in a zero-shot fashion (Zhang *et al.*, 2024a). Once the plan is complete, the controller LLM will have to re-asses the output, and decide if the desired result has been achieved, or further steps are needed. To improve results, researchers have also tried to use instruction-tuning of the controller LLM to improve the tool selection and planning abilities (Yang *et al.*, 2024). An obvious drawback of this approach is that it requires processing multiple rounds of instructions and multiple rounds of (possibly slow) "tool" foundation models to get to the desired result.

**Multimodal instruction tuning**   As seen in Section 4.7.2, instruction tuning is an important step to make LLMs useful task solvers. It requires the creation of datasets of instruction-formatted examples that will be used as training data for the model. These datasets are usually created by extending input-output pairs from multiple multi-modal datasets, such as COCO Captions (Chen *et al.*, 2015), LAION (Schuhmann *et al.*, 2022) or VQAv2 (Goyal *et al.*, 2017), with an instruction text defined using prompt templates for the different tasks (Dai *et al.*, 2024). Tasks can be created, for example, by using annotated bounding boxes to define a spatial relationship question, or using an existing image caption as supervision for an image description request. When no suitable datasets are available, and collecting them from other sources is impractical, researchers use generative models to create the examples through self-instruction (Brooks *et al.*, 2023; Wang *et al.*, 2022d). Models that extend instructions to multiple modalities have recently been proposed (Li *et al.*, 2023h; Wu *et al.*, 2023b). For example, a user could issue a query such as *"modify [image] to convey the feeling of [music]"*. With instruction tuning, MLLMs can be used in dialog systems (c.f. Sec 4.7), enriching the conversation with multimodal understanding and generation.

**Any-to-text MLLMs**   Recently, many MLLMs that add images to the accepted inputs have been developed: Alayrac *et al.* (2022) propose *Flamingo*, which use an LLM and a vision encoder trained with a contrastive loss similar to CLIP to build a model able to meet the state of the art performance in various image and video tasks. Liu *et al.* (2024a) propose *Llava*, an instruction-tuned multi-modal LLM that is able to accept input in both text and image format, and produce useful textual responses. The authors connect the grid visual embeddings from the last layers of the CLIP image encoder (Radford *et al.*, 2021) with the Vicuna language decoder (Chiang *et al.*, 2023) using a simple linear adaptation layer, and fine-tunning the model end to end (they keep the visual encoder weights frozen). To train the model, instruction-following training data is generated using GPT-4 (Achiam *et al.*, 2023) and image-caption datasets, such as COCO (Chen *et al.*, 2015). Liu *et al.* (2023a) change the connection layer from a linear projection to a two-layer MLP and obtain better results. Li *et al.* (2023c) propose *BLIP-2*, introducing a lightweight Query-Transformer (Q-Former), which consists of two transformer modules, to bridge the modality gap between the image encoder output and the LLM, and allow prompts to include both text and image.

Although image is the modality that received the most attention, some works have addressed adding audio (Deshmukh *et al.*, 2023; Kong *et al.*, 2024), and multiple modalities, like text, image, audio or video (Han *et al.*, 2023; Moon *et al.*, 2023; Lyu *et al.*, 2023). However, these models are still limited to generating only text output.

**Any-to-any MLLMs**   As mentioned earlier, to overcome the single-modality output limitation, authors have proposed systems that can both absorb information in multiple modalities, as well as generate response content in different modalities. For example, Next-GPT (Wu *et al.*, 2023b) attempts any-to-any modality conversion through an MM-LLM by using state-of-the-art encoders and decoders, connected to the LLM by thin adapter layers. Multi-modality switching instruction tunning is learned using a custom dataset of 5000 high quality samples. After warming up the adaptation layers, the whole system is trained using LoRA with the modality-switching dataset. For input, the authors use ImageBind (Girdhar *et al.*, 2023), which has been trained to produce aligned representations for image, audio and video, among other modalities, and adapt it using a linear layer to the Vicuna LLM, that does the core reasoning/instruction-following. For the output, small modality-specific transformers are trained to produce the input for three state-of-the-art decoder models, for audio, image and video. Tang *et al.* (2023) use a similar approach, with Llama2 (Touvron *et al.*, 2023) as the core LLM and state-of-the-art diffusion models to generate the multi-modal outputs.

Several companies have released proprietary generalist MLLM-powered chatbots, like OpenAI GPT-4 (Achiam *et al.*, 2023) and Google Gemini (Team *et al.*, 2023) or Anthropic Claude (Team, 2024). Even though these models are not explicitly trained as recommender systems, they are able to produce a variety of recommendation results, including shopping recommendations. For example, Gemini can receive images, audio and video as input, recognize objects in them, provide general advice for product understanding, and recommend products based on customer input.

## 5.4 Applications of Multimodal Recommendation Systems

Recent developments in multi-modal generative models open the door to many applications in recommender systems. In this section we review some of the most promising directions, in areas including e-commerce, in-context product visualization, marketing, online streaming, and travel and service recommendations.

**E-commerce** One of the most direct applications of generative multi-modal models for recommendation is e-commerce, where there is a large volume of product and customer data available that can be used to benefit the customer recommendations. Applications range from improving product images (Corneanu *et al.*, 2024), names and descriptions (Novgorodov *et al.*, 2019; Shao *et al.*, 2021), to generating reviews Truong and Lauw (2019) and review summaries (Schermerhorn, 2023), learning to generate better recommendation (Xiao *et al.*, 2022; Liu *et al.*, 2024b; Karra and Tulabandhula, 2024), and answering user questions (Deng *et al.*, 2022).

Karra and Tulabandhula (2024) propose to use multimodal large language models to improve recommendations by better understanding the behavior of users in e-commerce websites. As the user navigates during a browsing session, high-frequency screenshots are captured and provided to an MLLM together with specific prompts requesting to extract information such as price ranges, product categories and brand preferences, to generate a user behavioral summary. Next, this summary is provided to an LLM with tool-using abilities to derive features and constraints from the input, and use a recommender system to generate the final recommendation. Liu *et al.* (2024b) describe the limitations of current MLLM when used with multiple images as input in the prompt. To improve the performance, they propose to process the list of products interacted by the user as pairs of image and title to obtain text descriptions of the products. These descriptions can then be used in lieu of the images when using the interaction history as in-context-learning to generate new recommendations for a user with an MLLM. Truong and Lauw (2019) propose a system to generate multi-modal reviews. The proposed system uses item and user embeddings, obtained via matrix factorization, to predict the rating and compose a review text with a Long Short-Term Memory network. If a review image is available, it is also used to condition the text generation.

**In-context product visualization**    Applications such as "virtual try on" or "view in your room" augment an image or video with products such as clothes (Yuan *et al.*, 2013; Han *et al.*, 2018), sunglasses, or even makeup (Borges and Morimoto, 2019; Prinzivalli, 2019) to help users visualize how they would look in themselves, or how furniture or appliances would look in the context of their home (Reuksupasompon *et al.*, 2018; Perez, 2020; Berthiaume, 2023), before making a purchase decision. A traditional approach for these tasks is Augmented Reality (AR), that mixes real images, obtained from a camera feed, with virtual objects to generate novel views in real time. While AR has been used in numerous applications, ranging from education (Billinghurst, 2002) to assisting surgeons in medical operations (Dennler *et al.*, 2021), recent diffusion-based image generation models can be used to further improve virtual-try-on experiences (Wang *et al.*, 2024a; Xu *et al.*, 2024b; Wang and Ye, 2024), or generate outfits to try out (Xu *et al.*, 2024a), and make them more controllable (Seyfioglu *et al.*, 2024).

**Marketing** In marketing, multimodal generative models can be used to create personalized advertisement images and videos from product imagery and customer preferences to increase the probability of engagement (Wang *et al.*, 2023a; Chen *et al.*, 2021b). Wei *et al.* (2022b) generates personalized bundles and creates a customized image for display. Shilova *et al.* (2023) fine-tune a Stable Diffusion model to generate personalized images by outpainting input images without modifying the targeted object. For training, they leverage a $U^2$-Net segmentation network, and a BLIP model to generate masks and captions for a collection of training images, that the model will then learn to reconstruct. With adequate guardrails in place, generative models could also be used to synthesize personalized multimodal ad content like text (Loukili *et al.*, 2023), images (Mayahi and Vidrih, 2022) or video (Liu and Yu, 2023).

**Streaming services** Online video and audio streaming services strive to recommend the most valuable multimedia content to each user in order to maximize usage, ad revenue or click-through rate. Long and short-form video, music, audiobooks, podcasts and radio have different recommendation requirements, but the very content to recommend comes in multiple modalities that can be used to improve the suggestions. Even though most works on streaming content recommendation rely on user behavior and content metadata, recent works have applied multimodal learning to audio (Jones, 2023; Chen *et al.*, 2021c; Huang *et al.*, 2020; Deldjoo *et al.*, 2024b) and video (Lei *et al.*, 2021; Wei *et al.*, 2019; Yi *et al.*, 2022; Sun *et al.*, 2022) recommendation. Due to their extensive pre-training, large multi-modal generative models can further enhance the user experience in a content streaming recommendation setting by blending content understanding with personalization and generation, allowing them to complete tasks like answering to fine-grained content-related questions in natural language (e.g., "Does this movie contain a car chase scene that I will like?"), or generating personalized content to fulfill a user request (e.g., audio and music generation (Briot *et al.*, 2017; Lam *et al.*, 2024; Vyas *et al.*, 2023; Dhariwal *et al.*, 2020)).

**Travel and service recommendations** Services ranging from theme parks and concert venues to restaurants, auto mechanics, and laundry services receive customer ratings, reviews and clicks in many online platforms. Better understanding of contextual details such as location characteristics, services offered, past user experiences and popularity factors through multi-modal information, could lead to better and more personalized recommendations. Furthermore, future work could tackle comprehensive products such as interactive recommendation systems able to help users through the whole process of planning and booking complex multi-faceted events such as weddings (e.g., venue, menu, decoration, music) or travel (e.g., destination, transport, hotel, restaurant, activities, practical tips) in a conversational way (Xie *et al.*, 2024), accepting and incorporating user feedback and explaining the recommendations. Yan *et al.* (2023) generates explanations for recommendations focusing on making them informative and diverse. For that, they start by selecting a set of images for a given user and business using a detrimental point process that leverages CLIP features from the user history, and the business images. Then, they use a GPT-2-powered multi-modal decoder, trained with a personalized cross-modal contrastive loss, to generate natural language explanations. The results show that the proposed method produces more informative and diverse explanations compared to text-only alternatives.

# 6

## Evaluations and Benchmarks

Unlike traditional recommenders that predict ratings, rank items, or retrieve from a corpus of items, Gen-RecSys generates rich outputs. As we saw in Chapter 1, they could generate lists of recommendations, engage in conversations, and create visual content such as virtual try-ons. Not only that, the outputs can be personalized to the user's preferences. Moreover, they such systems can introduce entirely new capabilities that have not been seen before, such as on-demand item creation (including audio and video). This raises new challenges in assessing the performance, efficiency, and safety of these systems. This chapter explores how to evaluate and benchmark Generative Recommender Systems (Gen-RecSys).

We begin by examining the challenges in evaluating Gen-RecSys systems and highlighting the need to move beyond traditional metrics used widely today. We discuss how offline and online evaluation is done in Gen-RecSys, which covers new metrics, new methods adopted, and new challenges. We also discuss evaluating these systems for efficiency, which is an important factor to consider, given the computational demands of the generative models. Then, we discuss the existing and emerging benchmarks and what new benchmarks the community should build. Finally, we conclude by motivating the need for holistic evaluation of Gen-RecSys. We discuss the considerations around societal harms interleaved in these sections (a more thorough description is in the next chapter).

## 6.1   Challenges of Evaluating Gen-RecSys

Evaluation of traditional recommender systems is already challenging. For example, offline evaluation may not robustly predict online performance when models are exposed to end-users due to many factors. There can be a data distribution shift between the trained model and the new data that the model sees during its use (Ktena *et al.*, 2019; Zhao *et al.*, 2019). Position bias can be another factor that is challenging to capture in offline settings (Chen *et al.*, 2021a; Hofmann *et al.*, 2014). In general, several hard-to-measure factors constitute the user's context, such as their surroundings, mood, and previous content consumed elsewhere, which can influence recommendation acceptance. Also, see Chang *et al.*, 2024 for a survey on evaluating Large Language Models.

Evaluating Gen-RecSys presents unique challenges that differ significantly from traditional recommender system evaluation. These challenges stem from the following factors:

- **Output Complexity**: Gen-RecSys outputs can be more complex and diverse than simple ratings or rankings. Evaluating these outputs requires going beyond traditional metrics and developing new measures that capture the nuances of generated text, images, or other media.

- **Open-Ended Tasks**: Many Gen-RecSys applications are inherently open-ended, with no clear ground truth or predefined objectives. For example, training and evaluating a system that generates personalized shopping experiences or assists users in planning complex events requires defining and collecting complex instruction-following datasets. And how do we measure subjective outcomes like relevance, helpfulness, and user satisfaction in this context?

Evaluation of Gen-RecSys also differs from how evaluation is done for LLMs:

- **Task-specific metrics**: LLMs are evaluated on diverse tasks such as question-answering (MMLU by Hendrycks *et al.*, 2020, SQuAD by Rajpurkar *et al.*, 2016), code generation (Hendrycks *et al.*, 2021), mathematical ability (Hendrycks *et al.*, 2021) etc. Meanwhile, for Gen-RecSys, we must evaluate primarily based on the recommendation metrics. Moreover, a lot of the benchmarks for LLMs are closed-book (e.g. MMLU by Hendrycks *et al.*, 2020 measures closed-book question-answering), whereas Gen-RecSys might predominantly be open-book, restricted to a corpus.

- **User-Centric Evaluation and Personalization**: While LLMs are evaluated and trained to match human preferences (called alignment), this process produces an LLM aligned with society. In contrast, Gen-RecSys needs to consider each individual's preferences. While alignment is geared a lot towards safety, for Gen-RecSys, we need newer techniques to improve personalization.

- **Compound Systems**: Gen-RecSys often consist of multiple interconnected components, including generative models, traditional recommenders, and potentially other non-AI systems, thus making them compound systems (Zaharia *et al.*, 2024). Evaluating the performance of one component in isolation may not reflect the overall system behavior, highlighting the need for holistic evaluation methods.

- **Potential for Harm**: Gen-RecSys, like any powerful AI system, risks amplifying existing biases and generating harmful content. They can disseminate misinformation, manipulate and persuade users to change their behaviors, are hard to interpret due to their black-box nature, and can amplify echo chambers and filter bubbles. Evaluating these risks requires considering ethical considerations alongside performance metrics. They may also bring safety challenges that are not present in discriminative models. The next chapter (Chapter 7) will discuss this topic in detail.

We also note that modern large-scale recommender systems involve multiple stakeholders, and the evaluation aspects they care about will vary. For example, the system designers' offline metrics to evaluate the models are meaningless to end users. Still, they may evaluate the system's helpfulness, ease of use, novelty, and safety. Below, we list a few stakeholders and what they will be interested in evaluating:

- Data scientists and machine learning engineers will be interested in optimizing many of the above offline and online metrics, efficiency, and interpretability of the models.

- Business owners and product managers will be interested in long-term metrics, user retention, cost efficiency, business metrics, and competitive advantage.

- End users will consider subjective user experience, ease of use, relevance and personalization, and privacy and data usage.

- Government and civic organizations may care about the safety of such systems and will want to understand the societal implications.

## 6.2 Offline Evaluation of Gen-RecSys

Offline evaluation helps understand the performance and quality of recommender models in an isolated offline scenario, enabling quick rapid prototyping. It is one part of the puzzle and must be augmented with online evaluation (discussed in Section 6.5).

This section explores the key metrics and approaches used for the offline evaluation of Gen-RecSys.

### 6.2.1 Traditional Recommendations with Gen-RecSys

In traditional recommender models, the most common offline evaluation approach is to evaluate the model's ability to predict some events or user features. For example, the model may predict click-through rates, time spent, or conversion factors. And correspondingly, metrics such as recall@k, precision@k, NDCG@k, AUC, ROC, RMSE, MAE etc. may be used.

A similar approach may be used in Gen-RecSys for traditional discriminative recommendation tasks. For example, Geng *et al.*, 2022; Cao *et al.*, 2024b predict ratings, and Rajput *et al.*, 2023 use a generative model for retrieval.

### 6.2.2 Text generation

The following metrics are useful when Gen-Recsys generates textual output (e.g., review summarization or chat with the user to recommend movies).

**BLEU (Bilingual Evaluation Understudy, Papineni *et al.*, 2002)** BLEU is a widely used metric for evaluating machine translation. It measures the overlap between machine-generated text and reference text at different levels of n-grams. The metric provides a score reflecting the fluency and accuracy of the generated text. It is useful for evaluating explanations (Geng *et al.*, 2022), review generation and conversational recommendations (Liao *et al.*, 2019; Li *et al.*, 2018a).

**ROUGE (Recall-Oriented Understudy for Gisting Evaluation, Lin, 2004)** The ROUGE score measures the overlap of n-grams between a generated text and a reference, which means it measures the similarity of the generated text to the reference. Therefore, it is useful in evaluating explanations for recommendations and review summarization.

**Perplexity (Jelinek *et al.*, 1977)** Perplexity measures the uncertainty of a language model and quantifies how well the model predicts the next word in the text sequence. A lower perplexity generally indicates a more accurate and fluent model. For example, Hayati *et al.*, 2020 use perplexity as one of the metrics to show that their proposed method is better.

**Model-based evaluations** A model-based evaluation technique uses other models to assess the performance of LLMs. Strong LLMs can be used to evaluate the output of other LLMs based on some rubrics (these can be dynamic), or smaller specialized models can be used to evaluate on specific rubrics. One of the first methods to evaluate LLMs using judge LLMs is the Alpaca Eval framework Li *et al.*, 2023i. The framework collects the responses of different LLMs for a set of predefined 805 prompts and uses a strong LLM such as GPT-4 to generate the win rates. This also forms the basis of the Alpaca Leaderboard[1].

Zheng *et al.*, 2024 study the viability of using LLMs as a judge and find that GPT-4 judge agrees with human preferences 80% of the time on the two benchmarks they propose (MT-bench consisting of 80 multi-turn questions and ChatBot Arena[2] consisting of crowdsourced battles between LLMs). The paper also mentions the limitations of this method, such as Position bias (preferring generations positioned first when comparing multiple generations), verbosity bias (favoring long and verbose responses), and self-enhancement bias (preferring answers generated by themselves). Some of these limitations are fixed in newer versions of AlpacaEval: for example, Dubois *et al.*, 2024 correct for the length of the responses and find that the length-controlled AlpacaEval win-rates correlate with ChatBot Arena 98% of the time.

In contrast to Alpaca Eval, judge LLMs can also be used to evaluate a single (weaker) LLM. In this case, the judge LLM takes the prompt, the response of the weaker LLM, the reference answer, and a scoring rubric to output a number. For example, RAGAS (Es *et al.*, 2023) uses this method for evaluating Retrieval Augmented Generation (RAG) systems.

In addition to using strong LLMs as judge LLMs, it is possible to use small LLMs that are specifically fine-tuned for evaluation capabilities. Prometheus (Kim *et al.*, 2024; Kim *et al.*, 2023), Prometheus-Vision (Lee *et al.*, 2024), JudgeLM (Zhu *et al.*, 2023a), PandaLM (Wang *et al.*, 2023e) are some examples. ARES (Saad-Falcon *et al.*, 2023) trains three small specialized models for evaluating context relevance, answer faithfulness, and answer relevance (all of which are scalar outputs) for RAG systems.

---

[1]https://tatsu-lab.github.io/alpaca_eval/
[2]https://chat.lmsys.org/

**Offline Human evaluations**   While these metrics give a sense of the model's performance in an offline setting for open-ended generation, human evaluation may be necessary in many cases before deploying. This is different from A/B testing (covered in more detail in Section 6.5) in that human raters are specifically selected and trained to evaluate the system's performance before deploying to real users (at which point A/B testing can be performed). Human raters can evaluate many subjective and hard-to-quantify aspects such as fluency, naturalness, and overall quality, and catch corner case issues.

Let's, in particular, dig deeper into Conversational Recommender Systems (CRS), whose goal is to help users make decisions and find items of interest in a conversational way. Their effectiveness can be evaluated using known objective measures like recall@k, RMSE etc. There can be other objective measures based on completion or acceptance rates, such as add-to-cart actions (Mahmood and Ricci, 2009).

In addition to these, their efficiency can be measured in terms of how quickly and effectively CRS can help users find their desired items.

- Interaction Cycles: The number of interactions needed to reach a satisfactory recommendation, with shorter ones generally preferred (Mahmood and Ricci, 2009).

- Task Completion Time: How long it takes to complete a recommendation task, with shorter being better. An interesting finding by Iovine *et al.*, 2020 is that purely conversational recommendations that avoid button-based user interfaces have longer task completion times.

Lastly, it is also possible to measure the effectiveness that CRS helps complete subtasks (e.g. intent recognition, entity recognition, sentiment analysis).

### 6.2.3   Multimodal Content generation

As discussed in Chapter 5, VAEs, GANs, diffusion models and multimodal generative models are starting to be used to generate output images directly. These will have widespread use in many applications, such as fashion recommendations. For example Liu *et al.* (2021a) and (Zhou *et al.*, 2023a) use GANs to generate clothing images. While general-purpose image generation and video generation models are cropping up, we are yet to see the creation of models/tools that generate multimodal content while considering user preferences.

Below we list a few metrics specific to image generation that can be used to evaluate the quality of such systems.

**Inception Score (IS) (Salimans *et al.*, 2016)**   IS measures the diversity and quality of generated images. A higher IS suggests that the model generates more diverse and realistic images.

**Fréchet Inception Distance (FID) (Heusel *et al.*, 2017)**   FID compares the distribution of generated images to the distribution of real images. A lower FID indicates a better alignment between generated and real images.

**Precision and Recall (Sajjadi *et al.*, 2018)**   These are useful in measuring how well the generated and real images are close to each other. Precision measures the proportion of generated images that are similar to real images. Recall measures the proportion of real images that are similar to generated images.

**Model-based Evaluation**   Similar to the case of text generation, other models can be used as expert evaluators for evaluating multimodal content generation capabilities. Powerful image and video understanding models can be used to understand the content generated by image/video generation models. Prometheus-Vision (Lee *et al.*, 2024) is designed explicitly as a judge for evaluating text generated by VLMs (Vision-Language Models). TouchStone (Bai *et al.*, 2023) uses strong LLMs to evaluate VLMs across 27 tasks (such as Brand Recognition, Style appreciation etc.) from 5 categories (Visual Recognition, Visual Comprehension, Basic Description, Multi-Image Analysis, Visual Storytelling).

**Offline Human Evaluation**   Given the complexity of multimodal evaluations, open-source VLMs are not yet good evaluators and have low correlation with human evaluators (Lee *et al.*, 2024). Similar to text generation, a human-evaluation component is typically necessary for improving the confidence of evaluation. For example, model outputs in Koh *et al.*, 2021; Ramesh *et al.*, 2021 are rated by trained humans and the proportion of majority voting (amongst different image generation models) is used as a metric.

Video and other form of multimodal content generation models are quite new. Phenaki (Villegas *et al.*, 2022), which generates videos from textual prompts, uses FID as an offline metric and human evaluation. For such applications, especially personalized, new forms of evaluation will have to be designed.

## 6.3 Evaluating for Efficiency

Due to their large sizes and autoregressive nature, generative models are orders of magnitude more expensive in terms of computation than traditional machine-learning models. For example, Chowdhery *et al.*, 2023 use "6144 TPU v4 chips for 1200 hours and 3072 TPU v4 chips for 336 hours" to train a 540 billion parameter model, and Adler *et al.*, 2024 use 768 DGX H100 nodes to train a 340 billion parameter model.

They are expensive to train and serve and add user-noticeable latencies (Agrawal *et al.*, 2024a; Agrawal *et al.*, 2023a; Agrawal *et al.*, 2024b) when generating recommendations.

Thus, it is important to evaluate Gen-RecSys for efficiency. They need to be evaluated on the amount of computation they use for training and serving, the throughput and latency metrics, and the energy footprint they come with.

### Training Efficiency

We can evaluate training efficiency in terms of the total computational cost required to train a model and the overall training time (since this affects developer productivity).

The computational cost of training is typically measured in GPU hours (or CPU hours), RAM usage, storage requirements, or simply in terms of money spent. The chosen model architecture, hardware utilized, and size of the datasets and training algorithms influence the computational cost (Chowdhery *et al.*, 2023).

In industrial settings, where multiple users may simultaneously conduct hyperparameter tuning experiments, system designers must proactively provide additional resources to accommodate this increased usage. In practice, training resources are never enough and get shared across multiple users. So, to ensure fair scheduling, a given training job may be interrupted to allow other users' training jobs to run, which could increase training time (Verma *et al.*, 2015).

It is also helpful to consider the impact of optimization techniques on training efficiency. Techniques such as distributed training, mixed precision training, and early stopping can significantly reduce training time and costs, though they can impact the quality of the model (Adler *et al.*, 2024; Agarwal *et al.*, 2023a; Chowdhery *et al.*, 2023; Narayanan *et al.*, 2021).

**Inference efficiency**

Inference efficiency is assessed through the resources required (usually the number of GPUs or CPU cores or the money spent) and two primary performance metrics: throughput and latency (Agarwal *et al.*, 2023a; Agrawal *et al.*, 2024b). Throughput is generally measured in queries per second (QPS) that the system can handle.

Latency is a more complex metric. Median latencies give a rough overview of the system's performance in the steady state. But tail latencies (such as the 99th percentile) give a better sense of how the system behaves, especially during heavy loads and peak traffic conditions (Dean and Barroso, 2013).

System designers must evaluate balancing the training and serving costs (Amodei and Hernandez, 2018; Agrawal *et al.*, 2024a). In some cases, it is possible to use more computational resources for training (thus also extending training time) to reduce the inference costs. For example, smaller models can be trained longer with more data (e.g., LLaMA by Touvron *et al.*, 2023 and Gemma by Team *et al.*, 2024), or we can train much larger models and distill them into smaller ones (Hinton *et al.*, 2015; Agarwal *et al.*, 2023b). Or we can train with much more data to create better smaller models (e.g., LLaMA3). However, this may not necessarily work if the model needs to be updated often.

Given the computation costs, evaluating efficiency in Gen-RecSys is important. By carefully evaluating and optimizing training and inference resources, it is possible to create systems that deliver high-quality recommendations promptly and cost-effectively, ensuring a positive user experience and sustainable operational costs.

**Energy Consumption**

A related aspect of efficiency is energy use. Training and inference can both be quite energy-intensive, especially for large generative models (Spillo *et al.*, 2023; Luccioni *et al.*, 2024; You *et al.*, 2023; Chung *et al.*, 2023). Depending on the energy source used, higher energy usage can cause higher carbon dioxide emissions (Luccioni *et al.*, 2024).

Measuring and mitigating energy use can help reduce operational costs and ensure that the system is implemented in a climate-friendly manner. For instance, by running models at their maximum efficiency, utilizing energy-efficient hardware such as specialized processing units, and carefully optimizing the scheduling of training runs, one can make Gen-RecSys systems friendlier to the environment.

**Other efficiency factors**

Most industrial infrastructure today is built for discriminative models. When using Gen-RecSys, it is important to consider scalability. A scalable system should keep its accuracy and tail latencies bounded even as the number of requests increases.

Fault tolerance: Large models may require multiple hardware units (GPUs, CPUs, TPUs) to run due to their memory footprint, which can increase the system failure rate. In this context, recommender systems may be evaluated in terms of uptime, often expressed as a percentage (e.g., 99.9% uptime).

MFU, MBU: Introduced by Chowdhery *et al.*, 2023, MFU stands for Mean FLOPS utilization, and MBU stands for Mean Bandwidth Utilization. These two are useful metrics to understand how efficiently the underlying accelerators are used, which directly impacts the training/serving compute costs and energy costs.

By addressing these additional considerations, the evaluation of training and inference efficiency in generative recommender systems can be more comprehensive, leading to better-designed systems that are both effective and sustainable.

## 6.4   Benchmarks for Gen-RecSys

A benchmark usually consists of a task (e.g., rating prediction), a representative dataset (e.g., movie ratings and associated metadata), and one or more metrics (e.g., RMSE). This standardization allows experimenters to iterate on model development while testing (or benchmarking) the model on the dataset. Sometimes, the benchmark dataset can come with a training and testing split (or can be used that way). In other cases, the benchmark is solely reserved for testing purposes (e.g. MMLU by Hendrycks *et al.*, 2020).

In the case of traditional recommender models, numerous benchmarks have been established over time (e.g., MovieLens; elaborated below). These benchmarks typically revolve around user-item interaction datasets and traditional metrics discussed above (such as recall and precision). However, with the advent of Gen-RecSys, the need for improved metrics (as discussed earlier) and better datasets has become apparent. Gen-RecSys can generate open-ended text or other multimodal outputs and experiences. We first discuss some of the key characteristics we expect from Gen-RecSys benchmarks.

Below, we list some popular benchmarks used for evaluating traditional recommender models.

- Movielens (Harper and Konstan, 2015): released and maintained by the GroupLens research lab at the University of Minnesota, and is among the most popular benchmarks for evaluating recommender systems. They contain user ratings of movies and are available in various sizes (e.g., MovieLens 100K, MovieLens 1M, etc.).

- Amazon Reviews (He and McAuley, 2016): A large dataset comprising product reviews, user ratings, and product information.

- Yelp Challenge ("Yelp Dataset" 2014): A dataset containing user reviews, ratings, business information, and user data.

- Last.fm (Schedl, 2016): A dataset containing user music listening history and social network information.

- Book-Crossing (Ziegler *et al.*, 2005): A book ratings and user information dataset.

These primarily focus on user-item interaction data. They provide valuable ground truth for tasks like rating prediction or item ranking but offer limited utility for evaluating the generation of more complex outputs like textual explanations, personalized reviews, or multimodal recommendations.

Moreover, evaluating the quality of generated content (e.g., a review or an explanation) requires benchmarks that include human-generated content of varying styles and perspectives. This necessitates datasets beyond simple item or user ratings and incorporating rich textual, visual, or auditory information.

### 6.4.1  Emerging Gen-RecSys Benchmarks

- ReDial (Li *et al.*, 2018b): Designed for conversational recommendations, ReDial provides a collection of user-system dialogues about movie recommendations. It allows the evaluation of systems that engage in multi-turn interactions, understand user preferences, and provide personalized explanations. For example, this dataset can be used to evaluate for dialogue coherence in recommendations (i.e., assessing the fluency and relevance of the system's responses within the conversation).

- INSPIRED (Hayati *et al.*, 2020): The INSPIRED dataset includes 1001 human-to-human dialogues about movie recommendations, where users discuss their preferences and seek suggestions. It is designed to evaluate how well conversational recommender systems can generate recommendations through interactive dialogues. The focus is on the naturalness and engagement of the conversation, as well as the quality and relevance of the recommendations. This dataset helps assess systems that generate personalized recommendations based on user preferences and context.

- FaiRLLM (Zhang *et al.*, 2023b) proposes a benchmark called Fairness of Recommendation via LLM (FaiRLLM) to evaluate the fairness of recommendations produced by Gen-RecSys and highlights biases exhibited for sensitive user attributes (e.g. age, country, gender, race) in music and movie recommendations. The metrics they use to evaluate are Jaccard Similarity (Han *et al.*, 2022), SERP (SEarch Result Page Misinformation Score by Tomlein *et al.*, 2021) and PRAG (Pairwise Ranking Accuracy Gap by Beutel *et al.*, 2019) CFaiRLLM (Deldjoo and Di Noia, 2024) improves upon this by refining the methodology and the metrics used.

- BigBench(BIG-bench-authors, 2023): Although not specifically designed for recommendations, BigBench includes tasks that require reasoning about user preferences and generating personalized recommendations.

- Transferable Recommendations: NineRec (Zhang *et al.*, 2024c) introduces a benchmark for studying transferable recommendations, where the goal is to study how well a Gen-RecSys model trained for one domain transfers to other domains. The benchmark consists of a dataset comprising one source domain and nine target domains, where a descriptive text and a high-resolution cover image accompany each item in the dataset.

### 6.4.2   Holistic Evaluation of Gen-RecSys

Developing comprehensive evaluation frameworks and benchmarks for Gen-RecSys is a crucial area for future research. These frameworks should encompass various metrics and methods, including performance, efficiency, and safety metrics. The datasets used should be diverse and closely mimic real-world deployments.

Therefore, we encourage the community to devise a holistic evaluation suite to evaluate and comprehensively improve Gen-RecSys's transparency. We can take inspiration from HELM (Holistic Evaluation of Language Models, by Liang *et al.*, 2022) for this. HELM measures 7 metrics (accuracy, calibration, robustness, fairness, bias, toxicity, and efficiency) on each of the 16 scenarios they outline (87.5% of the time). Question answering, information retrieval, summarization, sentiment analysis, and toxicity detection are some scenarios they mention. It is similarly possible to develop scenarios and metrics in the case of Gen-RecSys. The scenarios should be broad and consider risks and societal implications. For example, scenarios in Gen-RecSys could include ranking, sequential recommendation, direct recommendation, review summarization, toxicity detection, explanation generation, etc. Like HELM, we should also recognize that the scenarios can be incomplete and clarify what is missing.

### 6.4.3 Future Benchmarks

Given the limitations of current benchmarks and the requirements for new ones, we outline a few future directions for new benchmarks for Gen-RecSys:

- Multimodal Datasets: These datasets should span various domains and include well-aligned text, images, audio, and video information.

- Preference Datasets: Mimicking the preference datasets collected for training human-preference-aligned LLMs, having similar datasets in the context of recommendations will be useful. For example, it will be good to have human preference datasets for review summarization, recommendation explanations, and shopping conversations.

- Red-teaming: In the context of recommendations, the models could be trained on a lot of sensitive data, and thus, we need benchmarks to evaluate how well the models can avoid divulging sensitive information. We also want benchmarks to test whether the models could provide harmful recommendations or fail due to adversarial attacks.

- Societal impact: It will be extremely useful to have benchmarks to measure generative recommender models' societal implications.

- Adaptability: How well do the models adapt to ever-changing user preferences?

By focusing on these future directions when developing benchmarks for generative recommender models, we can help system designers create helpful, robust, fair, and beneficial recommender systems.

## 6.5   Online and Longitudinal Evaluation of Gen-RecSys

While offline evaluation provides valuable insights, it does not fully capture the complexities of real-world deployment. For example, Gomez-Uribe and Hunt, 2015 mention the ineffectiveness of offline evaluations to predict online success in the Netflix recommender system. Therefore, online experiments (aka A/B testing) are critical for evaluating Gen-Recsys's true effectiveness (also see Rossetti *et al.*, 2016). Moreover, by extending the A/B experiments over a long duration, it is possible to study Gen-RecSys's longitudinal impact.

Moreover, suppose a generative recommender model replaces a traditional recommender model while providing newer abilities like conversational recommendations or multimodal outputs, for which no historical data exists for offline evaluation, the system designers might have to resort to A/B testing (after offline human evaluation).

### 6.5.1   A/B Testing

The standard way of online evaluation is to perform an A/B experiment, where similar traffic is diverted towards two arms or system versions: a control arm (the same as the production model) and a treatment arm (which carries the changes that must be tested).

Care must be taken to ensure that the cohort of users is as similar as possible between the two arms (Gunawardana *et al.*, 2012). For example, we do not want one arm to be served to users in one location and the other to users in another.

The length of the A/B experiment can depend on several factors, including how expensive it is to run the experiments. Checks must be in place so that a treatment arm can be shut down automatically if it produces excessively negative results (for example, the underlying model is broken). In most cases, A/B experiments should be run for at least 1 week to capture the effects of the days of the week. If the A/B experiment falls on holidays, those should be considered, especially if the user behavior is substantially different on those days (for example, this is the case for shopping).

Typically, users are randomly selected for every day of the experiment. However, if learning effects need to be captured, users will be randomly selected. Then the treatment and control will be exposed to the same set of users throughout the experiment.

A lot of these principles will apply to Gen-RecSys systems. However, there may be cases where a Gen-RecSys-based model may not be suitable for enrolling in an A/B experiment, which makes it harder to evaluate the true utility. For example, the model may be too expensive: either getting to a reasonable number of users for achieving statistical significance is hard, or the latencies are not tolerable. It may have been designed to be used purely as an offline model to generate data for distillation.

### 6.5.2 Longitudinal Evaluation

Longitudinal evaluation focuses on understanding Gen-RecSys's long-term impact on user behavior and business outcomes. It involves monitoring key metrics over extended periods to capture user preferences, learning effects, engagement patterns, and revenue generation changes. For example, system designers of shopping recommender models may track how well new merchants are faring over time, and content recommender systems may monitor whether the popularity bias is getting worse. We also note that recent works propose the use of offline simulations to measure the longitudinal impact (Zhang *et al.*, 2020 and Hazrati and Ricci, 2024).

### 6.5.3    Agent-based and Simulation-based evaluation

Recent work advances the measurement of longitudinal impact without relying solely on time-consuming user-facing A/B experiments. For example, Zhang *et al.*, 2020 uses agents to simulate user behavior, potentially offering insights into long-term effects and user preference shifts. Note that these agents are not related to LLM agents.

Given that LLMs understand natural language input very well, are highly adaptable to different scenarios, can understand several domains, and are capable at reasoning, they could be used to simulate user behavior. Based on these motivations, USimAgent (Zhang *et al.*, 2024b) uses LLMs to simulate users' search patterns, such as querying, clicking, and stopping behaviors.

User simulation can be taken one step further to simulate user interactions and thus mimic human behavior in a society. In fact, Park *et al.*, 2023 show that LLMs can be used to realistically simulate human behavior. In their scenario, each user is initialized as a prompt fed to a LLM, called a Generative Agent. They can show the diffusion of information across the small scale of users they simulate. Relatedly, Wang *et al.*, 2023b simulate user behavior in the context of recommender systems and social networks, and show their simulator (RecAgent) can realistically mimic human behavior and use it to study information cocoons and user conformity behaviors.

### 6.6    Conclusion

As we have seen so far, evaluating Gen-RecSys is challenging. From conducting offline and online evaluations to understanding longitudinal effects to estimating the efficiency and stability of such systems, there are many aspects to be considered. In addition, it is important to understand the safety and societal harm of such systems; this is especially important when using generative models that can amplify biases, create harmful responses, and be quite persuasive in altering human behavior.

# 7

## Societal Harms and Risks

### 7.1 Introduction

Generative recommender systems (Gen-RecSys) combine conventional recommendation techniques with generative AI models. While non-generative recommender systems (RecSys) typically suggest items from a fixed catalog of options, Gen-RecSys offer several key advancements. Most importantly, they can create new content for recommendations and engage in more flexible user interactions (for further discussion of the differences between non-generative RecSys and Gen-RecSys, see Chapter 1). Gen-RecSys enable more personalized and engaging recommendations, but also introduce unique societal risks that require careful consideration.

Non-generative RecSys have long been a subject of ethical scrutiny due to their potential societal impacts (Milano *et al.*, 2020; Chen *et al.*, 2023b). These ethical concerns include privacy infringements, threats to user autonomy, lack of algorithmic transparency, and societal-level effects such as amplification of filter bubbles and ideological polarization.

The recent integration of generative AI into RecSys has introduced a new layer of complexity to these ethical considerations. Researchers such as Bender *et al.* (2021), Weidinger *et al.* (2022), and Bird *et al.*

([2023]) have proposed comprehensive taxonomies of potential harms associated with generative AI technologies. These harms include a wide spectrum of actual and potential harms at both individual and societal levels, ranging from discrimination against marginalized communities, propagation of harmful content, information hazards, privacy violations, misinformation proliferation, malicious uses, environmental impacts, and security vulnerabilities.

When applied to the context of RecSys, existing concerns are amplified but also novel challenges are introduced due to various reasons. First, the increased complexity and unpredictability of RecSys outputs due to the integration of generative AI introduce challenges in anticipating and mitigating potential harms, as Gen-RecSys can create novel and unexpected outputs. Second, the generative nature of these RecSys blurs the lines of content accountability, raising questions about authorship and responsibility for the content they produce. Third, the interactive and adaptive capabilities of Gen-RecSys, while beneficial, open up new avenues for potential manipulation of user behavior and preferences on a larger scale than previously possible.

Building upon aforementioned research, this chapter examines the unique challenges and potential research directions associated with the societal risks of Gen-RecSys. Our analysis focuses on seven key areas: (1) contribution to disinformation and misinformation, (2) potential for manipulation and undue persuasion, (3) safety risks from reward misspecification, (4) interpretability and auditability issues, (5) fairness and societal bias concerns, (6) contribution to filter bubbles and echo chambers, (7) broader socio-political implications, and (8) privacy. Following our discussion of potential harms in each area, we briefly outline methods for harm evaluation and propose potential mitigation strategies.

## 7.2 Disinformation and misinformation

Gen-RecSys can represent a paradigm shift in the ways in which the information is created, curated, and disseminated, introducing novel challenges in the landscape of misinformation and disinformation (for a discussion of the novelty of Gen-RecSys in relation to content represen-tation, inference, and generation, see Chapter 4). Unlike non-generative RecSys that rank and suggest existing content, Gen-RecSys can dynamically generate new information for recommendation, blurring the lines between content curation and creation. This capability, while innovative, significantly complicates the structure and content of information ecosystems fueled by generative AI technologies (Kay *et al.*, 2024), particularly in relation to the challenges of misinformation and disinformation.

Misinformation, typically defined as unintentionally spread false information, and disinformation, deliberately fabricated and disseminated falsehoods, find fertile ground in Gen-RecSys.[1] The generative nature of Gen-RecSys not only amplifies the scale and speed of potential false information spread but also introduces the possibility of creating entirely new, highly personalized forms of misleading content that may be more difficult to detect and counter (Shin, 2024).

The psychological mechanisms underlying the propagation of misinformation and disinformation in Gen-RecSys contexts are complex and multifaceted (Pennycook and Rand, 2021; Roozenbeek and Van der Linden, 2024; Ariely, 2023). One primary reason for this complexity is that these systems' generative capabilities present dual risks: malicious actors could exploit them to orchestrate tailored disinformation campaigns, crafting highly convincing false narratives, while the systems themselves might inadvertently generate and recommend misinformation due to flawed training data or misinterpreted user preferences. The personalized recommendation algorithms of Gen-RecSys could amplify cognitive biases, potentially increasing users' susceptibility to false information that aligns with their existing beliefs or preferences (Shin *et al.*, 2024).

While the immediate and quantifiable impacts of misinformation and disinformation via RecSys on specific events such as elections, mental health crises, and addiction patterns is contested and remains subject to ongoing research (Tommasel and Menczer, 2022; Pathak *et al.*, 2023; Adam, 2024; Fernandez *et al.*, 2024; Garimella and Chauchard, n.d.), their potential to erode public trust, manipulate beliefs, and undermine democratic processes cannot be dismissed (Ecker *et al.*, 2024; Doorn,

---

[1]For a more comprehensive exploration of these concepts and related terms such as malinformation and fake news, see Ameur *et al.* (2023) and Shu *et al.* (2020).

2023; Kay *et al.*, 2024).

To fully understand and address these phenomena, we need to develop more sophisticated measurement methodologies. Traditional metrics used for static RecSys would not capture the complexities introduced by generative capabilities due to the dynamic and open-ended nature of engagement with end-users (Ktena *et al.*, 2019; Chen *et al.*, 2021a; Chang *et al.*, 2024) (See Chapter 6 for further discussion). Researchers should develop new evaluation methodologies to measure these impacts. These methodologies can include carefully-crafted longitudinal studies (e.g., Hazrati and Ricci (2024)) to assess changes in user beliefs and knowledge after prolonged exposure to Gen-RecSys recommendations, analyzing how generated misinformation propagates through user networks, and A/B testing with control groups to compare outcomes between users exposed to Gen-RecSys and those using traditional recommender systems, and algorithmic auditing for regular assessments of the Gen-RecSys algorithms to identify potential biases or vulnerabilities to misinformation (Sun, 2022).

## 7.3   Manipulation and persuasiveness

RecSys exert behavioral influence on their users through several channels: the selection of options and choice architecture; personalization of the recommendation context including how the user may interact with the recommender and personalization of recommended content; and the explanations accompanying the recommendations when available. Choice architecture refers to the way in which options are presented to users, and it includes the organization, structure, and framing of choices (Thaler *et al.*, 2014).[2]

RecSys, by their very nature, are architects of digital choice environments. They determine which items are shown to users, in what order, and often with what accompanying information. This structuring of the choice environment can be conceptualized as a form of digital nudging (Jesse and Jannach, 2021). A nudge is a form of behavioral influence that is both resistible and aligned with the user's underlying preferences (Thaler *et al.*, 2014). However, the power to shape choice architecture also opens the door to more problematic forms of influence. While nudging aims to be benign and beneficial, the same mechanisms can potentially be used for manipulation (influence that is harder to resist) or exploitation (influence not aligned with user preferences) (Collier *et al.*, 2022; Andrić and Kasirzadeh, 2023). The line between helpful guidance and undue influence can be thin and context-dependent.

---

[2]The concept of choice architecture comes from behavioral economics and has been extensively studied in relation to decision-making. It recognizes that the manner in which choices are presented can significantly impact the decisions people make. For example, the default option, the order of choices, or how information is framed can all influence user behavior.

While non-generative RecSys have long been associated with concerns about persuasiveness (Yoo *et al.*, 2012) and manipulation (Adomavicius *et al.*, 2013), Gen-RecSys can amplify these issues. This is due to several key factors that distinguish Gen-RecSys from their predecessors.

First, unlike non-generative RecSys that primarily filter and rank existing content, Gen-RecSys have the capacity to generate entirely new content tailored to individual users (Matz *et al.*, 2024). This personalized content creation can be far more persuasive than generic recommendations, as it can directly appeal to individual interests, preferences, and vulnerabilities. For instance, a Gen-RecSys could be used to create personalized news feeds that favor a particular political candidate or ideology, potentially influencing the outcome of elections or it could be used to generate targeted advertisements that exploit individual insecurities or desires, leading to impulsive purchases and over-consumption.

Second, Gen-RecSys leverage vast amounts of user data to create highly personalized recommendations that are far more targeted than traditional systems. This level of personalization can lead to a sense of hyper-relevance, making the recommendations seem more appealing and increasing the likelihood of user engagement. Gen-RecSys can employ sophisticated techniques such as natural language processing to subtly influence user behavior in sophisticated ways. For example, they can gradually nudge users towards certain preferences by subtly adjusting the content they recommend over time. This gradual manipulation can be difficult to detect and resist, as it often occurs below the level of conscious awareness.

Third, some Gen-RecSys can engage in interactive dialogue with users, further enhancing their persuasive power. These systems can build rapport and trust by simulating human-like conversation and responding to user input in real-time, making their recommendations more difficult to resist. For example, a Gen-RecSys could be used to create fake social media profiles that engage in conversations with real users, subtly manipulating their opinions and behaviors.

To evaluate the manipulative potential and persuasiveness of Gen-RecSys, researchers and policymakers need to develop comprehensive evaluation mechanisms and tools. These should consider multiple dimensions, including: (1) the system's impact on user autonomy and decision-making processes; (2) the system's ability to respect and maintain user preferences over time; and (3) the potential for creating or reinforcing echo chambers or filter bubbles. Longitudinal studies tracking changes in user behavior and attitudes over extended periods of interaction with these systems will be crucial. Additionally, developing robust metrics to quantify the degree of manipulation and persuasiveness, such as measuring deviations from baseline preferences or assessing the diversity of recommended content, will be essential for ongoing monitoring and regulation of these powerful technologies (Alslaity and Tran, 2021).

## 7.4 Reward misspecification and safety violation

While offering unprecedented levels of personalization, Gen-RecSys are susceptible to vulnerable reward functions. This vulnerability poses challenges to their safe development and deployment.

ML systems are usually trained to optimize specific goals, often defined by loss or reward functions. However, these functions can sometimes be imperfectly defined, leading to situations where the ML can achieve high-performance metrics without actually completing the intended task. This phenomenon is called reward hacking (Hutter, 2005; Everitt *et al.*, 2017; Pan *et al.*, 2022). In classical RecSys, reward hacking occurs when the system learns to exploit loopholes or biases in its reward function in order to maximize engagement metrics, even if it means sacrificing the true intended objective (Skalse *et al.*, 2022). For instance, a generative AI model intended to provide personalized news recommendations might prioritize engagement metrics and inadvertently amplify politically polarizing content, even if the ideal goal is to offer a balanced news diet.

The increasing complexity of generative AI models has raised concerns about new forms of reward hacking in RecSys. As these models interact more with the real world and incorporate feedback loops, unintended optimization and negative side effects can arise. For instance, feedback loops can induce in-context reward hacking, where models optimize for unintended objectives during real-time interactions (Pan *et al.*, 2024).

User tampering, a more insidious form of reward hacking, involves the system actively manipulating its reward function or feedback mechanisms to artificially inflate its performance metrics (Armstrong, 2015; Everitt *et al.*, 2021; Kasirzadeh and Evans, 2023). This could involve generating misleading or deceptive recommendations to elicit positive feedback, or even altering the underlying data used to evaluate the system's performance. Such tampering can undermine the integrity of the RecSys and lead to detrimental consequences for users. This issue could be exacerbated in Gen-RecSys as the model's ability to generate novel and creative recommendations could make it easier to disguise manipulative behavior. Additionally, the model's reliance on user feedback to learn and improve could make it more susceptible to manipulation by incentivizing users to inflate the system's performance.

To address these challenges of reward misspecification and potential safety violations in Gen-RecSys, several mitigation strategies can be employed. Multi-objective optimization techniques can be implemented to balance various goals, including user engagement, content diversity, and long-term user satisfaction, preventing over-optimization of a single metric (Stray *et al.*, 2021). Developing more sophisticated reward models that incorporate ethical constraints and long-term user welfare can help align system behavior with intended goals, potentially using inverse reinforcement learning to infer true user preferences (Leike *et al.*, 2018). Regular algorithmic audits and continuous monitoring systems, involving both automated checks and human oversight, can help detect anomalies or unintended behaviors (Raji *et al.*, 2020). While these strategies can contribute to creating more robust Gen-RecSys, these challenges are ongoing and evolving, necessitating continuous research, development, and vigilance in the field of AI ethics and safety.

## 7.5   Interpretability & auditability

RecSys are often characterized by a lack of transparency, operating as black boxes where the decision-making process remains opaque (Peake and Wang, 2018). The complex, non-linear transformations applied to input data within these models obstruct the underlying rationale for their recommendations. This opacity poses significant challenges for various stakeholders, including auditors, regulators, and even the engineers responsible for developing and maintaining these systems (Holzinger *et al.*, 2022).

Without a clear understanding of the recommendation mechanism, stakeholders may find it difficult to trust the system's outputs. This lack of trust can impede the legitimacy of the adoption and acceptance of these systems, particularly in high-stakes domains such as education, healthcare, or finance. Moreover, the opacity of these models can make it challenging to identify and mitigate potential biases or safety concerns, as the sources of these issues may be hidden within the complex internal workings of the model.

The purpose of RecSys from the users' standpoint is to personalise content to reduce information overload and improve user experience. In practice, this often means that the content with which an individual user interacts in a RecSys mediated environment will be different from that which is accessed by other users. For example, two users logging in to the same social media app will likely have different content appearing on their personalised news feed.

This feature of personalisation creates what Milano and Prunkl (2024) call an epistemically fragmented environment. In such an environment, users do not have shared information about a common context, but are each interacting with the system on an individual basis. This can have negative implications as it makes it more difficult to observe harmful patterns of recommendation (Milano *et al.*, 2021a). For example, a user who is negatively targeted by misinformation content might not become aware of this as the nature of the content has not been flagged or challenged by others in her community. In a different context, a user might be subject to negative price discrimination without realising that the recommendation algorithm is pushing for more expensive options based on their inferred characteristics.

Gen-RecSys further exacerbates these challenges. Unlike traditional RecSys that rely on explicit user-item interactions, generative models can create recommendations that do not directly stem from observed data. This can make it even more difficult to interpret the model's behavior and understand the factors that influence its recommendations.

To address these challenges, research in interpretable machine learning (Molnar *et al.*, 2020) and explainable AI (Dwivedi *et al.*, 2023) should make further progress. Explainable AI techniques such as saliency maps or counterfactual explanations can provide insights into the model's decision-making process, shedding light on the factors that contribute to specific recommendations. However, these techniques are often limited in their ability to explain complex, non-linear models, and may not be sufficient to fully address the opacity of Gen-RecSys (Sullivan and Kasirzadeh, 2024). Moving forward, a promising direction is developing more mechanistic interpretability techniques (Nanda *et al.*, 2023) that can effectively probe the underlying mechanisms of Gen-RecSys. This could involve leveraging techniques from causal inference to identify the causal relationships between input features and recommendations, or developing novel visualization methods that can effectively represent the complex decision boundaries learned by these models.

## 7.6 Fairness and societal bias

Research on fairness in machine learning and AI has intensified in recent years, with scholars documenting societal biases at all levels of the development pipeline: datasets, algorithms, and evaluation choices (Mehrabi *et al.*, 2021; Kasirzadeh and Klein, 2021; Madaio *et al.*, 2022). In the context of RecSys, defining fairness remains a debated issue, largely due to the nature of the recommendation task (Ekstrand *et al.*, 2021; Wu *et al.*, 2023c; Wang *et al.*, 2023f). Fairness in RecSys can be approached from two main perspectives: parity-based definitions, which focus on equalizing specific outcomes across different groups, and substantive notions, which consider broader concepts of justice and equity. The complexity of RecSys environments has led to various impossibility results, demonstrating that satisfying all desirable fairness criteria simultaneously is often infeasible. This has sparked further research into context-specific fairness definitions and trade-offs.

RecSys are inherently multistakeholder environments. While traditionally designed to predict user interactions, recommendations affect the interests of multiple parties (Milano *et al.*, 2021b; Abdollahpouri and Burke, 2021): users (receivers of recommendations), providers (e.g., sellers in e-commerce), system providers, and society at large. Each stakeholder may be affected differently by the system's behavior, necessitating a holistic approach to fairness and impact assessment (Deldjoo and Nazary, 2024; Deldjoo, 2024). This multistakeholder nature becomes even more pronounced in Gen-RecSys, where the system's ability to create novel content introduces additional complexities in balancing the interests of various stakeholders.

Diversity, closely related to fairness, is a key component in RecSys evaluation. Diversity can enhance the quality of recommendations by exposing users to a broader range of options. However, operationalizing diversity is challenging due to the multiple relevant interpretations in any given application. For example, in news recommender systems, diversity is crucial to ensure users are exposed to a broad range of information sources and opinions. Yet, measuring recommendation diversity remains a complex task without universally applicable solutions (Ada Lovelace Institute, 2022). Gen-RecSys add another layer of complexity to this issue, as they can not only select from existing items but also generate entirely new content, raising questions about how to ensure and measure diversity in generated recommendations.

A critical gap in the literature concerns the long-term impacts of Gen-RecSys on information diversity and societal fairness. While studies have started to examine the short-term effects of recommendation algorithms on user behavior (Chaney *et al.*, 2018), the potential for Gen-RecSys to create entirely unique information ecosystems for individual users remains rather underexplored. This challenge can raise questions about the applicability of traditional group fairness metrics in contexts where each user's experience is fundamentally distinct. As Gen-RecSys continue to evolve, several key research directions emerge as critical for addressing fairness and bias challenges.

First, developing novel fairness and diversity metrics specifically tailored to generative content is crucial (See Chapter 6 for a discussion of diversity metrics). These might include measures of demographic representation in generated content, semantic diversity across user experiences, and longitudinal fairness that captures how recommendations evolve over time. Second, attention should be given to intersectional fairness metrics that can capture the complex interplay of multiple protected attributes in generated recommendations (Foulds *et al.*, 2020). Third, researchers should work towards developing dynamic fairness measures that can adapt to changing user preferences, ensuring that Gen-RecSys remain aligned over time.

## 7.7 Filter bubbles and echo chambers

The potential for RecSys to facilitate the formation of filter bubbles and echo chambers has been a prominent concern since the proliferation of social media platforms. Filter bubbles, conceptualized as algorithmic structures that reinforce users' existing viewpoints by limiting exposure to diverse or contradictory content, have been hypothesized to contribute to increased polarization and decreased exposure to challenging ideas. (Pariser, 2011) originally hypothesised that social media feeds would insulate users, placing them in filter bubbles that limit their exposure to diversified viewpoints. Theoretical work studying the behaviour of traditional recommendation algorithms confirmed the tendency of these systems to give rise to malicious feedback loops (Jiang *et al.*, 2019), lending plausibility to the filter bubble hypothesis. Although empirical evidence is still the subject of discussion (Möller, 2021), the importance of recommender style and diversity has normative importance (Vrijenhoek *et al.*, 2021).

In contrast to filter bubbles, which limit the exposure of users to competing viewpoints (sometimes unintentionally), echo chambers are "social epistemic structure[s] from which other relevant voices have been actively excluded and discredited" (Nguyen, 2020). Empirical studies analysing data from social media have found evidence that recommender systems might be implicated in the creation of echo chambers. (Mønsted and Lehmann, 2022), for instance, found evidence that Twitter's recommender system drove the emergence of polarised factions functioning as echo chambers in the context of vaccine scepticism during the COVID-19 pandemic.

While the antidote to filter bubbles is to increase the diversity of recommendations, ensuring that users come into contact with options or opinions that may be different from what they originally entertained, the solution to the problem of echo chambers requires different strategies. Simply increasing diversity might, paradoxically, lead to the opposite effect of entrenching contrary views even further, as echo chambers provide ideologically charged spaces or 'epistemic bunkers' (Furman, 2023).

The next generation of Gen-RecSys have the potential to exacerbate the issues of filter bubbles and echo chambers, but could also present novel instruments to help combating them. For example, Gen-RecSys might exacerbate the risk of filter bubbles by expanding personalised content, generating new content on the fly that matches the predicted tastes and opinions of an individual user. If more of this type of personalised content becomes part of a user's media feed, it could proportionally crowd out other types of non-personalised content, making it more difficult for users to have a balanced picture of the content that is generally available.

Mis- and dis-information content produced by generative systems could also reinforce echo chambers, providing users with more arguments or (potentially fabricated) evidence to sustain ideologically motivated viewpoints. However, Gen-RecSys could also potentially be used to counteract these negative effects of recommendations. Personalised content could be used to break into echo chambers by tailoring the message in a way that helps to build common ground and depolarise, in keeping with the bridging approach proposed in (Stray, 2021).

## 7.8 Privacy

RecSys can deal with personal and sensitive data, raising privacy concerns (Müllner, 2023). Risks to privacy can come from different sources, including: i) data misuse – malicious use or dissemination of personal data; ii) data leaks – cyberattacks and de-anonymisation of user data, leaking of sensitive data for unauthorised purposes; iii) dangerous inferences – the recommender system behaviour reveals something that poses a privacy risk. This can occur, for example, when the recommendations given to a user could reveal information relative to their health or their political orientation to other users or observers.

Privacy concerns arise for traditional, non-generative Recommender Systems on three levels. Firstly, on the level of input data ingested by the system, data about users and item interactions can be sensitive and often subject to privacy regulations (such as EU GDPR). Secondly, on the level of processed or inferred data, the inferred user profiles or other characteristics associated to both users and items can encode private or sensitive information (Wachter and Mittelstadt, 2019). Thirdly, at the level of the system's output, the recommendations that are given to users can reveal, or be interpreted to reveal, potentially relevant sensitive data about them.

Generative models may introduce further concerns at all three levels (Slokom and Larson, 2021), but particularly at the level of inferred data, by expanding the range of inferred data and the ease of creating user profiles (see Chapter 4.2.3 – NL User Profiles). While these new techniques have potential to increase transparency (Radlinski *et al.*, 2022a), giving users a better control of their profiles and the data that is used for generating recommendations (e.g., "don't include my age in my profile"), they may also introduce privacy vulnerabilities as the increased transparency and interpretability could make them easier targets for potential attacks, and introduce bias in the recommendations.

## 7.9 Conclusion and future directions

We have discussed that Gen-RecSys amplifies and introduces novel societal risks that extend beyond those associated with traditional RecSys.

In particular, the capacity of Gen-RecSys to create entirely new content raises concerns about the propagation of misinformation, the reinforcement of harmful stereotypes, and the potential for manipulating user beliefs and behaviors. These systems can tailor persuasive narratives to individual users, introducing complex ethical questions about autonomy and the boundaries of acceptable influence. Moreover, the long-term exposure to highly personalized, generated content may have profound effects on users' beliefs, behaviors, and social interactions, potentially contributing to social fragmentation and the erosion of shared realities. At a broader level, Gen-RecSys may impact public discourse, social cohesion, and democratic processes in ways that are difficult to predict or control.

Evaluating these societal risks poses significant challenges. Developing content analysis tools that can assess the veracity and ethical implications of generated recommendations in real-time is a pressing need. Furthermore, measuring the degree of system influence on user agency requires new methodologies that can account for the subtle, cumulative effects of personalized content generation. The dynamic nature of Gen-RecSys complicates the application of static fairness metrics, necessitating adaptive measurement techniques that can evolve with the system and shifting user preferences. Longitudinal studies are crucial for understanding the long-term impacts of Gen-RecSys, but designing such studies presents methodological challenges. Traditional A/B testing may be insufficient to capture the cumulative impact of exposure to generated content over extended periods. Additionally, isolating the effects of Gen-RecSys from other influencing factors in users' information environments is particularly complex. Evaluating societal-level impacts, such as effects on public discourse and democratic processes, introduces further complications in establishing causal relationships and quantifying outcomes.

Addressing these evaluation challenges will require interdisciplinary collaboration, combining insights from computer science, ethics, philosophy, social psychology, and political science. Researchers must develop holistic assessment frameworks that can capture the multifaceted societal implications of Gen-RecSys, balancing the need for rigorous measurement with the social considerations of studying these powerful systems in real-world contexts.